\documentclass[review,authoryear,3p,times,10pt]{elsarticle}

\usepackage{multirow,setspace,times,amssymb,amsmath,graphicx,color,rotating,subfigure,url}
\usepackage{lineno,color}
\usepackage{natbib}
\usepackage{booktabs}%
\usepackage{longtable}%
\usepackage{rotating}
\usepackage{lscape}
\usepackage{ulem}
\usepackage{threeparttable}
\graphicspath{{Figures/}}
\usepackage[table]{xcolor}
\usepackage{tabularx}
\usepackage{graphicx} 
\usepackage{epstopdf}
\usepackage{mathrsfs}
\usepackage{makecell}
\usepackage[bookmarks=true,colorlinks,linkcolor=blue,anchorcolor=blue,citecolor=blue,unicode]{hyperref}
\usepackage{bookmark}

\usepackage{todonotes}
\usepackage{ragged2e}
\usepackage[font=small]{caption}

\hypersetup{CJKbookmarks=true}%
\makeatletter
\def\ps@pprintTitle{%
\let\@oddhead\@empty 
\let\@evenhead\@empty
}
\makeatother

\begin{document}

\begin{frontmatter}
\title{The impact of geopolitical risk on the international agricultural market: Empirical analysis based on the GJR-GARCH-MIDAS model} 

\author[SB]{Yun-Shi Dai}
\author[WHUT1,WHUT2]{Peng-Fei Dai}
\author[SB,RCE,DM]{Wei-Xing Zhou\corref{CorAuth}}
\ead{wxzhou@ecust.edu.cn}
\cortext[CorAuth]{Corresponding author.} 

\address[SB]{School of Business, East China University of Science and Technology, Shanghai 200237, China}
\address[WHUT1]{School of Management, Wuhan University of Technology, Wuhan 430070, China}
\address[WHUT2]{Research Institute of Digital Governance and Management Decision Innovation, Wuhan University of Technology, Wuhan 430070, China}
\address[RCE]{Research Center for Econophysics, East China University of Science and Technology, Shanghai 200237, China}
\address[DM]{School of Mathematics, East China University of Science and Technology, Shanghai 200237, China}

\begin{abstract}
The current international landscape is turbulent and unstable, with frequent outbreaks of geopolitical conflicts worldwide. Geopolitical risk has emerged as a significant threat to regional and global peace, stability, and economic prosperity, causing serious disruptions to the global food system and food security. Focusing on the international food market, this paper builds different dimensions of geopolitical risk measures based on the random matrix theory and constructs single- and two-factor GJR-GARCH-MIDAS models with fixed time span and rolling window, respectively, to investigate the impact of geopolitical risk on food market volatility. The findings indicate that modeling based on rolling window performs better in describing the overall volatility of the wheat, maize, soybean, and rice markets, and the two-factor models generally exhibit stronger explanatory power in most cases. In terms of short-term fluctuations, all four staple food markets demonstrate obvious volatility clustering and high volatility persistence, without significant asymmetry. Regarding long-term volatility, the realized volatility of wheat, maize, and soybean significantly exacerbates their long-run market volatility. Additionally, geopolitical risks of different dimensions show varying directions and degrees of effects in explaining the long-term market volatility of the four staple food commodities. This study contributes to the understanding of the macro-drivers of food market fluctuations, provides useful information for investment using agricultural futures, and offers valuable insights into maintaining the stable operation of food markets and safeguarding global food security.
\end{abstract}


\end{frontmatter}


\section{Introduction}

Food, being the most essential means of subsistence, supports the advancement of human society and prosperity. The issue of food security is therefore a major concern for both national economies and people's livelihoods, which has consistently garnered significant attention from the international community. However, the current food security situation is grim and complex, with global and local food security facing many threats and challenges. According to {\textit{The State of Food Security and Nutrition in the World 2023}}\footnote{\url{https://www.fao.org/publications/home/fao-flagship-publications}}, jointly published by the Food and Agriculture Organization of the United Nations (FAO), the International Fund for Agricultural Development (IFAD), and other international organizations, nearly 2.4 billion people, or 29.60\% of the world's population, were moderately or severely food insecure in 2022, with between 691 and 783 million facing hunger. Severe food insecurity is not caused by a single shock or hazard, but rather by a confluence of multiple factors, including conflict, extreme weather, and economic downturns. Among the diverse factors influencing global food security, persistent geopolitical conflicts are regarded as the foremost contributor to the ongoing exacerbation of the current global food crisis. The {\textit{Global Report on Food Crises 2023}}\footnote{\url{https://www.fsinplatform.org/global-report-food-crises-2023}} released by the Global Network Against Food Crises (GNAFC) also highlights that while economic shocks have had significant impacts on most countries experiencing food crises, and extreme weather events are important drivers of food insecurity in certain nations, conflict remains the primary cause of severe food insecurity for the majority of affected populations.

In recent years, the international political environment has been undergoing dynamic changes, marked by a proliferation of geopolitical conflicts. Geopolitical risk stands out as a main risk impacting international and regional peace and stability, and is widely acknowledged as a serious impediment to economic recovery and growth prospects. Geopolitical events such as the Sino-US trade war, the Russia-Ukraine conflict, and the Israel-Palestine conflict, as well as the economic sanctions and protectionist trade policies resulting from these events, not only pose threats to the international security system but also profoundly reshape the global political and economic landscape. With the continuous advancement of economic globalization and commodity financialization, increasingly frequent geopolitical events have further disrupted the commodity market. As the {\textit{World Economic Outlook: Navigating Global Divergences}}\footnote{\url{https://www.imf.org/en/Publications/WEO/Issues/2023/10/10/world-economic-outlook-october-2023}} issued by the International Monetary Fund (IMF) in 2023 points out, continued geopolitical tensions may exacerbate market fragmentation, and commodities are particularly vulnerable in the event of fragmentation due to their heavy production concentration, low elasticities of supply and demand, and sensitivity to geopolitics. Crucially, low-income countries with a high reliance on agricultural imports would face disproportionate impacts, intensifying fears of food insecurity. Furthermore, food prices, as a reflection of their value, have experienced drastic fluctuations in recent years. With the apparently growing global economic uncertainty, the factors influencing volatility in the international food market have become increasingly complicated.

The impact of adverse geopolitical events can be transmitted to the food market through multiple channels, as exemplified by the Russia-Ukraine conflict that broke out in 2022. Firstly, geopolitical conflicts have directly disrupted agricultural production in the countries involved and caused severe damage to arable land, agricultural equipment, and labour force, resulting in a significant decrease in agricultural productivity, output, and exports. Secondly, conflicts and sanctions, along with counter-sanctions, have reduced global fertilizer and energy supplies and driven up their prices, thereby increasing the production and transportation costs of agricultural products. Thirdly, conflict-induced border blockades and transportation disruptions have restricted the cross-regional flow of food and disrupted the food supply chain, which intensifies the tension between global food supply and demand. Moreover, due to geopolitical tensions, agricultural export restrictions aimed at lowering food prices and tackling domestic inflation have increased significantly, contributing to a resurgence of food trade protectionism and further exacerbating global food insecurity. In addition, the outbreak of geopolitical events has triggered financial market turmoil, with escalating capital speculation and rising risk-aversion sentiments, leading to frequent speculative behavior in the food futures market. Therefore, amid the increasing geopolitical risk, we focus on the international food market and try to elucidate the impact of geopolitical risk on food market volatility, which is crucial for understanding the drivers of food market volatility, maintaining the smooth operation of the food system, and ensuring global and local food security.

Using the MIDAS approach to deal with mixed-frequency data, we focus our research on wheat, maize, soybean, and rice to investigate the impact of geopolitical risk on the volatility of various international food submarkets. We first apply the random matrix theory (RMT) to build geopolitical risk indices of different dimensions, which are measured at the global level, across six major geographical regions, and for combinations of major producers, importers, and exporters. The findings suggest that the composite geopolitical risk index constructed by RMT is capable of effectively capturing changes in the overall geopolitical risk faced by multiple economies. Considering possible asymmetries in short-term fluctuations, we introduce the GJR-GARCH model into the MIDAS framework and construct single-factor GJR-GARCH-MIDAS models based on fixed time span and rolling window, respectively. In addition, we further develop two-factor GJR-GARCH-MIDAS models to simultaneously assess the impact of realized volatility and geopolitical risk on food market volatility. The results indicate that for all four staple crops, rolling-window-based modeling always yields better performance and thus enables a more accurate
description of market volatility. Furthermore, while there is no clear preference between single- and two-factor models, the latter generally demonstrates higher explanatory power for food market volatility in most cases. Moreover, in terms of short-term fluctuations, wheat, maize, soybean, and rice all show strong volatility clustering and persistence, without significant asymmetry. Concerning the long-run component of volatility, geopolitical risks of different dimensions exhibit varying directions and degrees of influence when used to explain the long-term volatility of different staple food markets.

The major contribution of our study can be summarized in the following three points. First, this paper further expands and enriches the existing literature on geopolitical risk. The world is currently undergoing unprecedented changes, with frequent geopolitical frictions and conflicts worldwide, which have exerted serious negative effects on international social stability and national economic development. Against this backdrop, scholars have begun to pay attention to the impact of geopolitical risk on financial markets. However, quantitative research on the impact of geopolitical risk on food market volatility is relatively limited, primarily due to the lack of reasonable geopolitical risk measures and the challenges associated with mixed-frequency data. Therefore, our paper can serve as a timely and valuable supplement to this topic. Second, to the best of our knowledge, this study is the first to focus on the food market, decompose its overall volatility into short- and long-term components, and examine the role of macro-factors in driving long-term volatility. Combining the random matrix theory and the GJR-GARCH-MIDAS model, we provide a comprehensive and in-depth investigation into the impact of geopolitical risk across multiple dimensions on the volatility of diverse food submarkets, which helps characterize and quantify the shocks of adverse geopolitical events. Third, our research has important implications for understanding the sources of food market volatility, stabilizing international and domestic food markets, as well as ensuring global and local food security. Furthermore, the findings of this paper also offer insights into using agricultural futures for risk diversification and asset allocation.

The remainder of this paper is structured as follows. Section~\ref{S1:LitRev} reviews the literature on volatility and geopolitical risk. Section~\ref{S1:Methodology} elaborates on the research methods employed, including the GJR-GARCH-MIDAS framework and random matrix theory. Section~\ref{S1:Data} introduces the data sources and provides the statistical description of the monthly geopolitical risk indices and daily international food prices. Section~\ref{S1:EmpAnal} presents the composite geopolitical risk indices across different dimensions, which are constructed by the random matrix theory, and discusses the empirical results of single-factor and two-factor GJR-GARCH-MIDAS models based on fixed time span and rolling window, respectively. Section~\ref{S1:Conclude} summarizes the paper and puts forward several suggestions.

\section{Literature review}
\label{S1:LitRev}

Modeling the volatility of financial assets has always been one of the most important research topics in financial theory and practice, especially in the fields of asset pricing and risk management \citep{Flannery-Protopapadakis-2002-RevFinancStud,Tse-Tsui-2002-JBusEconStat,Fu-2009-JFinancEcon,Girardi-Ergun-2013-JBankFinanc,Dai-Dai-Zhou-2023-JIntFinancMarkInstMoney}. Research on financial asset volatility can be traced back to the 1960s. \cite{Mandelbrot-1963-JBus, Mandelbrot-1967-JBus} reveals that the return series of financial assets are characterized by distributions with leptokurtosis, fat tail, and volatility clustering. \cite{Engle-1982-Econometrica} pioneers the autoregressive conditional heteroscedasticity (ARCH) model, which effectively describes the volatility clustering of financial time series. Considering the slow decay in AR parameter estimation, \cite{Bollerslev-1986-JEconom} adds the moving average term to the ARCH model and proposes the generalized autoregressive conditional heteroscedasticity (GARCH) model, which has been widely utilized and expanded due to its effectiveness in capturing the heteroscedasticity of financial returns. In order to examine the relationship between returns and risks, \cite{Engle-Lilien-Robins-1987-Econometrica} introduce conditional variance into the mean model and construct the ARCH in Mean (ARCH-M) model. Some scholars have noted that in financial markets, different information shocks often exhibit asymmetric effects, with bad news having a greater impact on volatility than good news, known as the leverage effect. \cite{Nelson-1991-Econometrica} proposes the exponential GARCH (EGARCH) model to explore the asymmetry of return volatility, relaxing the strict non-negativity assumption on parameters in the GARCH model. \cite{Glosten-Jagannathan-Runkle-1993-JFinanc} construct the GJR-GARCH model and verify its good performance in characterizing the leverage effect. \cite{Zakoian-1994-JEconDynControl} establishes the threshold GARCH (TGARCH) model to distinguish the impact of positive and negative information on volatility. After years of development and improvement, GARCH family models have become one of the mainstream methods for analyzing financial market volatility.

With the advancement of information technology and the diversification of statistical caliber, various frequencies of data have been further enriched and improved, which poses higher demands on econometric methods for mixed-frequency data. To leverage mixed-frequency data for research, the traditional practice is to homogenize data of different frequencies, such as aligning them to the lower frequency, and then analyze the processed data using relevant models. However, such data processing methods may result in the loss or omission of potential information in high-frequency data, thereby reducing the fitting effect of models and causing bias in estimated results. In this context, \cite{Ghysels-SantaClara-Valkanov-2005-JFinancEcon,Ghysels-SantaClara-Valkanov-2006-JEconom,Ghysels-Sinko-Valkanov-2007-EconomRev} propose the mixed data sampling (MIDAS) method and its extensions, the application of which offers a solution to the loss of useful information and the amplification of invalid information caused by the data preprocessing in traditional methods. Nevertheless, the MIDAS approach fails to accurately capture the stylized features of financial time series, including leptokurtosis, fat tail, and volatility clustering. Therefore, \cite{Engle-Ghysels-Sohn-2013-RevEconStat} develop the GARCH-MIDAS model by introducing the GARCH model into the MIDAS framework. The core idea of the GARCH-MIDAS method is to decompose volatility into short- and long-term components, allowing for separate modeling of each and incorporating low-frequency macro-variables when describing long-term volatility.

Given its ability to utilize information from data of different frequencies \citep{You-Liu-2020-JBankFinanc,Conrad-Schienle-2020-JBusEconStat,Francq-Kandji-Zakoian-2023-EconometTheory}, the GARCH-MIDAS model has been extensively applied to examine the impact of various low-frequency macro-variables on financial market volatility. \cite{Asgharian-Hou-Javed-2013-JForecast} discover that macroeconomic variables can enhance the predictive ability of the GARCH-MIDAS model for stock market volatility, particularly for the long-term component, a conclusion that is further supported by \cite{Pan-Bu-Liu-Wang-2020-QuantFinanc} and \cite{Fang-Lee-Su-2020-JEmpirFinanc}. \cite{Xu-Bo-Jiang-Liu-2019-Knowledge-BasedSyst} propose the MF-GARCH-MIDAS model to investigate the role of macroeconomic fundamentals and Google search index in predicting stock market volatility. \cite{Wang-Ma-Liu-Yang-2020-IntJForecast} make a series of extensions to the GARCH-MIDAS model to analyze asymmetry and extreme volatility effects. Focusing on the cryptocurrency market, \cite{Walther-Klein-Bouri-2019-JIntFinancMarkInstMoney} attempt to identify the most significant exogenous factors driving market volatility and find that global real economic activity outperforms the other factors examined, while \cite{Liang-Zhang-Li-Ma-2022-IntJFinancEcon} believe that the CBOE gold ETF volatility index exhibits stronger predictability than the other predictors. With the deepening of commodity financialization, the financial attributes of commodities have become increasingly prominent, making the commodity market an important research subject. \cite{Pan-Wang-Wu-Yin-2017-JEmpirFinanc} and \cite{Gong-Wang-Shao-2022-IntJFinancEcon} reach a consistent conclusion that macroeconomic fundamentals have strong explanatory power for oil price fluctuations. \cite{Mo-Gupta-Li-Singh-2018-EconModel} analyze the relationship between multiple macroeconomic factors and fluctuations of oil, metal, and agricultural futures, confirming the significant role played by macroeconomics in driving commodity futures volatility.

The world is currently undergoing unprecedented changes, with increasingly complex domestic and international environments marked by heightened levels of instability and uncertainty. The research on low-frequency factors in GARCH-MIDAS is no longer limited to macroeconomic variables, but extends to diverse uncertainties. \cite{Su-Fang-Yin-2017-EconLett} use the news implied volatility index to explore the effect of news uncertainty on U.S. financial market volatility. \cite{Li-Zhang-Li-2023-IntRevFinancAnal} pay attention to economic policy uncertainty (EPU) and emphasize its effectiveness as a predictor of stock market volatility, which is consistent with the findings of \cite{Salisu-Demirer-Gupta-2023-JForecast}. Based on the GARCH-MIDAS framework, \cite{Wu-Liu-2023-JEnvironManage} and \cite{Li-Bouri-Gupta-Fang-2023-JCleanProd} provide new insights into the linkage between climate policy uncertainty and volatility in green finance markets. \cite{Wei-Liu-Lai-Hu-2017-EnergyEcon} examine the roles of speculation, fundamentals, and uncertainties in predicting volatility in the crude oil market, demonstrating that EPU indices are major determinants of crude oil market volatility. \cite{Fang-Chen-Yu-Qian-2018-JFuturesMark}, \cite{Dai-Xiong-Zhang-Zhou-2022-ResourPolicy} and \cite{Raza-Masood-Benkraiem-Urom-2023-EnergyEcon} investigate the effects of global economic policy uncertainty on volatility in gold, crude oil, and precious metals markets, respectively, and prove the remarkable predictive power of EPU. \cite{Zhang-Hong-Ding-2023-Energy} employ a modified GARCH-MIDAS model to reveal the heterogeneous influences of climate policy uncertainty on fluctuations in crude oil and clean energy markets. \cite{Wang-Wu-Ma-Xu-2023-IntRevFinancAnal} analyze the performance of different weather variables in forecasting soybean market volatility, finding that weather indicators can provide valuable information for predicting soybean volatility.

In recent years, frequent geopolitical conflicts have brought considerable uncertainty to the global economy. Geopolitical risk, as a low-frequency macro-factor, has aroused wide concern among scholars, with its impact on various markets becoming a hot topic. \cite{Li-Ye-Bhuiyan-Huang-2023-JForecast} and \cite{Segnon-Gupta-Wilfling-2024-IntJForecast} construct extended GARCH-MIDAS models to investigate the role of geopolitical risk in forecasting stock market volatility. \cite{Liu-Ma-Tang-Zhang-2019-EnergyEcon} examine the predictive effect of geopolitical risk on oil volatility, highlighting its potential to provide valuable insights into oil fluctuations. \cite{Li-Wei-Chen-Ma-Liang-Chen-2022-IntJFinancEcon} endorse this opinion and emphasize the significant impact of geopolitical risk on oil market volatility. \cite{Liang-Ma-Wang-Zeng-2021-JForecast} compare the explanatory power of diverse uncertainty indices for the volatility of natural gas futures, showing stronger forecasting ability of geopolitical risk and stock market volatility index. \cite{Liu-Han-Xu-2021-IntRevFinancAnal} analyze the impact of geopolitical uncertainty on the volatility of energy commodity markets, and conclude that the noteworthy positive influence primarily transmits through the threat of adverse geopolitical events. Conversely, \cite{Zhang-Xiang-Zou-Guo-2024-IntRevFinancAnal} discover a significantly negative effect of geopolitical acts on the Chinese energy market. \cite{Gong-Xu-2022-EnergyEcon} explore the dynamic linkages among various commodity markets and further elucidate the influence of geopolitical risk on the interconnectedness between them. \cite{Abid-Dhaoui-Kaabia-Tarchella-2023-ResourPolicy} utilize the GARCH-MIDAS model to discuss the roles of geopolitical shocks on agricultural and energy prices, revealing that all commodity markets respond to geopolitical shocks, albeit with varying performance across different commodity types.

Food prices are the cornerstone of consumer prices. Stabilizing food prices is crucial for maintaining price stability, fostering economic development, and promoting social harmony. However, in recent years, the frequency and magnitude of food price volatility have notably increased due to various factors, including regional conflicts and extreme weather events. Recognizing the close relationship between agricultural commodities and geopolitical risk, a few studies on agricultural markets have begun to incorporate geopolitical factors into their analytical frameworks. Nevertheless, upon reviewing relevant literature, we find that most of the existing studies focus on examining the impact of specific adverse geopolitical events, such as the Russia-Ukraine conflict, on agricultural markets \citep{Zhou-Lu-Xu-Yan-Khu-Yang-Zhao-2023-ResourConservRecycl,Poertner-Lambrecht-Springmann-Bodirsky-Gaupp-Freund-LotzeCampen-Gabrysch-2022-OneEarth,Zhou-Dai-Duong-Dai-2024-JEconBehavOrgan}, or touching on agriculture products when exploring the influence of geopolitical risk on commodity markets \citep{Gong-Xu-2022-EnergyEcon,Abid-Dhaoui-Kaabia-Tarchella-2023-ResourPolicy}. In general, there is a notable lack of research specifically targeting the food market to quantitatively analyze the effects of geopolitical risk, and this is the gap that our paper attempts to fill. To achieve this, we adopt the GJR-GARCH-MIDAS framework, combined with the random matrix theory, to comprehensively and deeply investigate the impact of geopolitical risk across various dimensions on the global food market, so as to clarify the linkage between geopolitical risk and food market volatility.

\section{Methodology}
\label{S1:Methodology}

The geopolitical risk indices employed in this study are monthly data, while agricultural commodity prices are daily data. The use of traditional modeling methods for same-frequency data may lead to a potential loss of valuable information within high-frequency agricultural prices, resulting in bias in parameter estimation and volatility modeling. Therefore, we utilize the GARCH-MIDAS method proposed by \cite{Engle-Ghysels-Sohn-2013-RevEconStat}, and further develop the GJR-GARCH-MIDAS model to capture the effective information embedded in data of different frequencies.

\subsection{GJR-GARCH-MIDAS models with fixed time span} 

Since the low-frequency macro-variables are monthly data, let $t$ be fixed at the monthly frequency. For each agricultural commodity, its logarithmic return $r_{i,t}$ on day $i$ of month $t$ is defined as
\begin{subequations}
  \begin{equation}
    r_{i,t} = \mu + \sqrt{\tau_{t} g_{i,t}} \varepsilon_{i,t},\  i=1,\dots,N_{t},\ t=1,\dots,T,
    \label{Eq:function_return}
  \end{equation}
with
  \begin{equation}
    \varepsilon_{i,t} \big| \Omega_{i-1,t} \sim N(0,1),
    \label{Eq:function_error-term_normal}
  \end{equation}
  \label{Eq:function_GARCH-MIDAS_normal}%
\end{subequations}
where $N_{t}$ is the number of days in month $t$. $\Omega_{i-1,t}$ denotes the set of historical information available on day $i-1$ of month $t$, and the stochastic error term $\varepsilon_{i,t}$ obeys a standard normal distribution given $\Omega_{i-1,t}$. $\tau_{t}$ and $g_{i,t}$ denote the long- and short-term components of volatility, respectively.

The volatility of each agricultural commodity on day $i$ of month $t$, $\sigma_{i,t}^{2}$, can be decomposed into the product of the long-run component, $\tau_{t}$, and the short-run component, $g_{i,t}$, namely
\begin{equation}
    \sigma_{i,t}^{2} = \tau_{t} g_{i,t},
    \label{Eq:function_conditional_variance}
\end{equation}
where the short-term component $g_{i,t}$ is assumed to follow a GARCH(1,1) process \citep{Engle-Rangel-2008-RevFinancStud}. However, in order to take possible asymmetries in short-term fluctuations into account, we introduce GJR-GARCH into the model by supposing that the short-term component $g_{i,t}$ obeys a GJR-GARCH(1,1) process, which is defined as follows:
\begin{equation}
    g_{i,t} = \left( 1 - \alpha - \beta - 0.5\gamma \right) + \left( \alpha + 1_{\{r_{i-1,t}<0\}} \gamma \right) \times \frac{\left(r_{i-1,t}-\mu \right)^{2}}{\tau_{t}} + \beta g_{i-1,t},
    \label{Eq:function_short-term_component}
\end{equation}
where $\alpha > 0$, $\beta \geq 0$, and $\alpha + \beta + 0.5\gamma < 1$. The $1_{\{r_{i-1,t}<0\}}$ is the indicator function and takes the value of 1 when the return is negative and 0 otherwise.

Referring to \cite{Engle-Ghysels-Sohn-2013-RevEconStat}, \cite{Conrad-Loch-2015-JApplEconom} and \cite{Wei-Liu-Lai-Hu-2017-EnergyEcon}, we logarithmically transform the long-run component $\tau_{t}$ to ensure it remains positive. Given that $t$ is fixed at the monthly frequency, the $\tau_{t}$ measured by the realized volatility (RV) is denoted as
\begin{subequations}
\begin{equation}
   \log \tau_{t} = m + \theta \sum\limits_{k=1}^{K} \varphi_{k} \left(\omega_{1}, \omega_{2} \right) RV_{t-k},
    \label{Eq:function_log_long-term_component_RV_fixedspan}
\end{equation}
with
\begin{equation}
   RV_{t} = \sum\limits_{i=1}^{N_{t}} r_{i,t}^{2},
    \label{Eq:function_realized_volatility_fixedspan}
\end{equation}
  \label{Eq:function_long-term_component_RV_fixedspan}%
\end{subequations}
where $m$ and $K$ are the intercept term and the maximum lag order of realized volatility, respectively. $\theta$ represents the coefficient that captures the long-term impact of realized volatility on agricultural market volatility.

Considering that the beta lag based on the beta function is capable of flexibly describing different lag structures, the weighting function of the lagged variable in Eq. (\ref{Eq:function_log_long-term_component_RV_fixedspan}) is usually defined by the beta polynomial function, which can be expressed as
\begin{equation}
   \varphi_{k} \left(\omega_{1}, \omega_{2} \right) = \frac{\left(k/K \right)^{\omega_{1}-1} \left(1-k/K \right)^{\omega_{2}-1}}{\sum\nolimits_{j=1}^{K}\left(j/K \right)^{\omega_{1}-1}\left(1-j/K \right)^{\omega_{2}-1}}.
    \label{Eq:function_weight_beta}
\end{equation}
where all the weights sum to 1. Eqs. (\ref{Eq:function_GARCH-MIDAS_normal}) -- (\ref{Eq:function_weight_beta}) together form the GJR-GARCH-MIDAS model with fixed-span RV, where Eq. (\ref{Eq:function_long-term_component_RV_fixedspan}) is also known as the MIDAS filter.

Next, we replace RV with the geopolitical risk measure to directly incorporate geopolitical risk into the GJR-GARCH-MIDAS model, thereby examining the impact of geopolitical risk on volatility in agricultural markets. In this case, Eq. (\ref{Eq:function_log_long-term_component_RV_fixedspan}) is converted to
\begin{equation}
   \log \tau_{t} = m + \theta \sum\limits_{k=1}^{K} \varphi_{k} \left(\omega_{1}, \omega_{2} \right) MV_{t-k},
    \label{Eq:function_log_long-term_component_macrovariable_fixedspan}
\end{equation}
where $K$ is the maximum lag order, and $MV_{t-k}$ denotes the level of geopolitical risk lagged $k$ periods relative to the current period $t$. $\theta$ reflects the effect of geopolitical risk on the long-run volatility in agricultural markets. Eqs. (\ref{Eq:function_GARCH-MIDAS_normal}) -- (\ref{Eq:function_short-term_component}), Eq. (\ref{Eq:function_weight_beta}), and Eq. (\ref{Eq:function_log_long-term_component_macrovariable_fixedspan}) together form the single-factor GJR-GARCH-MIDAS model with a fixed-span macro-variable.

After analyzing RV and an individual macro-variable separately, we further construct the two-factor GJR-GARCH-MIDAS model to simultaneously explore the impact of realized volatility and geopolitical risk on market volatility. With both RV and a low-frequency macro-variable in the model, the long-term component $\tau_{t}$ is given by
\begin{equation}
  \log \tau_{t} = m + 
  \theta_{RV} \sum\limits_{k=1}^{K} \varphi_{k} \left(\omega_{1,RV}, \omega_{2,RV} \right) RV_{t-k} + \theta_{MV} \sum\limits_{k=1}^{K} \varphi_{k} \left(\omega_{1,MV}, \omega_{2,MV} \right) MV_{t-k},
    \label{Eq:function_log_long-term_component_multimacrovariable_fixedspan}
\end{equation}
Accordingly, Eqs. (\ref{Eq:function_GARCH-MIDAS_normal}) -- (\ref{Eq:function_short-term_component}), Eq. (\ref{Eq:function_weight_beta}), and Eq. (\ref{Eq:function_log_long-term_component_multimacrovariable_fixedspan}) together form the two-factor GJR-GARCH-MIDAS model with fixed-span RV and macro-variable.

\subsection{GJR-GARCH-MIDAS models with rolling window} 

By removing the constraint that the long-run volatility $\tau_{t}$ remains constant over the fixed time span $t$, we introduce the rolling window specification such that both the short-term component $g$ and the long-term component $\tau$ vary at the daily frequency. In this setting, the $\tau$ process with rolling-window RV is defined as
\begin{subequations}
\begin{equation}
   \log \tau_{i}^{(\text{rw})} = m^{(\text{rw})} + \theta^{(\text{rw})} \sum\limits_{k=1}^{K} \varphi_{k} \left(\omega_{1}, \omega_{2} \right) RV_{i-k}^{(\text{rw})},
    \label{Eq:function_log_long-term_component_RV_rollingwindow}
\end{equation}
with
\begin{equation}
   RV_{i}^{(\text{rw})} = \sum\limits_{j=1}^{N'} r_{i-j}^{2},
    \label{Eq:function_realized_volatility_rollingwindow}
\end{equation}
  \label{Eq:function_long-term_component_RV_rollingwindow}%
\end{subequations}
where $N'$ is the number of days in the rolling window. $N' = 22$ indicates that RV is calculated based on the monthly rolling window. Eqs. (\ref{Eq:function_GARCH-MIDAS_normal}) -- (\ref{Eq:function_short-term_component}), Eq. (\ref{Eq:function_weight_beta}), and Eq. (\ref{Eq:function_long-term_component_RV_rollingwindow}) together form the GJR-GARCH-MIDAS model with rolling-window RV.

Similarly, the long-term volatility $\tau$ calculated by the low-frequency macro-variable is denoted as
\begin{subequations}
\begin{equation}
   \log \tau_{i}^{(\text{rw})} = m^{(\text{rw})} + \theta^{(\text{rw})} \sum\limits_{k=1}^{K} \varphi_{k} \left(\omega_{1}, \omega_{2} \right) MV_{i-k}^{(\text{rw})},
    \label{Eq:function_log_long-term_component_macrovariable_rollingwindow}
\end{equation}
with
\begin{equation}
   MV_{i}^{(\text{rw})} = \frac{1}{N'} \sum\limits_{j=1}^{N'} MV_{i-j},
    \label{Eq:function_macrovariable_rollingwindow}
\end{equation}
  \label{Eq:function_long-term_component_macrovariable_rollingwindow}%
\end{subequations}
where $MV_{i}$ is a daily variable representing the level of the low-frequency macro-variable in the corresponding month. $MV_{i}^{(\text{rw})}$ denotes the level of the low-frequency macro-variable on day $i$ computed based on the rolling window, whose value is equal to the average of the daily values of the macro-variable within the rolling window of $N'$ days prior to day $i$. Eqs. (\ref{Eq:function_GARCH-MIDAS_normal}) -- (\ref{Eq:function_short-term_component}), Eq. (\ref{Eq:function_weight_beta}), and Eq. (\ref{Eq:function_long-term_component_macrovariable_rollingwindow}) together form the single-factor GJR-GARCH-MIDAS model with a rolling-window macro-variable.

The $\tau$ process based on both the rolling-window RV and low-frequency macro-variable is expressed as
\begin{equation}
  \log \tau_{i}^{(\text{rw})} = m^{(\text{rw})} + 
  \theta_{RV}^{(\text{rw})} \sum\limits_{k=1}^{K} \varphi_{k}\left(\omega_{1,RV}, \omega_{2,RV} \right) RV_{i-k}^{(\text{rw})} +  \theta_{MV}^{(\text{rw})} \sum\limits_{k=1}^{K} \varphi_{k} \left(\omega_{1,MV}, \omega_{2,MV} \right) MV_{i-k}^{(\text{rw})},
    \label{Eq:function_log_long-term_component_multimacrovariable_rollingwindow}
\end{equation}
Accordingly, Eqs. (\ref{Eq:function_GARCH-MIDAS_normal}) -- (\ref{Eq:function_short-term_component}), Eq. (\ref{Eq:function_weight_beta}), and Eq. (\ref{Eq:function_log_long-term_component_multimacrovariable_rollingwindow}) together form the two-factor GJR-GARCH-MIDAS model with rolling-window RV and macro-variable.

\subsection{Model estimation}

Generally, the impact of low-frequency macro-variables on long-term volatility diminishes as the lag period increases, which means that the larger the lag period, the smaller the influence of the macro-variables in the corresponding period on the volatility of the current period. Therefore, we impose a constraint on the weighting function $\varphi_{k}\left(\omega_{1}, \omega_{2} \right)$ by setting $\omega_{1}=1$, and determine the weighting coefficients of the low-frequency variables in each period only through the parameter $\omega_{2}$, ensuring that the weights of lagged terms decay. Consequently, Eq. (\ref{Eq:function_weight_beta}) can be simplified to
\begin{equation}
   \varphi_{k} \left(\omega_{2} \right) = \frac{\left(1-k/K \right)^{\omega_{2}-1}}{\sum\nolimits_{j=1}^{K}\left(1-j/K \right)^{\omega_{2}-1}}.
    \label{Eq:function_weight_beta_simplification}
\end{equation}

In addition, we introduce the variance ratio (VR) to measure the extent to which realized volatility and geopolitical risk contribute to the volatility of agricultural markets, respectively. The expression for VR is as follows:
\begin{equation}
   VR = \frac{\text{Var} \left( \log (\tau_{t}) \right)} {\text{Var} \left( \log(\tau_{t}g_{t}) \right)},
    \label{Eq:function_variance_ratio}
\end{equation}
where $g_{t}=\sum\nolimits_{i=1}^{N_{t}}g_{i,t}$. VR measures the explanatory power of different factors on the overall conditional variance and their relative importance.

Parameter estimation is performed by minimizing the negative log-likelihood function, namely $-$LLF, with the log-likelihood function for the GJR-GARCH-MIDAS model given by
\begin{equation}
    LLF \left( \Phi \right) = - \frac{1}{2} \sum\limits_{t=1}^{T} \left( \log \left(2\pi \right) + \log \tau_{t} \left( \Phi \right) g_{t} \left( \Phi \right) + \frac{ \left( r_{t}-\mu \right)^{2} }{\tau_{t} \left( \Phi \right) g_{t} \left( \Phi \right) } \right) 
    \label{Eq:Log_likelihood_density_function_normal}
\end{equation}
where the parameter set $\Phi = \left( \mu, \alpha, \beta, \gamma, m, \theta, \omega_{1}, \omega_{2} \right)$.

According to \cite{Engle-Ghysels-Sohn-2013-RevEconStat} and \cite{Asgharian-Hou-Javed-2013-JForecast}, setting the maximum lag order $K=36$ in modeling the long-term component yields favorable performance when the high-frequency variables are daily and the low-frequency variables are monthly. Furthermore, as noted by \cite{Conrad-Kleen-2020-JApplEconom}, measurement errors in the explanatory variables or misspecification of lag structures only have a minor effect on estimation. Therefore, drawing upon existing literature, we set the maximum lag order $K$ to 36 and employ the Akaike information criterion (AIC) and Bayesian information criterion (BIC) for a comprehensive assessment of the models' goodness-of-fit.

\subsection{Random matrix theory}

Referring to \cite{Plerou-Gopikrishnan-Rosenow-Amaral-Guhr-Stanley-2002-PhysRevE} and \cite{Dai-Xiong-Zhou-2021-FinancResLett}, we apply the random matrix theory to construct composite geopolitical risk indices for various combinations of economies. This approach aims to facilitate further research into the diverse effects of geopolitical risks of different dimensions.

Considering that the economy-specific geopolitical risk (GPRE) indices exhibit varying levels of volatility, we first standardize each index and then calculate the correlation coefficient $c_{xy}$ between the GPRE series of economy $x$ and economy $y$, which is given by
\begin{subequations}
\begin{equation}
    gpre_{x}(t) = \frac{GPRE_{x}(t) - \langle GPRE_{x} \rangle}{\sigma_{x}}
    \label{Eq:Standardization_GPRE}
\end{equation}
\begin{equation}
   c_{xy} = \left\langle gpre_{x}(t) \times gpre_{y}(t) \right\rangle,\ x, y=1, \cdots, N
   \label{Eq:Correlation_coefficient}
\end{equation}
  \label{Eq:function_GPRE_correlation_coefficient}%
\end{subequations}
where $N$ denotes the number of economies, and $\langle \cdots \rangle$ represents the average of a given time series. The standard deviation of $GPRE_{x}$ is $\sigma_{x} = \sqrt{\langle GPRE_{x}^{2} \rangle - \langle GPRE_{x} \rangle ^{2}}$. By definition, the correlation coefficient $c_{xy}$ takes values in the range $[-1,1]$, where $c_{xy}=1$ and $c_{xy}=-1$ indicate perfect positive and perfect negative correlation, respectively, and $c_{xy}=0$ corresponds to no correlation between $GPRE_{x}$ and $GPRE_{y}$.

For the correlation matrix $\mathbf{C}$ of geopolitical risk indices, we compute its eigenvalues based on the following equation:
\begin{equation}
   \mathbf{C} = \mathbf{U \Lambda U}^{\mathrm{T}},
   \label{Eq:Eigenvalue}
\end{equation}
where $\mathbf{U}$ is the $N \times N$ eigenvector matrix consisting of $N$ eigenvectors of $1 \times N$. $\mathbf{\Lambda}$ is the diagonal matrix containing the corresponding eigenvalues, and $\mathbf{U}^{\mathrm{T}}$ is the transpose of $\mathbf{U}$. With $N$ eigenvalues of the correlation matrix $\mathbf{C}$, we arrange them in descending order.

Finally, we construct a composite geopolitical risk index based on the eigenportfolio of the geopolitical risk indices of $N$ economies, whose value at time $t$ is obtained by
\begin{equation}
   GPR_{\text{RMT}}(t) = {\sum\limits_{n=1}^{N}u_{n}^{1}GPRE_{n}(t)} \Bigg/ {\sum\limits_{n=1}^{N}u_{n}^{1}}
   \label{Eq:Eigenportfolio}
\end{equation}
where $u_{n}^{1}$ denotes the $n$-th element of the eigenvector $\mathbf{u}^{1}$, and $\sum\limits_{n=1}^{N}u_{n}^{1}GPRE_{n}(t)$ is the projection of the geopolitical risk index $GPRE_{n}(t)$ of economy $n$ onto the eigenvector $\mathbf{u}^{1}$.

\section{Data description}
\label{S1:Data}

\subsection{Data sources}

We select wheat, maize, soybean, and rice as representatives of agricultural commodities. These four major staple crops provide essential energy and nutrients for human life activities, thus holding significant importance in the international agricultural market. Corresponding to the four staple food crops, we collect the daily closing prices of continuous contracts for wheat, corn, soybean, and rough rice futures traded on the Chicago Board of Trade (CBOT) through the Wind database. Given that the U.S. dollar serves as the primary pricing and settlement currency for international commodity trading and CBOT has global pricing power for agricultural commodities, the prices of CBOT agricultural futures are widely regarded as the most authoritative indicators, effectively reflecting trends in the international food market. 

Additionally, we utilize the geopolitical risk (GPR) index constructed by \cite{Caldara-Iacoviello-2022-AmEconRev}, along with two sub-indices, geopolitical threats (GPT) and geopolitical acts (GPA), as measures of global geopolitical risk. The GPR and its sub-indices are monthly indicators derived from automated text searches of 10 major newspapers, the values of which are calculated by the proportion of articles associated with adverse geopolitical events. The GPR index measures threats, realizations, and escalations of adverse events related to wars, terrorism, and tensions affecting world peace. The GPA and GPT indices quantify the realization or escalation of current adverse events and the anticipation or threat of future adverse events, respectively, distinguishing between direct negative consequences and potential risk implications of adverse geopolitical events.

Moreover, \cite{Caldara-Iacoviello-2022-AmEconRev} also develop individual GPR indices for 44 major economies to further assess the level of geopolitical risk specific to each economy. Overall, these indices have been widely applied in existing research and are considered effective in measuring the magnitude and variations of geopolitical risk \citep{Baumeister-Korobilis-Lee-2022-RevEconStat, Ahmed-Hasan-Kamal-2022-EurFinancManag, Lorente-Mohammed-CifuentesFaura-Shahzad-2023-RenewEnergy}. Regarding the sample period, we obtain monthly data for the GPR index and its sub-indices from January 2000 to October 2023, along with daily data for agricultural futures prices from January 3, 2000, to October 31, 2023, for empirical analysis.

\subsection{Statistical description}

\subsubsection{Agricultural futures}

\begin{figure}[!h]
  \centering
  \includegraphics[width=0.65\linewidth]{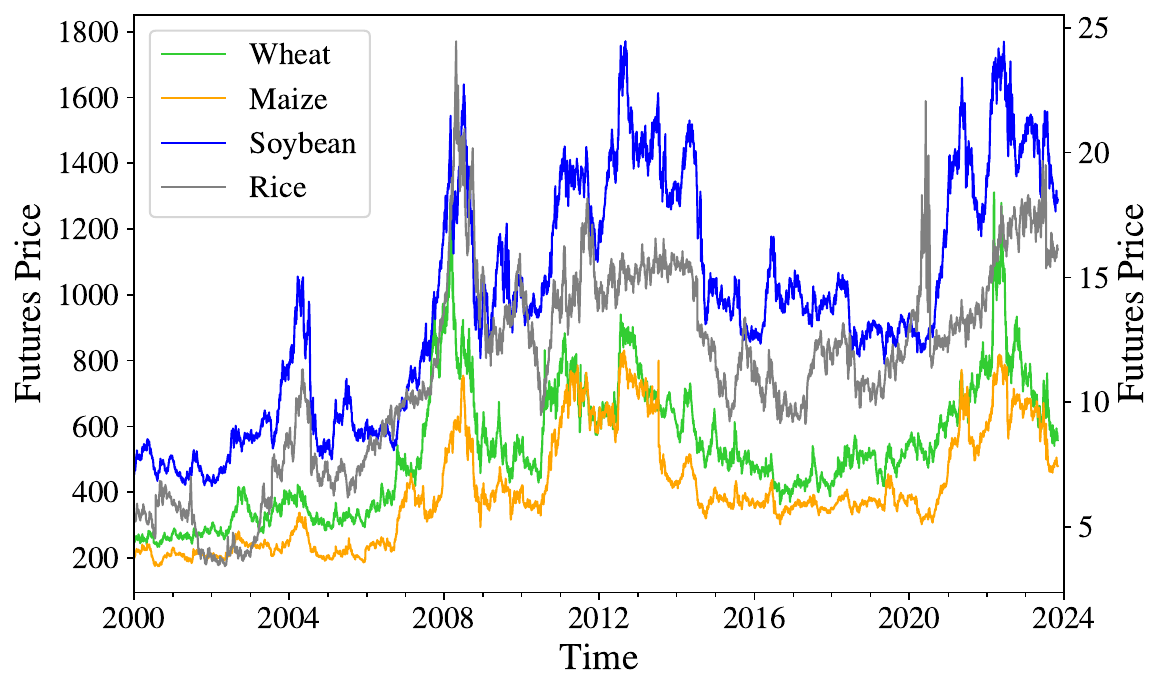}
  \caption{Evolution of futures prices for wheat, maize, soybean (left scale), and rice (right scale).}
\label{Fig:AgroPrice_evolution}
\end{figure}

Figure~\ref{Fig:AgroPrice_evolution} depicts the price movements of wheat, maize, soybean, and rice over the sample period, respectively. As can be seen, these four staple grains have experienced sudden and drastic price fluctuations during the three global food crises in the 21st century, namely during the periods of 2006--2008, 2010--2012, and from 2020 to the present.

Between 2006 and 2008, there was a surge in extreme weather events worldwide, such as Hurricane Katrina and heat waves in the United States of America, floods and extratropical storms in Europe, and massive floods in India. These extreme weather events severely affected grain production in the main grain-producing areas around the world, leading to a sharp increase in food prices. Additionally, the rise in energy prices and the development of biofuels further drove up the prices of food crops through substitution effects and cost channels. Moreover, changes in agricultural trade policies and market speculation also played a role in pushing up food prices. For instance, the Argentine government substantially increased export tariffs on soybeans and wheat in November 2007 to reduce food exports and maintain domestic food security.

During the period from 2010 to 2012, climate anomalies resulted in widespread reductions in food production, with the United States suffering blizzards and tornadoes, Europe encountering the fierce storm Xynthia, and India experiencing extreme heat and drought. In addition, regional conflicts, notably the Syrian Civil War that broke out in 2011, severely disrupted agricultural production in Syria, a major grain producer in West Asia, and reduced the global food supply to some extent. In anticipation of rising agricultural prices, several countries implemented relevant trade policies to restrict agricultural exports and safeguard domestic food security. For example, Russia announced a ban on wheat exports in August 2010, and Ukraine initiated grain export quotas starting in October 2010 and completely banned wheat exports in November 2012. These trade protection measures escalated supply shortages in the international food market and further drove up food prices.

From 2020 to the present, global warming and the prolonged La Niña phenomenon have led to frequent occurrences of extreme weather events worldwide. For instance, wildfires have erupted multiple times in Australia, Brazil, and Argentina, while Europe, India, and China have faced continuous heavy rainfall and flooding, along with record-breaking high temperatures and droughts in various regions. The reduced food production in major producing areas due to extreme weather has led to increased import demand, pushing up international grain prices as supply struggles to meet demand. Additionally, the supply chain disruption caused by the COVID-19 pandemic and the uneven global economic recovery have also been significant factors accounting for the rise in food prices. Furthermore, the ongoing Russia-Ukraine conflict, as a conflict between two major agricultural powers, has widened the gap between food supply and demand, contributed to volatile food prices, and heightened the current global food crisis.

Based on the daily prices of agricultural futures, we compute the logarithmic returns of agricultural commodity $q$ at time $i$ on time scale $\Delta i$ as follows:
\begin{equation}
  r_{q}(i) = \ln P_{q}(i) - \ln P_{q}(i - \Delta i)
  \label{Eq:Logarithmic_return}
\end{equation}
where $P_{q}(i)$ denotes the futures price of agricultural commodity $q$ at time $i$, and the time scale $\Delta i$ is 1 day.

Table~\ref{Tab:Agro_Stat_Test} presents the descriptive statistics of the return series for wheat, maize, soybean, and rice, respectively. It is observed that the means (Mean) of the return series for these four staple food crops are much smaller than their corresponding standard deviations (Std. Dev.), indicating that agricultural returns are widely dispersed and these agricultural futures markets exhibit significant volatility. Besides, the skewness (Skew.) of the return series for wheat and rice both exceed 0, suggesting right-skewed distributions of returns, while the skewness of the returns for maize and soybean are both less than 0, implying that their return series are left-skewed. This indicates that positive returns are more common in the wheat and rice markets, whereas negative returns are more prevalent in the maize and soybean markets. Moreover, the kurtosis (Kurt.) of each return series is notably higher than the kurtosis of a normal distribution, which is 3, verifying the presence of leptokurtosis and fat tail. Table~\ref{Tab:Agro_Stat_Test} further provides the test results of the agricultural return series, including diagnostic tests for normality, unit roots, white noise, and ARCH effects. The significantly positive J-B statistics for the Jarque-Bera test reflect that none of the return series follows a normal distribution. Meanwhile, the statistics for ADF and PP tests are all significant at the level of 1\%, demonstrating that these return series are stationary. The L-B and L-B$^{2}$ statistics represent the results of the Ljung-Box test for the return series and the squared return series, respectively, the statistical significance of which means that all return series are not white noise and there exist serial autocorrelation and conditional heteroscedasticity in each return series. Additionally, significant ARCH-LM statistics further confirm the presence of ARCH effects in these agricultural return series. Therefore, GARCH-class models can be constructed for the return series of wheat, maize, soybean, and rice for further research.

\begin{table}[!ht]
  \centering
  \setlength{\abovecaptionskip}{0pt}
  \setlength{\belowcaptionskip}{10pt}
  \caption{Descriptive statistics and diagnostic tests for agricultural return series}
  \setlength\tabcolsep{3pt}   \resizebox{\textwidth}{!}{ 
    \begin{tabular}{l c r@{.}l r@{.}l r@{.}l r@{.}l r@{.}l r@{.}l c r@{.}l r@{.}l r@{.}l r@{.}l r@{.}l r@{.}l}
    \toprule
         & \multicolumn{13}{c}{Descriptive statistics} && \multicolumn{12}{c}{Diagnostic tests}  \\
    \cline{2-14} \cline{16-27}
         & Obs. & \multicolumn{2}{c}{Max} & \multicolumn{2}{c}{Min} & \multicolumn{2}{c}{Mean} & \multicolumn{2}{c}{Std. Dev.} & \multicolumn{2}{c}{Skew.} & \multicolumn{2}{c}{Kurt.} && \multicolumn{2}{c}{J-B} & \multicolumn{2}{c}{ADF} & \multicolumn{2}{c}{PP} & \multicolumn{2}{c}{L-B} & \multicolumn{2}{c}{L-B$^{2}$} & \multicolumn{2}{c}{ARCH-LM}  \\
    \midrule
    Wheat & 6003 & 0&2581  &  $-$0&2289  & 0&0001  & 0&0210  & 0&1063  & 13&8397  && 29400&97$^{***}$  &  $-$18&78$^{***}$  &  $-$79&62$^{***}$  & 37&06$^{**}$  & 968&81$^{***}$  & 657&25$^{***}$ \\
    Maize & 6003 & 0&1356  &  $-$0&3986  & 0&0001  & 0&0187  &  $-$2&0343  & 45&1028  && 447523&23$^{***}$  &  $-$17&19$^{***}$  &  $-$77&30$^{***}$  & 29&11$^{*}$  & 40&40$^{***}$  & 36&92$^{**}$ \\
    Soybean & 6003 & 0&0763  &  $-$0&1583  & 0&0002  & 0&0156  &  $-$1&0302  & 10&9915  && 17035&78$^{***}$  &  $-$17&64$^{***}$  &  $-$77&03$^{***}$  & 31&86$^{**}$  & 672&64$^{***}$  & 342&63$^{***}$ \\
    Rice & 6003 & 0&3151  &  $-$0&3011  & 0&0002  & 0&0182  & 0&1155  & 50&1902  && 557021&27$^{***}$  &  $-$19&56$^{***}$  &  $-$73&58$^{***}$  & 57&56$^{***}$  & 741&01$^{***}$  & 651&35$^{***}$ \\
  \bottomrule
    \end{tabular}
    }%
  \begin{flushleft}
    \footnotesize
\justifying Note: This table presents the results of descriptive statistics and diagnostic tests for the return series of wheat, maize, soybean, and rice. The J-B statistic refers to the Jarque-Bera test, the significance of which indicates the non-normality of the series. The ADF and PP tests are unit-root tests, whose null hypothesis is that the series is non-stationary. The statistics of the Ljung-Box test for the series of returns and squared returns are denoted by L-B and L-B$^{2}$, respectively. The ARCH-LM statistic reflects the presence or absence of the ARCH effect. *, **, and *** represent statistical significance at the 10\%, 5\%, and 1\% levels, respectively.
\end{flushleft} 
  \label{Tab:Agro_Stat_Test}%
\end{table}%

\subsubsection{Geopolitical risk indices}

\begin{figure}[!h]
  \centering
  \includegraphics[width=0.65\linewidth]{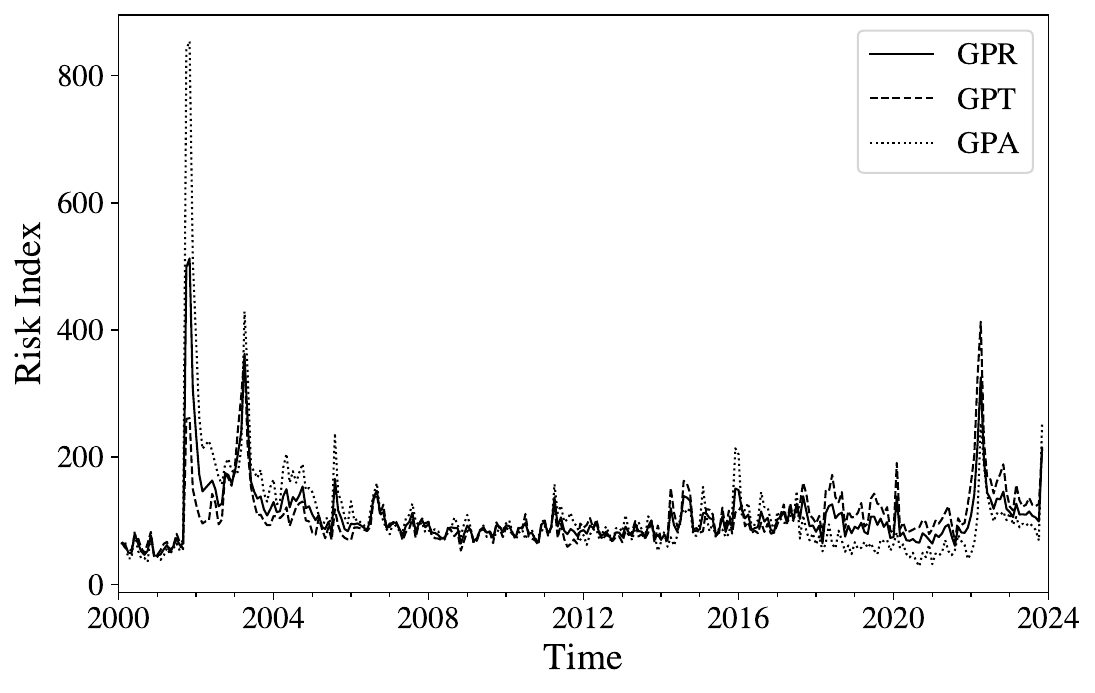}
  \caption{Evolution of global geopolitical risk (GPR), geopolitical threats (GPT), and geopolitical acts (GPA) indices.}
\label{Fig:GPRIndex_evolution}
\end{figure}

Figure~\ref{Fig:GPRIndex_evolution} describes the evolution of the GPR, GPT, and GPA indices over the period from January 2000 to October 2023, where the correlation coefficient between GPT and GPA over the entire sample period is 0.53. Comparison also reveals that the overall trends of the three indices are generally similar, but significant differences in their numerical levels are evident for specific events or during certain periods. For instance, following the September 11 terrorist attacks in 2001, the GPA index reached its peak, surpassing both the GPR and GPT indices by a considerable margin. Between 2018 and 2022, the GPA index remained at a relatively low level, while the GPT index experienced wide swings with relatively higher values. Moreover, as illustrated in Figure~\ref{Fig:GPRIndex_evolution}, the geopolitical risk faced by the world has undergone considerable changes, which correspond to some major international events. Specifically, the GPR index skyrocketed and quickly peaked after the 9/11 attacks in 2001, followed by a decline and another spike during the Iraq war in 2003. Subsequently, heightened levels of GPR were observed after the military intervention in Libya in 2011, Russia's annexation of Crimea around 2014, and the Paris terrorist attacks in 2015. In addition, the outbreak of the Russia-Ukraine conflict in 2022 led to another surge in the GPR index, which maintained relatively high levels until October 2023. Currently, the world is facing significant geopolitical risks on a global scale.

Table~\ref{Tab:GPRE_Stat} presents detailed information on the three global geopolitical risk indices and the individual geopolitical risk indices of 44 major economies, such as their corresponding labels (Label), abbreviated names (Abbr.), names of economies (Economies), and geographic regions (Region). Descriptive statistics for these risk indices are further reported, including the number of observations, maximum, minimum, mean, standard deviation, skewness, and kurtosis. As can be seen, global geopolitical risk indices significantly exceed economy-specific geopolitical risk indices in terms of level and volatility. Besides, different economies face varying levels of geopolitical risk, with the United States displaying the highest values across multiple statistical indicators, suggesting a substantial degree and scope of geopolitical risk concerning the country. Furthermore, countries such as Russia, Ukraine, and the United Kingdom also face considerable geopolitical risk. Moreover, both global and individual geopolitical risk indices exhibit positive skewness and kurtosis significantly larger than the corresponding value of 3 for a Gaussian normal distribution, which means that their distributions are right-skewed, leptokurtic, and fat-tailed.

\begin{table}[!ht]
  \centering
  \setlength{\abovecaptionskip}{0pt}
  \setlength{\belowcaptionskip}{10pt}
  \caption{Descriptive statistics of global and economy-specific geopolitical risk indices}
  \setlength\tabcolsep{3pt}   \resizebox{\textwidth}{!}{ 
    \begin{tabular}{ccccc r@{.}l r@{.}l r@{.}l r@{.}l r@{.}l r@{.}l}
    \toprule
     {Label} &  {Abbr.} & {Economies} & {Region} & {Obs.} & \multicolumn{2}{c}{Max} &  \multicolumn{2}{c}{Min} &  \multicolumn{2}{c}{Mean} &  \multicolumn{2}{c}{Std. Dev.} &  \multicolumn{2}{c}{Skew.} &  \multicolumn{2}{c}{Kurt.} \\
    \midrule
    1 & GPR & Global & Global & 286 & 512&5297  & 45&0606  & 105&4683  & 52&3943  & 4&2686  & 28&7064 \\
    2 & GPT & Global & Global & 286 & 412&9059  & 44&3577  & 105&2730  & 45&7425  & 2&9428  & 15&8275 \\
    3 & GPA & Global & Global & 286 & 854&0750  & 28&4546  & 107&6295  & 84&9558  & 5&4866  & 43&8773 \\
    4 & GPR\_ARG & Argentina & South America & 286 & 0&2628  & 0&0000  & 0&0276  & 0&0277  & 3&6343  & 23&8274 \\
    5 & GPR\_AUS & Australia & Asia and Oceania & 286 & 0&5275  & 0&0126  & 0&1055  & 0&0742  & 2&1272  & 9&4899 \\
    6 & GPR\_BEL & Belgium & Europe (South and West) & 286 & 1&0352  & 0&0185  & 0&1605  & 0&1455  & 2&9398  & 14&4339 \\
    7 & GPR\_BRA & Brazil & South America & 286 & 0&2252  & 0&0049  & 0&0536  & 0&0377  & 1&9077  & 7&5766 \\
    8 & GPR\_CAN & Canada & North America & 286 & 1&7243  & 0&0565  & 0&2229  & 0&1730  & 4&6838  & 34&8634 \\
    9 & GPR\_CHE & Switzerland & Europe (South and West) & 286 & 0&5102  & 0&0042  & 0&0602  & 0&0590  & 4&2860  & 28&5662 \\
    10 & GPR\_CHL & Chile & South America & 286 & 0&2855  & 0&0000  & 0&0164  & 0&0224  & 7&1593  & 78&3492 \\
    11 & GPR\_CHN & China & Asia and Oceania & 286 & 2&5631  & 0&1613  & 0&5561  & 0&3133  & 1&8245  & 8&7863 \\
    12 & GPR\_COL & Colombia & South America & 286 & 0&3352  & 0&0000  & 0&0394  & 0&0379  & 3&2277  & 18&9626 \\
    13 & GPR\_DEU & Germany & Europe (South and West) & 286 & 2&7472  & 0&0868  & 0&4258  & 0&3177  & 3&4846  & 21&4383 \\
    14 & GPR\_DNK & Denmark & Europe (North and East) & 286 & 0&1706  & 0&0000  & 0&0354  & 0&0291  & 1&9893  & 7&9960 \\
    15 & GPR\_EGY & Egypt & Middle East and Africa & 286 & 1&5983  & 0&0140  & 0&1845  & 0&1589  & 4&0018  & 28&2746 \\
    16 & GPR\_ESP & Spain & Europe (South and West) & 286 & 1&1496  & 0&0173  & 0&1068  & 0&1113  & 5&2924  & 39&6342 \\
    17 & GPR\_FIN & Finland & Europe (North and East) & 286 & 0&5505  & 0&0000  & 0&0371  & 0&0652  & 5&1961  & 34&6965 \\
    18 & GPR\_FRA & France & Europe (South and West) & 286 & 2&7991  & 0&1698  & 0&5438  & 0&3393  & 3&0010  & 15&4443 \\
    19 & GPR\_GBR & United Kingdom & Europe (North and East) & 286 & 5&9946  & 0&4041  & 1&1202  & 0&6996  & 4&0634  & 24&7621 \\
    20 & GPR\_HKG & Hong Kong & Asia and Oceania & 286 & 0&4816  & 0&0000  & 0&0609  & 0&0729  & 2&7360  & 11&8635 \\
    21 & GPR\_HUN & Hungary & Europe (North and East) & 286 & 0&5873  & 0&0000  & 0&0318  & 0&0553  & 5&8451  & 47&8761 \\
    22 & GPR\_IDN & Indonesia & Asia and Oceania & 286 & 0&5101  & 0&0000  & 0&0527  & 0&0541  & 3&8349  & 26&0449 \\
    23 & GPR\_IND & India & Asia and Oceania & 286 & 0&9460  & 0&0640  & 0&2211  & 0&1319  & 2&9287  & 14&7033 \\
    24 & GPR\_ISR & Israel & Middle East and Africa & 286 & 4&2877  & 0&0541  & 0&4006  & 0&3523  & 5&7080  & 55&5876 \\
    25 & GPR\_ITA & Italy & Europe (South and West) & 286 & 0&6619  & 0&0281  & 0&1450  & 0&0975  & 2&2497  & 9&7905 \\
    26 & GPR\_JPN & Japan & Asia and Oceania & 286 & 1&2368  & 0&0587  & 0&2410  & 0&1678  & 2&2132  & 9&5600 \\
    27 & GPR\_KOR & South Korea & Asia and Oceania & 286 & 1&8159  & 0&0649  & 0&3129  & 0&2480  & 2&5829  & 11&9941 \\
    28 & GPR\_MEX & Mexico & North America & 286 & 0&4574  & 0&0162  & 0&0983  & 0&0624  & 2&4625  & 11&6776 \\
    29 & GPR\_MYS & Malaysia & Asia and Oceania & 286 & 0&9728  & 0&0000  & 0&0403  & 0&0660  & 10&3286  & 141&4349 \\
    30 & GPR\_NLD & Netherlands & Europe (South and West) & 286 & 0&4472  & 0&0105  & 0&0837  & 0&0588  & 2&8285  & 14&4776 \\
    31 & GPR\_NOR & Norway & Europe (North and East) & 286 & 0&4877  & 0&0037  & 0&0533  & 0&0478  & 3&9584  & 28&8192 \\
    32 & GPR\_PER & Peru & South America & 286 & 0&1275  & 0&0000  & 0&0197  & 0&0208  & 2&5767  & 11&4280 \\
    33 & GPR\_PHL & Philippines & Asia and Oceania & 286 & 0&2624  & 0&0000  & 0&0456  & 0&0430  & 2&2835  & 8&8994 \\
    34 & GPR\_POL & Poland & Europe (North and East) & 286 & 2&1201  & 0&0038  & 0&0953  & 0&1797  & 7&0116  & 67&2361 \\
    35 & GPR\_PRT & Portugal & Europe (South and West) & 286 & 0&2446  & 0&0000  & 0&0224  & 0&0229  & 4&3820  & 36&3819 \\
    36 & GPR\_RUS & Russia & Europe (North and East) & 286 & 8&9733  & 0&2178  & 0&8516  & 0&8321  & 4&8562  & 38&8348 \\
    37 & GPR\_SAU & Saudi Arabia & Middle East and Africa & 286 & 1&4385  & 0&0286  & 0&2297  & 0&1798  & 2&4140  & 12&1207 \\
    38 & GPR\_SWE & Sweden & Europe (North and East) & 286 & 0&5505  & 0&0065  & 0&0558  & 0&0590  & 4&0522  & 26&7578 \\
    39 & GPR\_THA & Thailand & Asia and Oceania & 286 & 0&1941  & 0&0000  & 0&0398  & 0&0316  & 1&8378  & 7&2249 \\
    40 & GPR\_TUN & Tunisia & Middle East and Africa & 286 & 0&4758  & 0&0000  & 0&0328  & 0&0575  & 4&6407  & 29&8402 \\
    41 & GPR\_TUR & Turkey & Middle East and Africa & 286 & 1&2001  & 0&0200  & 0&2559  & 0&1857  & 1&7496  & 7&7852 \\
    42 & GPR\_TWN & Taiwan & Asia and Oceania & 286 & 0&7708  & 0&0000  & 0&0729  & 0&0931  & 3&0568  & 16&0876 \\
    43 & GPR\_UKR & Ukraine & Europe (North and East) & 286 & 8&8737  & 0&0000  & 0&3443  & 0&8885  & 5&0666  & 37&8505 \\
    44 & GPR\_USA & United States & North America & 286 & 13&2290  & 0&8201  & 2&4264  & 1&3621  & 4&4591  & 31&3968 \\
    45 & GPR\_VEN & Venezuela & South America & 286 & 0&5240  & 0&0000  & 0&0551  & 0&0615  & 3&7564  & 21&7476 \\
    46 & GPR\_VNM & Vietnam & Asia and Oceania & 286 & 0&1580  & 0&0000  & 0&0294  & 0&0239  & 2&0796  & 9&0910 \\
    47 & GPR\_ZAF & South Africa & Middle East and Africa & 286 & 0&2419  & 0&0042  & 0&0547  & 0&0327  & 1&8198  & 8&2905 \\
  \bottomrule
    \end{tabular}
    }%
  \begin{flushleft}
    \footnotesize
    \justifying Note: This table provides detailed information on three global geopolitical risk indices and 44 economy-specific geopolitical risk indices, along with their respective descriptive statistics.
\end{flushleft} 
  \label{Tab:GPRE_Stat}%
\end{table}%

\section{Empirical analysis}
\label{S1:EmpAnal}

\subsection{Construction of composite geopolitical risk indices}

We apply the random matrix theory to analyze the individual geopolitical risk indices of 44 economies developed by \cite{Caldara-Iacoviello-2022-AmEconRev}. Composite geopolitical risk indices for various combinations of economies are further constructed to examine the impact of geopolitical risks faced by six major geographical regions, as well as for combinations of economies with significant influence in production, import, and export, on the volatility of the international food market, respectively.

\begin{figure}[!ht]
\centering
\includegraphics[height=0.26\textheight]{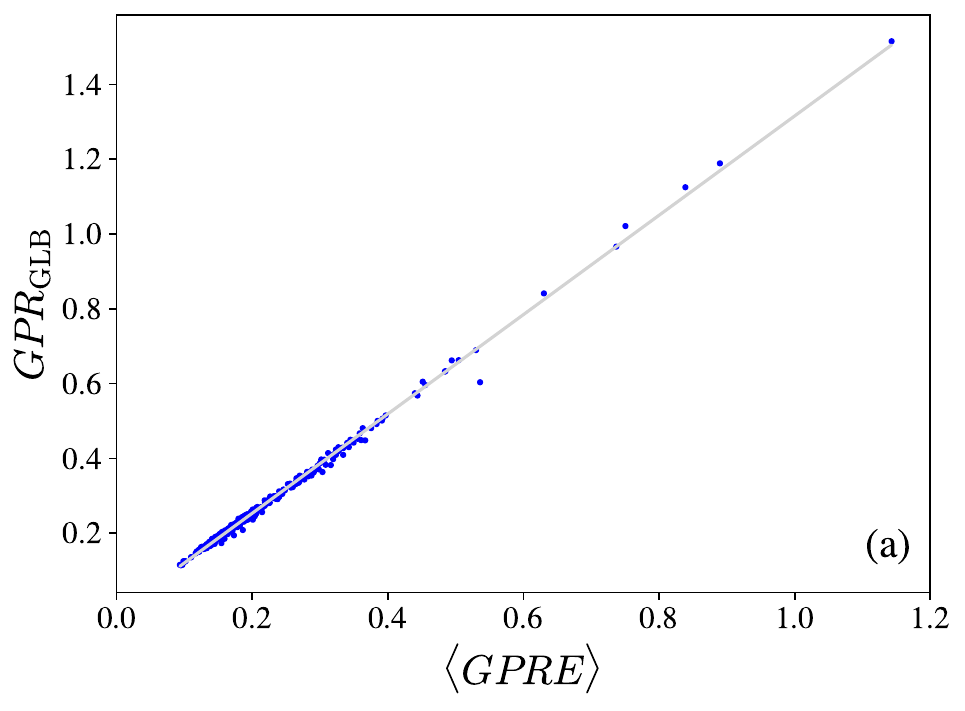}
\includegraphics[height=0.26\textheight]{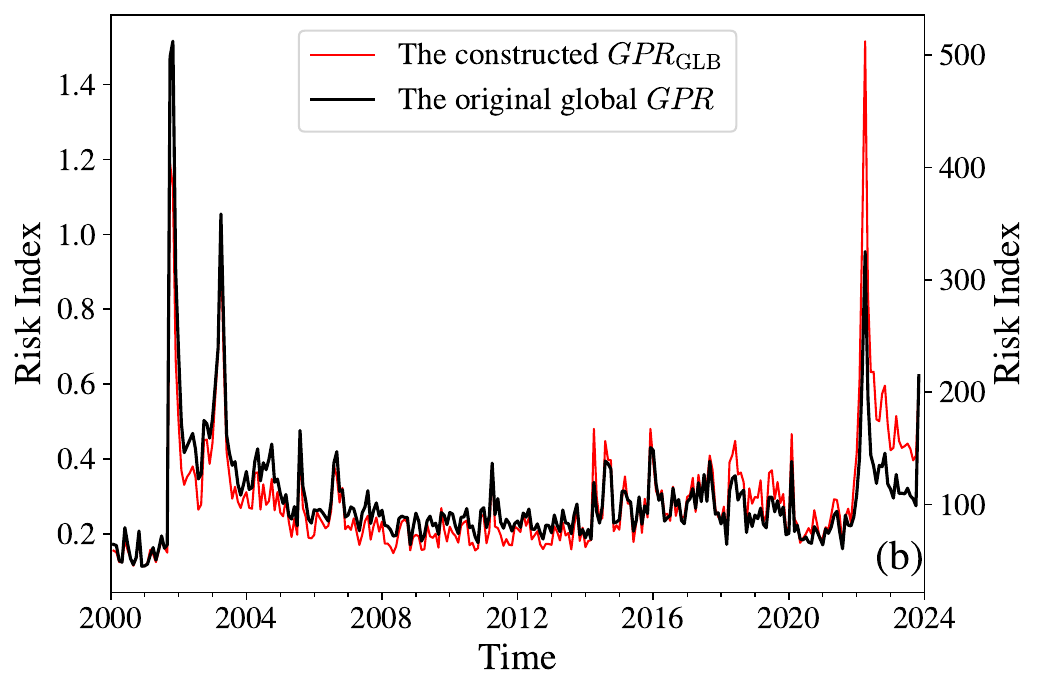}
\caption{Comparison of the global geopolitical risk index $GPR_{\text{GLB}}$ constructed by this paper with the mean GPRE index $\langle{GPRE}\rangle$ (a) and with the original global geopolitical risk index $GPR$ constructed by \cite{Caldara-Iacoviello-2022-AmEconRev} (b).}
\label{Fig:GPRE_umaxcomponent}
\end{figure}

We first construct the RMT-based global geopolitical risk index $GPR_{\text{GLB}}$ using the economy-specific geopolitical risk indices. Figure~\ref{Fig:GPRE_umaxcomponent}(a) illustrates the relationship between the constructed global geopolitical risk index $GPR_{\text{GLB}}$ and the average geopolitical risk index $\langle GPRE \rangle$ of the 44 economies. A clear linear relationship between $GPR_{\text{GLB}}$ and $\langle GPRE \rangle$ is observed, in which the coefficient of determination R$^{2}$ reaches 0.997. Hence, the global geopolitical risk index $GPR_{\text{GLB}}$, which is constructed based on the random matrix theory, can be regarded as an effective indicator reflecting the overall geopolitical risk level of these 44 major economies. Similar conclusions have been reported in studies of the U.S. stock market \citep{Plerou-Gopikrishnan-Rosenow-Amaral-Guhr-Stanley-2002-PhysRevE}, global crude oil market \citep{Dai-Xie-Jiang-Jiang-Zhou-2016-EmpirEcon}, and global agricultural futures market \citep{Dai-Huynh-Zheng-Zhou-2022-ResIntBusFinanc}, as well as the construction of global economic policy uncertainty indices \citep{Dai-Xiong-Zhou-2021-FinancResLett}. Figure~\ref{Fig:GPRE_umaxcomponent}(b) further compares the evolution of the RMT-based geopolitical risk index $GPR_{\text{GLB}}$ with the global geopolitical risk index GPR developed by \cite{Caldara-Iacoviello-2022-AmEconRev}. Despite numerical differences, the trends of these two indices are highly similar, with a correlation coefficient of 0.881, which also supports that the composite geopolitical risk index constructed by the random matrix theory performs well in characterizing the movements of the overall geopolitical risk faced by multiple economies.

Next, we refer to \cite{Caldara-Iacoviello-2022-AmEconRev} to divide these 44 major economies into six regions based on geographical location, including North America, South America, Europe (North and East), Europe (South and West), Middle East and Africa, and Asia and Oceania. With the random matrix theory, we then construct the composite geopolitical risk index associated with each region, namely $GPR_{\text{NA}}$, $GPR_{\text{SA}}$, $GPR_{\text{ENE}}$, $GPR_{\text{ESW}}$, $GPR_{\text{MEA}}$, and $GPR_{\text{AO}}$, to explore the different effects of geopolitical risks across diverse geographical regions on the international food market. The economies included in each region are shown in Table~\ref{Tab:GPRE_Stat}. Furthermore, considering the varying roles played by different economies in global food production, import, and export, we collect data on yields, import volumes, and export volumes of wheat, maize, soybean, and rice for these 44 important economies during the period 2000--2021 from the statistics released by the U.S. Department of Agriculture (USDA)\footnote{\url{https://apps.fas.usda.gov/psdonline/app/index.html\#/app/downloads}}. With the rankings of total yields, total import volume, and total export volume, we then construct composite geopolitical risk indices, denoted as $GPR_{\text{prod}}$, $GPR_{\text{imp}}$, and $GPR_{\text{exp}}$, based on the eigenportfolios of the individual geopolitical risk indices of the top ten producers, importers, and exporters of the four staple crops, respectively. The top ten producing, importing, and exporting economies of a major grain are considered to have significant influence on the global production, supply, and demand of that particular grain. Table~\ref{Tab:Components_of_Constructed_Macro_Variables} presents the components of the composite geopolitical risk indices constructed in our paper.

\begin{table}[!ht]
  \centering
  \setlength{\abovecaptionskip}{0pt}
  \setlength{\belowcaptionskip}{10pt}
  \caption{Components of different geopolitical risk indices constructed by the random matrix theory}
  \setlength\tabcolsep{3pt}   \resizebox{\textwidth}{!}{ 
    \begin{tabular}{ll}
    \toprule
    \multicolumn{1}{c}{Variable} & \multicolumn{1}{c}{Components} \\
    \midrule
    $GPR_{\text{NA}}$ & GPR\_CAN, GPR\_MEX, GPR\_USA \\
    $GPR_{\text{SA}}$ & GPR\_ARG, GPR\_BRA, GPR\_CHL, GPR\_COL, GPR\_PER, GPR\_VEN \\
    $GPR_{\text{ENE}}$ & GPR\_DNK, GPR\_FIN, GPR\_HUN, GPR\_NOR, GPR\_POL, GPR\_RUS, GPR\_SWE, GPR\_UKR, GPR\_GBR \\
    $GPR_{\text{ESW}}$ & GPR\_BEL, GPR\_FRA, GPR\_DEU, GPR\_ITA, GPR\_NLD, GPR\_PRT, GPR\_ESP, GPR\_CHE \\
    $GPR_{\text{MEA}}$ & GPR\_EGY, GPR\_ISR, GPR\_SAU, GPR\_ZAF, GPR\_TUN, GPR\_TUR \\
    $GPR_{\text{AO}}$ & GPR\_AUS, GPR\_CHN, GPR\_HKG, GPR\_JPN, GPR\_KOR, GPR\_PHL, GPR\_TWN, GPR\_IDN, GPR\_IND, GPR\_MYS, GPR\_THA, GPR\_VNM \\
    $GPR_{\text{wh,prod}}$ & GPR\_CHN, GPR\_IND, GPR\_RUS, GPR\_USA, GPR\_FRA, GPR\_CAN, GPR\_DEU, GPR\_AUS, GPR\_UKR, GPR\_TUR \\
    $GPR_{\text{wh,imp}}$ & GPR\_USA, GPR\_CHN, GPR\_BRA, GPR\_ARG, GPR\_MEX, GPR\_IND, GPR\_UKR, GPR\_IDN, GPR\_FRA, GPR\_ZAF \\
    $GPR_{\text{wh,exp}}$ & GPR\_USA, GPR\_BRA, GPR\_ARG, GPR\_CHN, GPR\_IND, GPR\_CAN, GPR\_UKR, GPR\_RUS, GPR\_IDN, GPR\_ITA \\
    $GPR_{\text{mz,prod}}$ & GPR\_CHN, GPR\_IND, GPR\_IDN, GPR\_VNM, GPR\_THA, GPR\_PHL, GPR\_BRA, GPR\_JPN, GPR\_USA, GPR\_KOR \\
    $GPR_{\text{mz,imp}}$ & GPR\_EGY, GPR\_ITA, GPR\_IDN, GPR\_BRA, GPR\_JPN, GPR\_ESP, GPR\_NLD, GPR\_CHN, GPR\_KOR, GPR\_MEX \\
    $GPR_{\text{mz,exp}}$ & GPR\_JPN, GPR\_MEX, GPR\_KOR, GPR\_CHN, GPR\_EGY, GPR\_ESP, GPR\_TWN, GPR\_VNM, GPR\_NLD, GPR\_COL \\
    $GPR_{\text{sb,prod}}$ & GPR\_CHN, GPR\_NLD, GPR\_MEX, GPR\_JPN, GPR\_DEU, GPR\_ESP, GPR\_TWN, GPR\_THA, GPR\_IDN, GPR\_EGY \\
    $GPR_{\text{sb,imp}}$ & GPR\_MEX, GPR\_VEN, GPR\_TUR, GPR\_BRA, GPR\_ITA, GPR\_COL, GPR\_PRT, GPR\_SAU, GPR\_ESP, GPR\_FRA \\
    $GPR_{\text{sb,exp}}$ & GPR\_USA, GPR\_CAN, GPR\_RUS, GPR\_FRA, GPR\_AUS, GPR\_UKR, GPR\_ARG, GPR\_DEU, GPR\_HUN, GPR\_GBR \\
    $GPR_{\text{rc,prod}}$ & GPR\_USA, GPR\_ARG, GPR\_BRA, GPR\_UKR, GPR\_FRA, GPR\_CHN, GPR\_HUN, GPR\_IND, GPR\_RUS, GPR\_ZAF \\
    $GPR_{\text{rc,imp}}$ & GPR\_BRA, GPR\_USA, GPR\_ARG, GPR\_CAN, GPR\_NLD, GPR\_UKR, GPR\_RUS, GPR\_CHN, GPR\_BEL, GPR\_IND \\
    $GPR_{\text{rc,exp}}$ & GPR\_USA, GPR\_BRA, GPR\_IND, GPR\_ARG, GPR\_RUS, GPR\_FRA, GPR\_ESP, GPR\_CHN, GPR\_ITA, GPR\_AUS \\
  \bottomrule
    \end{tabular}
    }%
  \begin{flushleft}
    \footnotesize
    \justifying Note: This table lists the components of each of the composite geopolitical risk indices built on the basis of random matrix theory.
    \end{flushleft} 
  \label{Tab:Components_of_Constructed_Macro_Variables}%
\end{table}%

\subsection{Estimation of GJR-GARCH-MIDAS models with single factor}

To ensure numerical stability during the process of parameter optimization, we follow the approach mentioned in \cite{Conrad-Loch-Rittler-2014-JEmpirFinanc}, \cite{Conrad-Kleen-2020-JApplEconom}, and \cite{Caldara-Iacoviello-2022-AmEconRev}, by magnifying the agricultural return series by a factor of 100 and taking the logarithm of the geopolitical risk index series for model estimation.

Tables~\ref{Tab:Single_Factor_Estimation_Wheat_Futures}--\ref{Tab:Single_Factor_Estimation_Rice_Futures} report the parameter estimates and goodness-of-fit statistics of different single-factor GJR-GARCH-MIDAS models for wheat, maize, soybean, and rice, where Panels A and B display the estimated results of models with fixed time span and rolling window, respectively. After comparing Panels A and B, we note that the estimates of parameters and their statistical significance in the single-factor GJR-GARCH-MIDAS models with fixed time span and with rolling window are broadly similar, but the latter generally outperform the former in terms of goodness-of-fit, as evidenced by higher LLF values and lower AIC and BIC values. For these four staple crops, rolling-window models with a single factor are comparatively accurate in modeling their market volatility, the estimated results of which in Panel B are therefore the focus of our subsequent analysis.

In terms of short-run fluctuations, the estimated coefficients of the ARCH and GARCH terms, that is, the estimates of $\alpha$ and $\beta$, for wheat, maize, soybean, and rice, are all significant at the 1\% level, which suggests that the agricultural returns exhibit strong volatility clustering. Furthermore, although the estimates of $\gamma$ in the single-factor models corresponding to wheat, maize, and soybean are consistently negative, and positive for rice, they are all statistically insignificant, implying the absence of significant asymmetry in the short-term volatility for these four staple grains. Moreover, the sum of the estimates of $\alpha$, $\beta$, and 0.5$\gamma$ in each model is less than 1, meeting the requirement for model stability, and the sum is quite close to 1, indicating high persistence of short-term fluctuations in agricultural markets. As far as long-run volatility is concerned, except for rice, the $\theta$ coefficient for each of the other three staple crops is significantly positive when using its realized volatility as the explanatory factor. This reflects the positive impact of realized volatility on the long-term fluctuations in the wheat, maize, and soybean markets. However, when geopolitical risk, a macro-factor, is taken as the explanatory variable, the results for the four staple grains vary, as analyzed below.

Combined with Table~\ref{Tab:Single_Factor_Estimation_Wheat_Futures}, it can be found that for wheat, most of the $\theta$ estimates in single-factor models incorporating different geopolitical risk indicators are statistically insignificant, which implies that geopolitical risk factors have a limited long-term impact on wheat market volatility. In addition, the variance ratio, denoted as VR, for each single-factor model is relatively small, further proving the limited ability of individual geopolitical risk factors in explaining the long-term volatility of the international wheat market.

\begin{table}[!ht]
  \centering
  \scriptsize
  \setlength{\abovecaptionskip}{0pt}
  \setlength{\belowcaptionskip}{10pt}
  \caption{Estimation of single-factor models with fixed time span and rolling window for wheat}
  \setlength\tabcolsep{3pt}   \resizebox{\textwidth}{!}{ 
    \begin{tabular}{l r@{.}l r@{.}l r@{.}l r@{.}l r@{.}l r@{.}l r@{.}l rccc}
    \toprule
         & \multicolumn{2}{c}{$\mu$} & \multicolumn{2}{c}{$\alpha$} & \multicolumn{2}{c}{$\beta$} & \multicolumn{2}{c}{$\gamma$} & \multicolumn{2}{c}{$m$} & \multicolumn{2}{c}{$\theta$} & \multicolumn{2}{c}{$\omega$} & VR (\%) & LLF & AIC & BIC \\
    \midrule
    \multicolumn{19}{l}{\textit{Panel A: Single-factor models with fixed time span}} \\
    $RV_{\text{wh}}$ & 0&0144 & 0&1078$^{***}$ & 0&8307$^{***}$ & $-$0&0457 & 0&8813$^{***}$ & 0&0055$^{***}$ & 6&4272$^{***}$ & 47.9222 & $-$11066.39 & 22145.84 & 22192.74 \\
    & (0&0263) & (0&0160) & (0&0250) & (0&0285) & (0&0955) & (0&0008) & (1&4543) &  \multicolumn{4}{c}{} \\
    $GPR$ & 0&0133 & 0&0666$^{***}$ & 0&9236$^{***}$ & $-$0&0132 & 1&3374 & 0&0295 & 1&5848 & 0.0109 & $-$11082.50 & 22178.07 & 22224.97 \\
    & (0&0263) & (0&0182) & (0&0251) & (0&0144) & (2&4722) & (0&5382) & (2&3881) &  \multicolumn{4}{c}{} \\ 
    $GPT$ & 0&0117 & 0&0671$^{***}$ & 0&9225$^{***}$ & $-$0&0127 & $-$0&0016 & 0&3205$^{***}$ & 2&1773 & 1.3560 & $-$11082.11 & 22177.29 & 22224.19 \\ 
    & (0&0263) & (0&0188) & (0&0262) & (0&0146) & (0&0174) & (0&0334) & (3&1342) &  \multicolumn{4}{c}{} \\ 
    $GPA$ & 0&0130 & 0&0664$^{***}$ & 0&9235$^{***}$ & $-$0&0128 & 1&4864$^{***}$ & $-$0&0034 & 1&0001 & 0.0004 & $-$11082.50 & 22178.05 & 22224.95 \\ 
    & (0&0263) & (0&0179) & (0&0248) & (0&0144) & (0&5083) & (0&1093) & (7&2401) &  \multicolumn{4}{c}{} \\
    $GPR_{\text{GLB}}$ & 0&0099 & 0&0478 & 0&9559$^{***}$ & $-$0&0075 & $-$0&2621 & 0&0427 & 3&3377 & 0.0326 & $-$11103.86 & 22220.79 & 22267.69 \\
	& (0&6913) & (0&1685) & (0&0385) & (0&3187) & (75&6928) & (30&6270) & (792&9951) &   \multicolumn{4}{c}{} \\
    $GPR_{\text{NA}}$ & 0&0128 & 0&0475$^{***}$ & 0&9563$^{***}$ & $-$0&0076 & $-$0&2723 & $-$0&1004 & 2&7679 & 0.1380 & $-$11103.87 & 22220.80 & 22267.70 \\
    & (0&0355) & (0&0080) & (0&0081) & (0&0108) & (0&7234) & (1&2229) & (15&6459) &  \multicolumn{4}{c}{} \\
    $GPR_{\text{SA}}$ & 0&0115 & 0&0677$^{***}$ & 0&9209$^{***}$ & $-$0&0127 & 2&2260$^{***}$ & 0&2154 & 1&5079 & 1.6951 & $-$11081.84 & 22176.74 & 22223.64 \\
    & (0&0264) & (0&0189) & (0&0266) & (0&0148) & (0&8293) & (0&2290) & (3&0523) &  \multicolumn{4}{c}{} \\
    $GPR_{\text{ENE}}$ & 0&0104 & 0&0488$^{***}$ & 0&9547$^{***}$ & $-$0&0070 & $-$0&6386 & $-$0&2839 & 5&8159 & 4.8193 & $-$11103.80 & 22220.65 & 22267.55 \\
    & (0&0321) & (0&0087) & (0&0086) & (0&0095) & (1&9359) & (0&4555) & (18&9504) &  \multicolumn{4}{c}{} \\
    $GPR_{\text{ESW}}$ & 0&0105 & 0&0481$^{***}$ & 0&9561$^{***}$ & $-$0&0083 & 0&3749 & 0&5274 & 8&5666 & 8.1664 & $-$11102.31 & 22217.67 & 22264.57 \\
    & (0&0392) & (0&0095) & (0&0098) & (0&0119) & (1&4152) & (0&6507) & (9&4764) &  \multicolumn{4}{c}{} \\
    $GPR_{\text{MEA}}$ & 0&0097 & 0&0478$^{***}$ & 0&9560$^{***}$ & $-$0&0075 & $-$1&0592 & $-$0&5679 & 2&4380 & 4.1868 & $-$11103.60 & 22220.25 & 22267.15 \\
    & (0&0320) & (0&0083) & (0&0080) & (0&0086) & (1&9905) & (1&1760) & (3&0341) &  \multicolumn{4}{c}{} \\
    $GPR_{\text{AO}}$ & 0&0084 & 0&0472$^{***}$ & 0&9563$^{***}$ & $-$0&0069 & 1&5425 & 0&9441 & 1&3623 & 17.5424 & $-$11103.52 & 22220.10 & 22267.00 \\
    & (0&0321) & (0&0079) & (0&0080) & (0&0091) & (5&4545) & (2&2713) & (3&2234) &  \multicolumn{4}{c}{} \\
    $GPR_{\text{wh,prod}}$ & 0&0100 & 0&0476$^{***}$ & 0&9561$^{***}$ & $-$0&0075 & 0&1445 & 0&6429 & 1&0594 & 4.8844 & $-$11103.70 & 22220.47 & 22267.37 \\
    & (0&0319) & (0&0082) & (0&0080) & (0&0089) & (2&7412) & (2&1049) & (2&1060) &  \multicolumn{4}{c}{} \\
    $GPR_{\text{wh,imp}}$ & 0&0093 & 0&0475$^{***}$ & 0&9563$^{***}$ & $-$0&0074 & 1&6484 & 1&1107 & 1&9743 & 19.0923 & $-$11103.15 & 22219.35 & 22266.25 \\
    & (0&0367) & (0&0090) & (0&0081) & (0&0121) & (10&9030) & (3&7806) & (5&9914) &  \multicolumn{4}{c}{} \\
    $GPR_{\text{wh,exp}}$ & 0&0092 & 0&0478$^{***}$ & 0&9561$^{***}$ & $-$0&0079 & $-$0&3207 & $-$0&0658 & 4&0627 & 0.0944 & $-$11103.86 & 22220.79 & 22267.69 \\
    & (0&0338) & (0&0086) & (0&0081) & (0&0104) & (2&2000) & (1&8300) & (24&4089) &  \multicolumn{4}{c}{} 
    \vspace{2mm}\\
    
    \multicolumn{19}{l}{\textit{Panel B: Single-factor models with rolling window}} \\
    $RV_{\text{wh}}^{(\text{rw})}$ & 0&0161 & 0&1051$^{***}$ & 0&8382$^{***}$ & $-$0&0408 & 0&9134$^{***}$ & 0&0051$^{***}$ & 6&0271$^{***}$ & 36.5707 & $-$11000.33 & 22013.67 & 22060.57 \\
    & (0&0265) & (0&0174) & (0&0297) & (0&0303) & (0&1156) & (0&0008) & (1&5499) &  \multicolumn{4}{c}{} \\
    $GPR^{(\text{rw})}$ & 0&0153 & 0&0655$^{***}$ & 0&9260$^{***}$ & $-$0&0134 & 2&0838 & $-$0&1303 & 1&3031 & 0.1755 & $-$11008.92 & 22030.84 & 22077.74 \\
    & (0&0264) & (0&0172) & (0&0231) & (0&0143) & (2&4766) & (0&5399) & (2&5379) &  \multicolumn{4}{c}{} \\
    $GPT^{(\text{rw})}$ & 0&0145 & 0&0660$^{***}$ & 0&9248$^{***}$ & $-$0&0128 & 0&5173 & 0&2102 & 1&9117 & 0.4811 & $-$11008.83 & 22030.67 & 22077.57 \\ 
    & (0&0264) & (0&0183) & (0&0249) & (0&0144) & (2&7025) & (0&5858) & (3&0457) &  \multicolumn{4}{c}{} \\ 
    $GPA^{(\text{rw})}$ & 0&0153 & 0&0655$^{***}$ & 0&9259$^{***}$ & $-$0&0135 & 2&2086$^{*}$ & $-$0&1596 & 1&0001 & 0.8672 & $-$11008.67 & 22030.34 & 22077.24 \\ 
    & (0&0265) & (0&0175) & (0&0235) & (0&0144) & (1&2423) & (0&2739) & (1&3254) &  \multicolumn{4}{c}{} \\
    $GPR_{\text{GLB}}^{(\text{rw})}$ & 0&0102 & 0&0479$^{***}$ & 0&9560$^{***}$ & $-$0&0078 & $-$0&1708 & $-$0&3687 & 3&8939 & 2.3894 & $-$11032.89 & 22078.79 & 22125.69 \\
    & (0&0328) & (0&0080) & (0&0080) & (0&0093) & (1&0919) & (0&8552) & (3&8763) &  \multicolumn{4}{c}{} \\
    $GPR_{\text{NA}}^{(\text{rw})}$ & 0&0151 & 0&0652$^{***}$ & 0&9264$^{***}$ & $-$0&0132 & 1&4744$^{***}$ & $-$0&0847 & 3&8445 & 0.1365 & $-$11008.91 & 22030.83 & 22077.73 \\
    & (0&0265) & (0&0221) & (0&0339) & (0&0142) & (0&2272) & (1&3174) & (35&5426) &  \multicolumn{4}{c}{} \\
    $GPR_{\text{SA}}^{(\text{rw})}$ & 0&0131 & 0&0665$^{***}$ & 0&9230$^{***}$ & $-$0&0123 & 2&2196$^{*}$ & 0&2103 & 1&0004 & 1.3246 & $-$11008.39 & 22029.79 & 22076.69 \\
    & (0&0265) & (0&0187) & (0&0258) & (0&0152) & (1&1629) & (0&3224) & (4&9588) &  \multicolumn{4}{c}{} \\
    $GPR_{\text{ENE}}^{(\text{rw})}$ & 0&0123 & 0&0481$^{***}$ & 0&9562$^{***}$ & $-$0&0086 & $-$0&1645 & $-$0&2680 & 6&0698 & 4.0935 & $-$11032.75 & 22078.51 & 22125.41 \\
    & (0&0325) & (0&0081) & (0&0081) & (0&0091) & (0&8643) & (0&5123) & (8&9046) &  \multicolumn{4}{c}{} \\
    $GPR_{\text{ESW}}^{(\text{rw})}$ & 0&0118 & 0&0477$^{***}$ & 0&9564$^{***}$ & $-$0&0082 & 0&6041 & 0&4940 & 13&6844$^{**}$ & 7.5051 & $-$11031.11 & 22075.23 & 22122.13 \\
    & (0&0398) & (0&0092) & (0&0114) & (0&0126) & (2&4004) & (1&2591) & (6&7575) &  \multicolumn{4}{c}{} \\
    $GPR_{\text{MEA}}^{(\text{rw})}$ & 0&0109 & 0&0476$^{***}$ & 0&9560$^{***}$ & $-$0&0073 & $-$1&0866 & $-$0&9225 & 1&9970 & 9.5475 & $-$11032.58 & 22078.17 & 22125.07 \\
    & (0&0326) & (0&0082) & (0&0083) & (0&0092) & (2&4763) & (1&4948) & (1&8378) &  \multicolumn{4}{c}{} \\
    $GPR_{\text{AO}}^{(\text{rw})}$ & 0&0169 & 0&0633$^{***}$ & 0&9301$^{***}$ & $-$0&0132 & 1&1441$^{**}$ & $-$0&1866 & 69&8142$^{***}$ & 2.0042 & $-$11008.23 & 22029.47 & 22076.37 \\
    & (0&0264) & (0&0167) & (0&0232) & (0&0140) & (0&5433) & (0&2837) & (21&9186) &  \multicolumn{4}{c}{} \\
    $GPR_{\text{wh,prod}}^{(\text{rw})}$ & 0&0142 & 0&0480$^{***}$ & 0&9560$^{***}$ & $-$0&0080 & 0&1007 & $-$0&3084 & 4&2069 & 2.0353 & $-$11032.91 & 22078.83 & 22125.73 \\
    & (0&0328) & (0&0080) & (0&0080) & (0&0093) & (0&5358) & (0&8219) & (4&5523) &  \multicolumn{4}{c}{} \\
    $GPR_{\text{wh,imp}}^{(\text{rw})}$ & 0&0114 & 0&0478$^{***}$ & 0&9560$^{***}$ & $-$0&0076 & 1&2571 & 0&6397 & 1&6582 & 5.6447 & $-$11032.87 & 22078.74 & 22125.64 \\
    & (0&0325) & (0&0079) & (0&0081) & (0&0092) & (3&6983) & (1&9840) & (2&7934) &  \multicolumn{4}{c}{} \\
    $GPR_{\text{wh,exp}}^{(\text{rw})}$ & 0&0146 & 0&0659$^{***}$ & 0&9250$^{***}$ & $-$0&0130 & 1&5116$^{***}$ & 0&0506 & 5&3885 & 0.0815 & $-$11008.93 & 22030.88 & 22077.78 \\
    & (0&0270) & (0&0174) & (0&0235) & (0&0145) & (0&3935) & (0&6681) & (80&2215) &  \multicolumn{4}{c}{} \\
  \bottomrule
    \end{tabular}
    }%
  \begin{flushleft}
    \footnotesize
    \justifying Note: This table reports the estimated results of different single-factor GJR-GARCH-MIDAS models for the wheat market based on fixed time span (Panel A) and rolling window (Panel B), with the standard errors shown in parentheses. VR, LLF, AIC, and BIC denote the values of the explanatory power, log-likelihood function, Akaike information criterion, and Bayesian information criterion, respectively.
    \end{flushleft} 
  \label{Tab:Single_Factor_Estimation_Wheat_Futures}%
\end{table}%

As can be seen from Table~\ref{Tab:Single_Factor_Estimation_Maize_Futures}, geopolitical risk factors have greater effects on maize market volatility than on wheat market volatility. From a global perspective, the GPT index $GPT^{(\text{rw})}$ and our constructed global geopolitical risk index $GPR_{\text{GLB}}^{(\text{rw})}$ correspond to single-factor models with significantly negative estimates of $\theta$, indicating the prominently negative impact of global geopolitical threats and composite geopolitical risk based on 44 major economies on the long-term volatility of the international maize market. In other words, as global geopolitical threats intensify or composite geopolitical risks increase, the long-run component of maize market volatility tends to decrease, a finding that is somewhat counterintuitive. One possible explanation is that when geopolitical events occur globally, both the global geopolitical risk and threat rise significantly, but uncertainty decreases as the events have already taken place. Given the essential nature of food as a survival necessity, traders can roughly predict future trends in food prices through the available current information, leading to a relative reduction in long-term volatility in the food market. Furthermore, the occurrence of adverse geopolitical events also fosters global risk aversion, thus reducing long-run volatility in the maize market.

Similar puzzling phenomena have been observed between stock market volatility and economic policy uncertainty in the United States and the United Kingdom, where periods of high economic policy uncertainty correspond to low stock market volatility \citep{Bialkowski-Dang-Wei-2022-JFinancEcon}. Possible reasons given for the negative relationship between economic policy uncertainty and stock market volatility include low-quality political signals, high disagreement among investors, and abnormal stock market performance. In our study, the data we use to measure geopolitical risk is also built on the basis of newspapers, so the accuracy of political signals, the amount of information, and the level of noise contained in news reports may also exert a certain influence on investors in agricultural futures markets, thereby affecting the volatility of the food market. Additionally, in line with the significance of $\theta$, single-factor models incorporating $GPT^{(\text{rw})}$ and $GPR_{\text{GLB}}^{(\text{rw})}$ both exhibit large variance ratios, where the VR of the model corresponding to $GPR_{\text{GLB}}^{(\text{rw})}$ implies that approximately 23.42\% of expected volatility in the international maize market can be explained by our constructed global geopolitical risk index based on the random matrix theory.

After further examining the effects of geopolitical risks faced by the six major geographic regions on the long-term component of volatility, we find that, except for South America, the corresponding $\theta$ coefficients for geopolitical risk factors of the other five regions are all significantly negative. This indicates that geopolitical risks in North America, Europe, Asia, and other regions all show a significant negative impact on the long-term volatility of the international maize market. In particular, the single-factor GJR-GARCH-MIDAS model containing $GPR_{\text{ESW}}^{(\text{rw})}$ exhibits the highest LLF and the lowest AIC and BIC, suggesting the best fitting effect of this model. Meanwhile, the single-factor model with $GPR_{\text{ENE}}^{(\text{rw})}$ has the largest VR, meaning that the geopolitical risk in Europe (North and East) displays the strongest explanatory power for the overall volatility of the international maize market compared to other regions, accounting for about 33.03\% of the volatility. Moreover, the significantly negative estimates of $\theta$ associated with $GPR_{\text{mz,prod}}^{(\text{rw})}$, $GPR_{\text{mz,imp}}^{(\text{rw})}$, and $GPR_{\text{mz,exp}}^{(\text{rw})}$ imply that the long-term volatility of the maize market is likely to be negatively and noticeably affected by the composite geopolitical risks from combinations of economies with global influence in maize production, import, and export. Among these three single-factor models, the GJR-GARCH-MIDAS model containing $GPR_{\text{mz,exp}}^{(\text{rw})}$ has the largest LLF, the smallest AIC and BIC, as well as the highest VR, which demonstrates that the composite geopolitical risk of major maize exporters explains the expected volatility of the international maize market to a greater extent. This finding aligns with the strong explanatory power of geopolitical risk in Europe (North and East) for maize market volatility, as four of the top ten maize exporters are located in Europe (North and East), namely Ukraine, Hungary, Russia, and Poland.

\begin{table}[!ht]
  \centering
  \scriptsize
  \setlength{\abovecaptionskip}{0pt}
  \setlength{\belowcaptionskip}{10pt}
  \caption{Estimation of single-factor models with fixed time span and rolling window for maize}
  \setlength\tabcolsep{3pt}   \resizebox{\textwidth}{!}{ 
    \begin{tabular}{l r@{.}l r@{.}l r@{.}l r@{.}l r@{.}l r@{.}l r@{.}l rccc}
    \toprule
         & \multicolumn{2}{c}{$\mu$} & \multicolumn{2}{c}{$\alpha$} & \multicolumn{2}{c}{$\beta$} & \multicolumn{2}{c}{$\gamma$} & \multicolumn{2}{c}{$m$} & \multicolumn{2}{c}{$\theta$} & \multicolumn{2}{c}{$\omega$} & VR (\%) & LLF & AIC & BIC \\
    \midrule
    \multicolumn{19}{l}{\textit{Panel A: Single-factor models with fixed time span}} \\
    $RV_{\text{mz}}$ & 0&0234 & 0&1786$^{***}$ & 0&7776$^{***}$ & $-$0&0411 & 0&9398$^{***}$ & 0&0065$^{**}$ & 1&2765 & 14.0911 & $-$10358.55 & 20730.17 & 20777.07 \\
    & (0&0222) & (0&0587) & (0&0555) & (0&0468) & (0&2734) & (0&0027) & (0&9380) &  \multicolumn{4}{c}{} \\
    $GPR$ & 0&0212 & 0&1718$^{***}$ & 0&7903$^{***}$ & $-$0&0362 & 4&9147$^{**}$ & $-$0&7540 & 6&4027$^{**}$ & 8.5093 & $-$10363.70 & 20740.45 & 20787.35 \\
    & (0&0225) & (0&0609) & (0&0570) & (0&0461) & (2&2040) & (0&4774) & (3&1772) &  \multicolumn{4}{c}{} \\ 
    $GPT$ & 0&0215 & 0&1680$^{***}$ & 0&7874$^{***}$ & $-$0&0335 & 5&4693$^{***}$ & $-$0&8816$^{***}$ & 64&6345 & 23.4209 & $-$10350.34 & 20713.74 & 20760.64 \\ 
    & (0&0225) & (0&0595) & (0&0587) & (0&0445) & (1&5668) & (0&3375) & (60&3481) &  \multicolumn{4}{c}{} \\ 
    $GPA$ & 0&0204 & 0&1697$^{***}$ & 0&8018$^{***}$ & $-$0&0367 & 0&9710 & 0&1165 & 1&0000 & 0.4463 & $-$10370.31 & 20753.68 & 20800.58 \\ 
    & (0&0229) & (0&0603) & (0&0535) & (0&0446) & (1&6762) & (0&3609) & (4&7975) &  \multicolumn{4}{c}{} \\
    $GPR_{\text{GLB}}$ & 0&0218 & 0&1719$^{***}$ & 0&7740$^{***}$ & $-$0&0324 & $-$0&0253 & $-$1&0649$^{***}$ & 4&4977 & 28.7972 & $-$10347.79 & 20708.65 & 20755.55 \\
    & (0&0222) & (0&0600) & (0&0594) & (0&0468) & (0&5687) & (0&4069) & (3&1659) &  \multicolumn{4}{c}{} \\
    $GPR_{\text{NA}}$ & 0&0210 & 0&1718$^{***}$ & 0&7871$^{***}$ & $-$0&0359 & 1&3417$^{***}$ & $-$0&7647$^{**}$ & 7&9084 & 12.2051 & $-$10361.16 & 20735.37 & 20782.27 \\
    & (0&0225) & (0&0612) & (0&0586) & (0&0465) & (0&1923) & (0&3878) & (5&0221) &  \multicolumn{4}{c}{} \\
    $GPR_{\text{SA}}$ & 0&0200 & 0&1689$^{***}$ & 0&7983$^{***}$ & $-$0&0357 & 0&5755 & $-$0&2566 & 18&2477 & 4.0335 & $-$10367.30 & 20747.66 & 20794.56 \\
    & (0&0229) & (0&0600) & (0&0550) & (0&0444) & (0&5849) & (0&1584) & (12&6619) &  \multicolumn{4}{c}{} \\
    $GPR_{\text{ENE}}$ & 0&0227 & 0&1740$^{***}$ & 0&7658$^{***}$ & $-$0&0304 & $-$0&5277 & $-$1&1705$^{***}$ & 1&1050 & 38.3245 & $-$10339.14 & 20691.33 & 20738.23 \\
    & (0&0221) & (0&0574) & (0&0555) & (0&0458) & (0&5486) & (0&3210) & (0&6784) & \multicolumn{4}{c}{} \\
    $GPR_{\text{ESW}}$ & 0&0230 & 0&1753$^{***}$ & 0&7609$^{***}$ & $-$0&0324 & $-$0&7003 & $-$1&2580$^{***}$ & 2&2159 & 37.8152 & $-$10335.04 & 20683.14 & 20730.04 \\
    & (0&0219) & (0&0582) & (0&0573) & (0&0470) & (0&5768) & (0&3384) & (1&3776) &  \multicolumn{4}{c}{} \\
    $GPR_{\text{MEA}}$ & 0&0190 & 0&1715$^{***}$ & 0&7801$^{***}$ & $-$0&0345 & $-$0&6797 & $-$1&2717$^{***}$ & 1&3145 & 21.5630 & $-$10353.17 & 20719.39 & 20766.29 \\
    & (0&0226) & (0&0614) & (0&0601) & (0&0472) & (0&7978) & (0&4681) & (1&2014) &  \multicolumn{4}{c}{} \\
    $GPR_{\text{AO}}$ & 0&0217 & 0&1719$^{***}$ & 0&7877$^{***}$ & $-$0&0363 & 0&4367 & $-$0&5341$^{***}$ & 5&9603$^{**}$ & 10.8630 & $-$10361.84 & 20736.74 & 20783.64 \\
    & (0&0226) & (0&0598) & (0&0558) & (0&0460) & (0&4280) & (0&2045) & (2&4926) &  \multicolumn{4}{c}{} \\
    $GPR_{\text{mz,prod}}$ & 0&0221 & 0&1732$^{***}$ & 0&7790$^{***}$ & $-$0&0344 & 0&7515$^{**}$ & $-$0&9429$^{**}$ & 6&1565$^{*}$ & 19.4103 & $-$10355.91 & 20724.88 & 20771.78 \\
    & (0&0224) & (0&0608) & (0&0593) & (0&0469) & (0&3240) & (0&3887) & (3&3059) &  \multicolumn{4}{c}{} \\
    $GPR_{\text{mz,imp}}$ & 0&0221 & 0&1734$^{***}$ & 0&7830$^{***}$ & $-$0&0358 & 0&3142 & $-$0&6340$^{***}$ & 5&6171$^{**}$ & 13.7296 & $-$10358.93 & 20730.93 & 20777.83 \\
    & (0&0225) & (0&0602) & (0&0565) & (0&0466) & (0&4356) & (0&2231) & (2&6969) &  \multicolumn{4}{c}{} \\
    $GPR_{\text{mz,exp}}$ & 0&0223 & 0&1730$^{***}$ & 0&7709$^{***}$ & $-$0&0322 & 0&5958 & $-$1&1348$^{**}$ & 2&6187 & 32.0645 & $-$10345.99 & 20705.04 & 20751.94 \\
    & (0&0222) & (0&0594) & (0&0584) & (0&0465) & (0&3702) & (0&4646) & (2&4191) &  \multicolumn{4}{c}{} 
    \vspace{2mm}\\
    
    \multicolumn{19}{l}{\textit{Panel B: Single-factor models with rolling window}} \\
    $RV_{\text{mz}}^{(\text{rw})}$ & 0&0238 & 0&1779$^{***}$ & 0&7760$^{***}$ & $-$0&0407 & 0&9512$^{***}$ & 0&0060$^{**}$ & 1&1577 & 11.0585 & $-$10306.95 & 20626.91 & 20673.81 \\
    & (0&0224) & (0&0592) & (0&0588) & (0&0467) & (0&2681) & (0&0026) & (0&9580) &  \multicolumn{4}{c}{} \\
    $GPR^{(\text{rw})}$ & 0&0211 & 0&1705$^{***}$ & 0&7892$^{***}$ & $-$0&0342 & 4&6253$^{**}$ & $-$0&6916 & 6&6161$^{**}$ & 5.9025 & $-$10312.38 & 20637.77 & 20684.67 \\
    & (0&0227) & (0&0626) & (0&0621) & (0&0459) & (2&2853) & (0&4957) & (3&2252) &  \multicolumn{4}{c}{} \\ 
    $GPT^{(\text{rw})}$ & 0&0227 & 0&1684$^{***}$ & 0&7790$^{***}$ & $-$0&0309 & 6&9626$^{***}$ & $-$1&2093$^{**}$ & 5&0820 & 20.1519 & $-$10298.93 & 20610.86 & 20657.76 \\ 
    & (0&0225) & (0&0622) & (0&0659) & (0&0462) & (2&3455) & (0&5086) & (4&5075) &  \multicolumn{4}{c}{} \\ 
    $GPA^{(\text{rw})}$ & 0&0206 & 0&1679$^{***}$ & 0&7999$^{***}$ & $-$0&0347 & 0&6948 & 0&1736 & 1&0001 & 0.8406 & $-$10317.45 & 20647.91 & 20694.81 \\ 
    & (0&0234) & (0&0625) & (0&0585) & (0&0454) & (2&7561) & (0&5911) & (8&3436) &  \multicolumn{4}{c}{} \\
    $GPR_{\text{GLB}}^{(\text{rw})}$ & 0&0224 & 0&1711$^{***}$ & 0&7719$^{***}$ & $-$0&0307 & $-$0&0084 & $-$1&0492$^{**}$ & 4&3639 & 23.4249 & $-$10296.34 & 20605.68 & 20652.58 \\
    & (0&0224) & (0&0616) & (0&0643) & (0&0470) & (0&6171) & (0&4406) & (3&5358) &  \multicolumn{4}{c}{}  \\
    $GPR_{\text{NA}}^{(\text{rw})}$ & 0&0214 & 0&1701$^{***}$ & 0&7865$^{***}$ & $-$0&0339 & 1&3421$^{***}$ & $-$0&7214$^{*}$ & 7&7293$^{**}$ & 8.9148 & $-$10310.12 & 20633.25 & 20680.15 \\
    & (0&0227) & (0&0629) & (0&0638) & (0&0462) & (0&1863) & (0&4046) & (3&7284) &  \multicolumn{4}{c}{} \\
    $GPR_{\text{SA}}^{(\text{rw})}$ & 0&0205 & 0&1686$^{***}$ & 0&7957$^{***}$ & $-$0&0346 & 0&7087 & $-$0&2161 & 15&3763 & 2.2242 & $-$10316.22 & 20645.44 & 20692.34 \\
    & (0&0231) & (0&0620) & (0&0604) & (0&0446) & (0&6260) & (0&1708) & (10&3821) &  \multicolumn{4}{c}{} \\
    $GPR_{\text{ENE}}^{(\text{rw})}$ & 0&0208 & 0&1744$^{***}$ & 0&7592$^{***}$ & $-$0&0299 & $-$0&6414 & $-$1&2290$^{***}$ & 1&0152$^{*}$ & 33.0257 & $-$10284.48 & 20581.97 & 20628.87 \\
    & (0&0223) & (0&0582) & (0&0588) & (0&0463) & (0&5045) & (0&2936) & (0&5813) &  \multicolumn{4}{c}{} \\
    $GPR_{\text{ESW}}^{(\text{rw})}$ & 0&0220 & 0&1772$^{***}$ & 0&7560$^{***}$ & $-$0&0326 & $-$0&7661 & $-$1&2967$^{***}$ & 1&8615 & 31.6843 & $-$10281.97 & 20576.95 & 20623.85 \\
    & (0&0221) & (0&0592) & (0&0596) & (0&0476) & (0&6159) & (0&3591) & (1&5767) &  \multicolumn{4}{c}{} \\
    $GPR_{\text{MEA}}^{(\text{rw})}$ & 0&0178 & 0&1756$^{***}$ & 0&7688$^{***}$ & $-$0&0366 & $-$0&9365 & $-$1&4185$^{***}$ & 1&0000 & 20.9797 & $-$10298.76 & 20610.52 & 20657.42 \\
    & (0&0227) & (0&0616) & (0&0614) & (0&0484) & (0&6886) & (0&4036) & (0&6183) &  \multicolumn{4}{c}{} \\
    $GPR_{\text{AO}}^{(\text{rw})}$ & 0&0222 & 0&1708$^{***}$ & 0&7857$^{***}$ & $-$0&0347 & 0&4393 & $-$0&5283$^{***}$ & 6&5235$^{**}$ & 9.3911 & $-$10308.99 & 20630.98 & 20677.88 \\
    & (0&0228) & (0&0615) & (0&0608) & (0&0461) & (0&4152) & (0&1989) & (3&1132) &  \multicolumn{4}{c}{} \\
    $GPR_{\text{mz,prod}}^{(\text{rw})}$ & 0&0216 & 0&1711$^{***}$ & 0&7792$^{***}$ & $-$0&0322 & 0&7851$^{**}$ & $-$0&8923$^{**}$ & 6&4183$^{*}$ & 14.7098 & $-$10304.56 & 20622.14 & 20669.04 \\
    & (0&0226) & (0&0622) & (0&0640) & (0&0466) & (0&3255) & (0&3927) & (3&4669) &  \multicolumn{4}{c}{} \\
    $GPR_{\text{mz,imp}}^{(\text{rw})}$ & 0&0220 & 0&1707$^{***}$ & 0&7825$^{***}$ & $-$0&0336 & 0&3182 & $-$0&6261$^{***}$ & 5&8153$^{**}$ & 11.7039 & $-$10306.11 & 20625.24 & 20672.14 \\
    & (0&0227) & (0&0615) & (0&0612) & (0&0464) & (0&4241) & (0&2177) & (2&8404) &  \multicolumn{4}{c}{} \\
    $GPR_{\text{mz,exp}}^{(\text{rw})}$ & 0&0226 & 0&1733$^{***}$ & 0&7669$^{***}$ & $-$0&0314 & 0&5465 & $-$1&1895$^{**}$ & 2&1084 & 26.1937 & $-$10293.55 & 20600.12 & 20647.02 \\
    & (0&0223) & (0&0605) & (0&0620) & (0&0467) & (0&4266) & (0&5400) & (2&6843) &  \multicolumn{4}{c}{} \\
  \bottomrule
    \end{tabular}
    }%
  \begin{flushleft}
    \footnotesize
    \justifying Note: This table presents the parameter estimates and goodness-of-fit statistics of different single-factor GJR-GARCH-MIDAS models for the maize market based on fixed time span (Panel A) and rolling window (Panel B), respectively.
    \end{flushleft} 
  \label{Tab:Single_Factor_Estimation_Maize_Futures}%
\end{table}%

As shown in Table~\ref{Tab:Single_Factor_Estimation_Soybean_Futures}, similar to the maize market, significantly negative estimates of $\theta$ in the single-factor models corresponding to $GPT^{(\text{rw})}$ and $GPR_{\text{GLB}}^{(\text{rw})}$ indicate their considerably negative impact on the long-run volatility of the international soybean market. Further research based on diverse geographic regions reveals that geopolitical risks in North and South America, as well as in Europe, exhibit negative influences on the long-term soybean market volatility, and these effects are statistically significant. In particular, the GJR-GARCH-MIDAS model with $GPR_{\text{NA}}^{(\text{rw})}$ demonstrates the optimal fitting effect, as evidenced by its highest LLF value and lowest AIC and BIC values. Furthermore, the maximum VR associated with this model implies that the geopolitical risk in North America has an explanatory power of up to 64.74\% for the expected soybean market volatility. Moreover, the $\theta$ coefficients in single-factor models with $GPR_{\text{sb,prod}}^{(\text{rw})}$, $GPR_{\text{sb,imp}}^{(\text{rw})}$, and $GPR_{\text{sb,exp}}^{(\text{rw})}$ are all significantly negative, which reflects the substantial impact of geopolitical risks faced by important soybean producers, importers, and exporters on the long-term volatility of the soybean market. Notably, the model including $GPR_{\text{sb,prod}}^{(\text{rw})}$ has the largest VR, suggesting that compared to importing and exporting economies, the geopolitical risk from the combination of major soybean-producing economies provides a better explanation for the overall volatility of the international soybean market.

\begin{table}[!ht]
  \centering
  \scriptsize
  \setlength{\abovecaptionskip}{0pt}
  \setlength{\belowcaptionskip}{10pt}
  \caption{Estimation of single-factor models with fixed time span and rolling window for soybean}
  \setlength\tabcolsep{3pt}   \resizebox{\textwidth}{!}{ 
    \begin{tabular}{l r@{.}l r@{.}l r@{.}l r@{.}l r@{.}l r@{.}l r@{.}l rccc}
    \toprule
         & \multicolumn{2}{c}{$\mu$} & \multicolumn{2}{c}{$\alpha$} & \multicolumn{2}{c}{$\beta$} & \multicolumn{2}{c}{$\gamma$} & \multicolumn{2}{c}{$m$} & \multicolumn{2}{c}{$\theta$} & \multicolumn{2}{c}{$\omega$} & VR (\%) & LLF & AIC & BIC \\
    \midrule
    \multicolumn{19}{l}{\textit{Panel A: Single-factor models with fixed time span}} \\
    $RV_{\text{sb}}$ & 0&0311$^{*}$ & 0&0787$^{***}$ & 0&9176$^{***}$ & $-$0&0136 & 0&8153$^{***}$ & 0&0054$^{*}$ & 1&0000$^{**}$ & 3.2516 & $-$9404.32 & 18821.71 & 18868.61 \\
    & (0&0188) & (0&0108) & (0&0089) & (0&0165) & (0&2721) & (0&0029) & (0&4386) &  \multicolumn{4}{c}{} \\
    $GPR$ & 0&0308$^{*}$ & 0&0790$^{***}$ & 0&9205$^{***}$ & $-$0&0153 & 6&4809 & $-$1&1559 & 3&2977$^{**}$ & 13.4956 & $-$9402.30 & 18817.65 & 18864.55 \\
    & (0&0187) & (0&0106) & (0&0084) & (0&0159) & (5&1427) & (1&0910) & (1&4839) &  \multicolumn{4}{c}{} \\ 
    $GPT$ & 0&0306 & 0&0783$^{***}$ & 0&9183$^{***}$ & $-$0&0131 & 7&3063$^{**}$ & $-$1&3458$^{*}$ & 3&5475 & 19.7857 & $-$9400.78 & 18814.62 & 18861.52 \\ 
    & (0&0187) & (0&0106) & (0&0086) & (0&0161) & (3&7210) & (0&7995) & (3&7568) &  \multicolumn{4}{c}{} \\ 
    $GPA$ & 0&0310$^{*}$ & 0&0786$^{***}$ & 0&9187$^{***}$ & $-$0&0140 & 1&0941$^{***}$ & 0&0018 & 1&7525 & 0.0001 & $-$9405.49 & 18824.04 & 18870.94 \\ 
    & (0&0188) & (0&0108) & (0&0088) & (0&0163) & (0&2362) & (0&0024) & (1&9212) &  \multicolumn{4}{c}{} \\
    $GPR_{\text{GLB}}$ & 0&0307$^{*}$ & 0&0782$^{***}$ & 0&9179$^{***}$ & $-$0&0125 & $-$0&4694 & $-$1&1780$^{*}$ & 2&5432$^{**}$ & 21.7968 & $-$9400.12 & 18813.30 & 18860.20 \\
    & (0&0186) & (0&0107) & (0&0086) & (0&0161) & (0&8166) & (0&6236) & (1&1529) &  \multicolumn{4}{c}{} \\
    $GPR_{\text{NA}}$ & 0&0322$^{*}$ & 0&0789$^{***}$ & 0&9215$^{***}$ & $-$0&0152 & 0&9926$^{***}$ & $-$1&8850 & 3&2240$^{*}$ & 47.8715 & $-$9398.78 & 18810.62 & 18857.52 \\
    & (0&0185) & (0&0104) & (0&0080) & (0&0156) & (0&2962) & (1&1581) & (1&7346) &  \multicolumn{4}{c}{} \\
    $GPR_{\text{SA}}$ & 0&0320$^{*}$ & 0&0764$^{***}$ & 0&9189$^{***}$ & $-$0&0113 & $-$1&2591 & $-$0&6716$^{*}$ & 6&8536 & 16.3796 & $-$9398.50 & 18810.06 & 18856.96 \\
    & (0&0185) & (0&0104) & (0&0085) & (0&0158) & (1&3147) & (0&3957) & (9&9494) &  \multicolumn{4}{c}{} \\
    $GPR_{\text{ENE}}$ & 0&0319$^{*}$ & 0&0798$^{***}$ & 0&9142$^{***}$ & $-$0&0117 & $-$0&3744 & $-$0&8959$^{*}$ & 1&0004 & 17.3995 & $-$9399.83 & 18812.72 & 18859.62 \\
    & (0&0187) & (0&0112) & (0&0095) & (0&0169) & (0&9075) & (0&4692) & (1&7111) &  \multicolumn{4}{c}{} \\
    $GPR_{\text{ESW}}$ & 0&0309$^{*}$ & 0&0793$^{***}$ & 0&9164$^{***}$ & $-$0&0140 & $-$0&5049 & $-$0&9622$^{**}$ & 1&9267$^{*}$ & 17.1240 & $-$9401.17 & 18815.41 & 18862.31 \\
    & (0&0186) & (0&0109) & (0&0089) & (0&0162) & (0&7309) & (0&4376) & (1&0642) &  \multicolumn{4}{c}{} \\
    $GPR_{\text{MEA}}$ & 0&0309 & 0&0785$^{***}$ & 0&9191$^{***}$ & $-$0&0141 & $-$0&1641 & $-$0&7878 & 3&5176$^{***}$ & 8.9274 & $-$9402.98 & 18819.02 & 18865.92 \\
    & (0&0188) & (0&0111) & (0&0087) & (0&0164) & (0&8423) & (0&4962) & (1&2614) &  \multicolumn{4}{c}{} \\
    $GPR_{\text{AO}}$ & 0&0316$^{*}$ & 0&0781$^{***}$ & 0&9182$^{***}$ & $-$0&0137 & $-$0&0626 & $-$0&6163 & 7&2543 & 11.6471 & $-$9401.62 & 18816.29 & 18863.19 \\
    & (0&0191) & (0&0118) & (0&0084) & (0&0179) & (1&9345) & (0&9831) & (55&9963) &  \multicolumn{4}{c}{} \\
    $GPR_{\text{sb,prod}}$ & 0&0313$^{*}$ & 0&0782$^{***}$ & 0&9187$^{***}$ & $-$0&0134 & 0&2056 & $-$1&2685$^{*}$ & 2&2806$^{*}$ & 25.5031 & $-$9400.37 & 18813.79 & 18860.69 \\
    & (0&0186) & (0&0106) & (0&0085) & (0&0159) & (0&5102) & (0&7095) & (1&2395) &  \multicolumn{4}{c}{} \\
    $GPR_{\text{sb,imp}}$ & 0&0313$^{*}$ & 0&0781$^{***}$ & 0&9173$^{***}$ & $-$0&0127 & $-$0&3987 & $-$0&8943$^{*}$ & 3&4552$^{**}$ & 15.9681 & $-$9401.41 & 18815.87 & 18862.77 \\
    & (0&0186) & (0&0106) & (0&0087) & (0&0161) & (0&8150) & (0&4936) & (1&6725) &  \multicolumn{4}{c}{} \\
    $GPR_{\text{sb,exp}}$ & 0&0308$^{*}$ & 0&0780$^{***}$ & 0&9177$^{***}$ & $-$0&0123 & 0&2210 & $-$1&1592$^{*}$ & 2&0597$^{*}$ & 21.9188 & $-$9400.41 & 18813.88 & 18860.78 \\
    & (0&0186) & (0&0106) & (0&0087) & (0&0161) & (0&4897) & (0&6205) & (1&1015) &  \multicolumn{4}{c}{} 
    \vspace{2mm}\\
    
    \multicolumn{19}{l}{\textit{Panel B: Single-factor models with rolling window}} \\
    $RV_{\text{sb}}^{(\text{rw})}$ & 0&0302 & 0&0779$^{***}$ & 0&9178$^{***}$ & $-$0&0128 & 0&7709$^{***}$ & 0&0054$^{*}$ & 1&0058 & 3.1876 & $-$9349.34 & 18711.70 & 18758.60 \\
    & (0&0190) & (0&0109) & (0&0091) & (0&0165) & (0&2706) & (0&0033) & (0&8045) &  \multicolumn{4}{c}{} \\
    $GPR^{(\text{rw})}$ & 0&0304 & 0&0775$^{***}$ & 0&9227$^{***}$ & $-$0&0143 & 11&6841 & $-$2&2843 & 3&0591$^{**}$ & 44.0718 & $-$9345.35 & 18703.70 & 18750.60 \\
    & (0&0189) & (0&0101) & (0&0078) & (0&0155) & (7&2130) & (1&5371) & (1&2916) &  \multicolumn{4}{c}{} \\
    $GPT^{(\text{rw})}$ & 0&0307 & 0&0770$^{***}$ & 0&9199$^{***}$ & $-$0&0127 & 8&6524$^{**}$ & $-$1&6370$^{**}$ & 3&9018 & 27.0810 & $-$9344.47 & 18701.94 & 18748.84 \\ 
     & (0&0188) & (0&0103) & (0&0082) & (0&0159) & (3&7547) & (0&8017) & (3&6956) &  \multicolumn{4}{c}{} \\
    $GPA^{(\text{rw})}$ & 0&0299 & 0&0771$^{***}$ & 0&9205$^{***}$ & $-$0&0135 & 2&1381 & $-$0&2305 & 4&0927$^{***}$ & 1.4017 & $-$9350.32 & 18713.64 & 18760.54 \\ 
    & (0&0190) & (0&0106) & (0&0093) & (0&0160) & (2&3275) & (0&4966) & (1&2938) &  \multicolumn{4}{c}{} \\
    $GPR_{\text{GLB}}^{(\text{rw})}$ & 0&0301 & 0&0766$^{***}$ & 0&9195$^{***}$ & $-$0&0119 & $-$0&8279 & $-$1&4360$^{**}$ & 2&8823$^{**}$ & 30.3854 & $-$9344.09 & 18701.18 & 18748.08 \\
    & (0&0188) & (0&0103) & (0&0083) & (0&0158) & (0&8649) & (0&6672) & (1&3554) &  \multicolumn{4}{c}{} \\
    $GPR_{\text{NA}}^{(\text{rw})}$ & 0&0314$^{*}$ & 0&0771$^{***}$ & 0&9227$^{***}$ & $-$0&0136 & 0&8966$^{***}$ & $-$2&3348$^{**}$ & 3&5123$^{**}$ & 64.7375 & $-$9341.34 & 18695.69 & 18742.59 \\
    & (0&0187) & (0&0099) & (0&0076) & (0&0153) & (0&2875) & (1&1092) & (1&6046) &  \multicolumn{4}{c}{} \\
    $GPR_{\text{SA}}^{(\text{rw})}$ & 0&0308$^{*}$ & 0&0750$^{***}$ & 0&9200$^{***}$ & $-$0&0106 & $-$1&6053 & $-$0&7620$^{**}$ & 5&8168 & 17.9079 & $-$9342.83 & 18698.67 & 18745.57 \\
    & (0&0187) & (0&0101) & (0&0084) & (0&0154) & (1&2829) & (0&3842) & (5&2831) &  \multicolumn{4}{c}{} \\
    $GPR_{\text{ENE}}^{(\text{rw})}$ & 0&0301 & 0&0769$^{***}$ & 0&9166$^{***}$ & $-$0&0116 & $-$0&4935 & $-$0&9354$^{*}$ & 1&0000 & 16.6966 & $-$9344.89 & 18702.79 & 18749.69 \\
    & (0&0187) & (0&0108) & (0&0091) & (0&0160) & (0&8486) & (0&4888) & (1&2643) &  \multicolumn{4}{c}{} \\
    $GPR_{\text{ESW}}^{(\text{rw})}$ & 0&0303 & 0&0776$^{***}$ & 0&9173$^{***}$ & $-$0&0127 & $-$0&6447 & $-$1&0372$^{**}$ & 2&2937$^{*}$ & 19.1267 & $-$9345.46 & 18703.94 & 18750.84 \\
    & (0&0188) & (0&0107) & (0&0088) & (0&0161) & (0&7216) & (0&4367) & (1&2380) &  \multicolumn{4}{c}{} \\
    $GPR_{\text{MEA}}^{(\text{rw})}$ & 0&0301 & 0&0771$^{***}$ & 0&9198$^{***}$ & $-$0&0133 & $-$0&1558 & $-$0&7602 & 4&0925$^{**}$ & 7.7856 & $-$9347.95 & 18708.91 & 18755.81 \\
    & (0&0189) & (0&0109) & (0&0087) & (0&0162) & (0&9290) & (0&5446) & (1&6844) &  \multicolumn{4}{c}{} \\
    $GPR_{\text{AO}}^{(\text{rw})}$ & 0&0307 & 0&0763$^{***}$ & 0&9190$^{***}$ & $-$0&0117 & $-$0&2469 & $-$0&7002 & 7&0810 & 13.7517 & $-$9346.01 & 18705.04 & 18751.94 \\
    & (0&0190) & (0&0106) & (0&0084) & (0&0161) & (0&9203) & (0&4892) & (20&9388) &   \multicolumn{4}{c}{} \\
    $GPR_{\text{sb,prod}}^{(\text{rw})}$ & 0&0301 & 0&0763$^{***}$ & 0&9196$^{***}$ & $-$0&0115 & 0&1055 & $-$1&3825$^{*}$ & 2&5188$^{**}$ & 28.4244 & $-$9344.62 & 18702.25 & 18749.15 \\
    & (0&0188) & (0&0103) & (0&0084) & (0&0157) & (0&5066) & (0&7152) & (1&2563) &  \multicolumn{4}{c}{} \\
    $GPR_{\text{sb,imp}}^{(\text{rw})}$ & 0&0305 & 0&0769$^{***}$ & 0&9180$^{***}$ & $-$0&0116 & $-$0&6145 & $-$1&0182$^{**}$ & 4&0338$^{*}$ & 19.8323 & $-$9345.43 & 18703.87 & 18750.77 \\
    & (0&0188) & (0&0105) & (0&0086) & (0&0160) & (0&8152) & (0&5009) & (2&4019) &  \multicolumn{4}{c}{} \\
    $GPR_{\text{sb,exp}}^{(\text{rw})}$ & 0&0297 & 0&0764$^{***}$ & 0&9188$^{***}$ & $-$0&0113 & 0&1014 & $-$1&2849$^{**}$ & 2&2795$^{**}$ & 25.4342 & $-$9344.78 & 18702.57 & 18749.47 \\
    & (0&0187) & (0&0104) & (0&0085) & (0&0158) & (0&4864) & (0&6299) & (1&1536) &  \multicolumn{4}{c}{} \\
  \bottomrule
    \end{tabular}
    }%
  \begin{flushleft}
    \footnotesize
    \justifying Note: This table gives the estimated results of different single-factor GJR-GARCH-MIDAS models for the soybean market with fixed time span (Panel A) and rolling window (Panel B), respectively.
    \end{flushleft} 
  \label{Tab:Single_Factor_Estimation_Soybean_Futures}%
\end{table}%

Based on Table~\ref{Tab:Single_Factor_Estimation_Rice_Futures}, in addition to the different results regarding realized volatility, the estimated results of the single-factor models with geopolitical risk factors for rice also differ from those for the other three grains. Specifically, the estimates of $\theta$ associated with geopolitical risk factors at the global level are all positive, with the coefficients for $GPR^{(\text{rw})}$ and $GPA^{(\text{rw})}$ being statistically significant, meaning that global geopolitical risk and geopolitical acts tend to exacerbate the long-term rice market volatility markedly. Additionally, the two single-factor models exhibit higher values for VR and LLF, as well as lower values for AIC and BIC, which suggests that these models have better fitting performance and can effectively account for the expected volatility of the international rice market. Besides, from the estimated results for geopolitical risks in the six major geographic regions, we note that although the estimates of $\theta$ in these models are consistently positive, only the corresponding estimate of $GPR_{\text{NA}}^{(\text{rw})}$ is statistically significant. Furthermore, the model containing $GPR_{\text{NA}}^{(\text{rw})}$ also demonstrates the best fitting effect and the largest VR, indicating that the geopolitical risk in North America can serve as an important macro-factor in explaining the overall volatility of the rice market. The United States, situated in North America, has consistently been one of the world's leading rice exporters for many years, whereas Mexico has maintained its position as a major rice importer on the global stage, making the above finding of great relevance. Moreover, the $\theta$ coefficients of $GPR_{\text{rc,prod}}^{(\text{rw})}$, $GPR_{\text{rc,imp}}^{(\text{rw})}$, and $GPR_{\text{rc,exp}}^{(\text{rw})}$ are all insignificantly positive, implying that the composite geopolitical risks faced by different combinations of economies influential in rice production, import, and export may intensify the long-run volatility of the international rice market, but these effects are not statistically significant.

\begin{table}[!ht]
  \centering
  \scriptsize
  \setlength{\abovecaptionskip}{0pt}
  \setlength{\belowcaptionskip}{10pt}
  \caption{Estimation of single-factor models with fixed time span and rolling window for rice}
  \setlength\tabcolsep{3pt}   \resizebox{\textwidth}{!}{ 
    \begin{tabular}{l r@{.}l r@{.}l r@{.}l r@{.}l r@{.}l r@{.}l r@{.}l rccc}
    \toprule
         & \multicolumn{2}{c}{$\mu$} & \multicolumn{2}{c}{$\alpha$} & \multicolumn{2}{c}{$\beta$} & \multicolumn{2}{c}{$\gamma$} & \multicolumn{2}{c}{$m$} & \multicolumn{2}{c}{$\theta$} & \multicolumn{2}{c}{$\omega$} & VR (\%) & LLF & AIC & BIC \\
    \midrule
    \multicolumn{19}{l}{\textit{Panel A: Single-factor models with fixed time span}} \\
    $RV_{\text{rc}}$ & 0&0534$^{***}$ & 0&0678$^{***}$ & 0&9189$^{***}$ & 0&0066 & 1&5179$^{***}$ & $-$0&0038 & 5&8063$^{**}$ & 14.4246 & $-$9669.79 & 19352.64 & 19399.54 \\
    & (0&0198) & (0&0121) & (0&0130) & (0&0194) & (0&4797) & (0&0038) & (2&5899) &  \multicolumn{4}{c}{} \\
    $GPR$ & 0&0473$^{**}$ & 0&0709$^{***}$ & 0&9065$^{***}$ & 0&0075 & $-$7&3742$^{*}$ & 1&8474$^{**}$ & 1&3873$^{***}$ & 25.9881 & $-$9663.12 & 19339.31 & 19386.21 \\
    & (0&0195) & (0&0135) & (0&0119) & (0&0204) & (3&8745) & (0&8567) & (0&4670) &  \multicolumn{4}{c}{} \\ 
    $GPT$ & 0&0511$^{***}$ & 0&0713$^{***}$ & 0&9103$^{***}$ & 0&0066 & $-$3&0435 & 0&9213 & 1&2433$^{***}$ & 5.5882 & $-$9672.85 & 19358.76 & 19405.66 \\ 
    & (0&0195) & (0&0128) & (0&0112) & (0&0199) & (3&8162) & (0&8493) & (0&4289) &  \multicolumn{4}{c}{} \\ 
    $GPA$ & 0&0507$^{***}$ & 0&0705$^{***}$ & 0&9046$^{***}$ & 0&0077 & $-$2&5444 & 0&8011$^{*}$ & 2&0249 & 17.5057 & $-$9665.50 & 19344.06 & 19390.96 \\ 
    & (0&0196) & (0&0134) & (0&0127) & (0&0210) & (1&8725) & (0&4147) & (2&6839) &  \multicolumn{4}{c}{} \\
        $GPR_{\text{GLB}}$ & 0&0525$^{***}$ & 0&0709$^{***}$ & 0&9104$^{***}$ & 0&0072 & 1&4565$^{*}$ & 0&2061 & 10&0062 & 1.0455 & $-$9674.64 & 19362.34 & 19409.24 \\
    & (0&0197) & (0&0126) & (0&0110) & (0&0203) & (0&7662) & (0&5232) & (97&2675) &  \multicolumn{4}{c}{} \\
    $GPR_{\text{NA}}$ & 0&0481$^{**}$ & 0&0716$^{***}$ & 0&9061$^{***}$ & 0&0074 & 1&3063$^{***}$ & 1&4786$^{**}$ & 1&4203$^{***}$ & 24.3209 & $-$9664.78 & 19342.62 & 19389.52 \\
    & (0&0194) & (0&0137) & (0&0121) & (0&0205) & (0&2866) & (0&7276) & (0&5142) &  \multicolumn{4}{c}{} \\
    $GPR_{\text{SA}}$ & 0&0497$^{**}$ & 0&0707$^{***}$ & 0&9091$^{***}$ & 0&0079 & 2&5701$^{*}$ & 0&4006 & 1&4164 & 3.7826 & $-$9673.00 & 19359.06 & 19405.96 \\
    & (0&0195) & (0&0132) & (0&0114) & (0&0207) & (1&4638) & (0&4135) & (1&0878) &  \multicolumn{4}{c}{} \\
    $GPR_{\text{ENE}}$ & 0&0521$^{***}$ & 0&0712$^{***}$ & 0&9109$^{***}$ & 0&0066 & 1&5707 & 0&2310 & 1&0001 & 1.1413 & $-$9674.53 & 19362.11 & 19409.01 \\
    & (0&0196) & (0&0125) & (0&0112) & (0&0201) & (1&0757) & (0&5700) & (1&8660) &  \multicolumn{4}{c}{} \\
    $GPR_{\text{ESW}}$ & 0&0521$^{***}$ & 0&0717$^{***}$ & 0&9109$^{***}$ & 0&0060 & 2&1966$^{*}$ & 0&6114 & 1&5445 & 6.2258 & $-$9672.94 & 19358.95 & 19405.85 \\
    & (0&0196) & (0&0127) & (0&0112) & (0&0197) & (1&1459) & (0&6020) & (1&2511) &  \multicolumn{4}{c}{} \\
    $GPR_{\text{MEA}}$ & 0&0520$^{***}$ & 0&0701$^{***}$ & 0&9118$^{***}$ & 0&0088 & 0&6867 & $-$0&3249 & 12&7015$^{**}$ & 2.1980 & $-$9673.79 & 19360.63 & 19407.53 \\
    & (0&0197) & (0&0125) & (0&0109) & (0&0202) & (0&5427) & (0&3167) & (5&2591) &  \multicolumn{4}{c}{} \\
    $GPR_{\text{AO}}$ & 0&0519$^{***}$ & 0&0713$^{***}$ & 0&9109$^{***}$ & 0&0063 & 1&4098$^{**}$ & 0&1148 & 8&6841 & 0.4139 & $-$9674.97 & 19362.99 & 19409.89 \\
    & (0&0196) & (0&0125) & (0&0109) & (0&0202) & (0&6184) & (0&2902) & (11&7754) &  \multicolumn{4}{c}{} \\
    $GPR_{\text{rc,prod}}$ & 0&0491$^{**}$ & 0&0717$^{***}$ & 0&9084$^{***}$ & 0&0071 & 2&1337$^{***}$ & 1&2055 & 1&3668$^{**}$ & 15.1012 & $-$9669.05 & 19351.16 & 19398.06 \\
    & (0&0195) & (0&0133) & (0&0117) & (0&0202) & (0&7931) & (0&8200) & (0&6569) &  \multicolumn{4}{c}{} \\
    $GPR_{\text{rc,imp}}$ & 0&0519$^{***}$ & 0&0718$^{***}$ & 0&9102$^{***}$ & 0&0058 & 2&2664 & 0&5980 & 2&3699 & 5.2011 & $-$9673.39 & 19359.84 & 19406.75 \\
    & (0&0196) & (0&0129) & (0&0115) & (0&0199) & (1&7267) & (0&8658) & (5&3882) &  \multicolumn{4}{c}{} \\
    $GPR_{\text{rc,exp}}$ & 0&0507$^{***}$ & 0&0719$^{***}$ & 0&9101$^{***}$ & 0&0058 & 1&9172$^{**}$ & 1&1063 & 1&0512$^{**}$ & 11.1702 & $-$9670.81 & 19354.69 & 19401.59 \\
    & (0&0195) & (0&0131) & (0&0114) & (0&0198) & (0&7688) & (0&8980) & (0&4614) &  \multicolumn{4}{c}{} 
    \vspace{2mm}\\
    
    \multicolumn{19}{l}{\textit{Panel B: Single-factor models with rolling window}} \\
    $RV_{\text{rc}}^{(\text{rw})}$ & 0&0512$^{***}$ & 0&0658$^{***}$ & 0&9231$^{***}$ & 0&0063 & 1&6624$^{***}$ & $-$0&0048 & 6&1661$^{**}$ & 23.6831 & $-$9593.91 & 19200.83 & 19247.73 \\
    & (0&0198) & (0&0129) & (0&0147) & (0&0189) & (0&5241) & (0&0051) & (2&9494) &  \multicolumn{4}{c}{} \\
    $GPR^{(\text{rw})}$ & 0&0474$^{**}$ & 0&0711$^{***}$ & 0&9061$^{***}$ & 0&0075 & $-$6&0912 & 1&5704$^{*}$ & 1&5470$^{***}$ & 17.4217 & $-$9590.24 & 19193.49 & 19240.39 \\
    & (0&0195) & (0&0137) & (0&0119) & (0&0208) & (3&9130) & (0&8610) & (0&5176) &  \multicolumn{4}{c}{} \\
    $GPT^{(\text{rw})}$ & 0&0491$^{**}$ & 0&0708$^{***}$ & 0&9106$^{***}$ & 0&0079 & $-$2&3032 & 0&7641 & 1&2581$^{**}$ & 3.3969 & $-$9598.71 & 19210.42 & 19257.32 \\ 
     & (0&0195) & (0&0130) & (0&0110) & (0&0204) & (3&8053) & (0&8467) & (0&6299) &  \multicolumn{4}{c}{} \\
    $GPA^{(\text{rw})}$ & 0&0481$^{**}$ & 0&0700$^{***}$ & 0&9054$^{***}$ & 0&0085 & $-$2&8841 & 0&8782$^{**}$ & 2&1166 & 18.6340 & $-$9591.45 & 19195.91 & 19242.81 \\ 
    & (0&0196) & (0&0135) & (0&0123) & (0&0212) & (1&8549) & (0&4123) & (1&8736) &  \multicolumn{4}{c}{} \\
    $GPR_{\text{GLB}}^{(\text{rw})}$ & 0&0487$^{**}$ & 0&0710$^{***}$ & 0&9102$^{***}$ & 0&0081 & 2&1965$^{*}$ & 0&7396 & 1&2040$^{**}$ & 5.2362 & $-$9598.17 & 19209.35 & 19256.25 \\
    & (0&0195) & (0&0130) & (0&0110) & (0&0203) & (1&1288) & (0&7314) & (0&5193) &  \multicolumn{4}{c}{} \\
    $GPR_{\text{NA}}^{(\text{rw})}$ & 0&0466$^{**}$ & 0&0713$^{***}$ & 0&9065$^{***}$ & 0&0083 & 1&3100$^{***}$ & 1&4144$^{*}$ & 1&5256$^{***}$ & 20.0716 & $-$9591.31 & 19195.63 & 19242.53 \\
    & (0&0195) & (0&0138) & (0&0119) & (0&0207) & (0&2928) & (0&7230) & (0&5010) &  \multicolumn{4}{c}{} \\
    $GPR_{\text{SA}}^{(\text{rw})}$ & 0&0468$^{**}$ & 0&0694$^{***}$ & 0&9089$^{***}$ & 0&0121 & 2&3483 & 0&3298 & 1&6729 & 2.3367 & $-$9598.55 & 19210.10 & 19257.00 \\
    & (0&0196) & (0&0131) & (0&0114) & (0&0213) & (1&4916) & (0&4244) & (1&6532) &  \multicolumn{4}{c}{} \\
    $GPR_{\text{ENE}}^{(\text{rw})}$ & 0&0503$^{**}$ & 0&0705$^{***}$ & 0&9112$^{***}$ & 0&0081 & 1&4692$^{**}$ & 0&1571 & 1&1195$^{**}$ & 0.4884 & $-$9600.16 & 19213.33 & 19260.24 \\
    & (0&0196) & (0&0127) & (0&0108) & (0&0205) & (0&6630) & (0&3192) & (0&5149) &  \multicolumn{4}{c}{} \\
    $GPR_{\text{ESW}}^{(\text{rw})}$ & 0&0499$^{**}$ & 0&0713$^{***}$ & 0&9108$^{***}$ & 0&0076 & 2&0620$^{**}$ & 0&5165 & 1&7144$^{*}$ & 4.1030 & $-$9598.87 & 19210.75 & 19257.65 \\
    & (0&0196) & (0&0129) & (0&0109) & (0&0203) & (1&0081) & (0&5268) & (0&9277) &  \multicolumn{4}{c}{} \\
    $GPR_{\text{MEA}}^{(\text{rw})}$ & 0&0491$^{**}$ & 0&0693$^{***}$ & 0&9125$^{***}$ & 0&0100 & 0&5675 & $-$0&4048 & 12&2168$^{**}$ & 2.8953 & $-$9599.00 & 19211.02 & 19257.92\\
    & (0&0196) & (0&0128) & (0&0109) & (0&0207) & (0&6287) & (0&3789) & (5&9337) &  \multicolumn{4}{c}{} \\
    $GPR_{\text{AO}}^{(\text{rw})}$ & 0&0490$^{**}$ & 0&0704$^{***}$ & 0&9108$^{***}$ & 0&0093 & 1&3981$^{*}$ & 0&0911 & 1&9745 & 0.1631 & $-$9600.30 & 19213.62 & 19260.52 \\
    & (0&0196) & (0&0127) & (0&0110) & (0&0207) & (0&8059) & (0&3658) & (3&5774) &  \multicolumn{4}{c}{} \\
    $GPR_{\text{rc,prod}}^{(\text{rw})}$ & 0&0474$^{**}$ & 0&0715$^{***}$ & 0&9088$^{***}$ & 0&0077 & 2&0971$^{***}$ & 1&1371 & 1&4454$^{**}$ & 12.1294 & $-$9595.33 & 19203.67 & 19250.57 \\
    & (0&0195) & (0&0136) & (0&0115) & (0&0206) & (0&7886) & (0&8067) & (0&6759) &  \multicolumn{4}{c}{} \\
    $GPR_{\text{rc,imp}}^{(\text{rw})}$ & 0&0500$^{**}$ & 0&0707$^{***}$ & 0&9104$^{***}$ & 0&0084 & 1&9961 & 0&4375 & 2&1509 & 2.3945 & $-$9599.31 & 19211.63 & 19258.53 \\
    & (0&0196) & (0&0128) & (0&0111) & (0&0204) & (1&4144) & (0&7219) & (2&3091) &  \multicolumn{4}{c}{} \\
    $GPR_{\text{rc,exp}}^{(\text{rw})}$ & 0&0483$^{**}$ & 0&0716$^{***}$ & 0&9098$^{***}$ & 0&0071 & 1&8144$^{***}$ & 0&9376 & 1&2676$^{**}$ & 7.5309 & $-$9597.32 & 19207.65 & 19254.55 \\
    & (0&0195) & (0&0132) & (0&0111) & (0&0203) & (0&7020) & (0&8074) & (0&5111) &  \multicolumn{4}{c}{} \\
  \bottomrule
    \end{tabular}
    }%
  \begin{flushleft}
    \footnotesize
    \justifying Note: This table shows the parameter estimates and goodness-of-fit statistics of different single-factor GJR-GARCH-MIDAS models for the rice market with fixed time span (Panel A) and rolling window (Panel B), respectively.
    \end{flushleft} 
  \label{Tab:Single_Factor_Estimation_Rice_Futures}%
\end{table}%

\subsection{Estimation of GJR-GARCH-MIDAS models with two factors}

In order to simultaneously evaluate the impact of realized volatility and geopolitical risk on the long-term component of volatility in the international food market, we further construct two-factor GJR-GARCH-MIDAS models that include both RV and geopolitical risk factors. Tables~\ref{Tab:Two_Factor_Estimation_Wheat_Futures}--\ref{Tab:Two_Factor_Estimation_Rice_Futures} present the estimated results of various two-factor GJR-GARCH-MIDAS models for wheat, maize, soybean, and rice, respectively, where Panels A and B correspond to fixed-span and rolling-window models. The comparison reveals that two-factor models with fixed time span exhibit larger variance ratios, thus possessing stronger explanatory power for overall fluctuations. However, two-factor models based on rolling window consistently perform better, showing larger LLF values, as well as smaller AIC and BIC values. Therefore, considering the goodness of fit of these models, we continue to focus on the estimated results of two-factor GJR-GARCH-MIDAS models with rolling window in the following analysis.

Similar to the modeling results of short-term fluctuations in single-factor models, the coefficients of the ARCH and GARCH terms in all the two-factor GJR-GARCH-MIDAS models are statistically significant at the 1\% level, corroborating the finding that futures returns of the four staple food grains exhibit apparent volatility clustering. This means that positive shocks typically increase short-run volatility in agricultural futures markets. Additionally, the estimates of $\gamma$ are positive for rice and negative for wheat, maize, and soybean, but none are statistically significant. Accordingly, there is no significant asymmetry in the short-term market volatility for these four major grains. Furthermore, the sum of the estimates of $\alpha$, $\beta$, and 0.5$\gamma$ in each model is less than and very close to 1, indicating the stability of the model and the high persistence of short-term fluctuations.

From the modeling results of the long-term volatility, it can be found that the estimates of the parameters $\omega_{RV}$ and $\omega_{MV}$ in all the models are greater than 1, which suggests that the weighting functions corresponding to realized volatility and geopolitical risk are both decreasing functions. That is to say, the closer the time period is to the current period, the greater the impact of these explanatory variables on food market volatility. Moreover, $\theta_{RV}$ and $\theta_{MV}$ represent the coefficients that measure the effects of realized volatility and geopolitical risk on the long-run component of volatility, respectively. Due to the different performance of models for wheat, maize, soybean, and rice, we continue the analytical framework of single-factor models and discuss the estimated results of the long-term volatility in two-factor models for each grain market individually.

As shown in Table~\ref{Tab:Two_Factor_Estimation_Wheat_Futures}, the influence of realized volatility is pronounced for wheat, with the estimate of $\theta_{RV}$ being significantly positive at the 1\% level in each model. This is consistent with the findings of single-factor models incorporating RV as the explanatory variable, reaffirming that realized volatility significantly increases the long-term volatility in the wheat market. In contrast, the estimates of $\theta_{MV}$ only show statistical significance in a few models. Specifically, $GPR_{\text{SA}}^{(\text{rw})}$ and $GPR_{\text{MEA}}^{(\text{rw})}$ correspond to significantly positive and negative estimates, respectively, indicating that after excluding the effect of realized volatility, heightened geopolitical risk in South America contributes to the increase in long-run wheat market fluctuations, whereas the opposite is true for geopolitical risk in the Middle East and Africa. On the one hand, South America is one of the most important food-producing regions in the world, playing a crucial role in the production and supply of grains. On the other hand, the Middle East and Africa face severe challenges in agricultural production and development due to climate conditions, regional conflicts, and weak agricultural foundations, resulting in low food self-sufficiency and heavy reliance on imports. According to USDA statistics, wheat yields in the Middle East and Africa represent only 8.35\% of the total global output, while its wheat imports account for 44.18\% of the global total. This partly explains why elevated geopolitical risks faced by South America as well as the Middle East and Africa have equally significant but opposite effects on the overall volatility in the international wheat market. Furthermore, the two-factor models containing both $RV_{\text{wh}}^{(\text{rw})}$ and $GPR_{\text{SA}}^{(\text{rw})}$ or $GPR_{\text{MEA}}^{(\text{rw})}$ demonstrate strong explanatory power and high goodness-of-fit in line with statistical significance.

\begin{table}[!ht]
  \centering
  \setlength{\abovecaptionskip}{0pt}
  \setlength{\belowcaptionskip}{10pt}
  \caption{Estimation of two-factor models with fixed time span and rolling window for wheat}
  \setlength\tabcolsep{3pt}   \resizebox{\textwidth}{!}{ 
    \begin{tabular}{l r@{.}l r@{.}l r@{.}l r@{.}l r@{.}l r@{.}l r@{.}l r@{.}l r@{.}l rccc}
    \toprule
         & \multicolumn{2}{c}{$\mu$} & \multicolumn{2}{c}{$\alpha$} & \multicolumn{2}{c}{$\beta$} & \multicolumn{2}{c}{$\gamma$} & \multicolumn{2}{c}{$m$} & \multicolumn{2}{c}{$\theta_{RV}$} & \multicolumn{2}{c}{$\theta_{MV}$} & \multicolumn{2}{c}{$\omega_{RV}$} & \multicolumn{2}{c}{$\omega_{MV}$} & VR (\%) & LLF & AIC & BIC \\
    \midrule
    \multicolumn{23}{l}{\textit{Panel A: Two-factor models with fixed time span}} \\
    $RV_{\text{wh}} + GPR$ & 0&0136 & 0&1080$^{***}$ & 0&8301$^{***}$ & $-$0&0449 & 0&0023 & 0&0056$^{***}$ & 0&1890$^{***}$ & 6&3031$^{***}$ & 2&4266 & 47.8220 & $-$11065.96 & 22148.71 & 22209.01 \\
    & (0&0263) & (0&0159) & (0&0247) & (0&0283) & (0&1107) & (0&0008) & (0&0331) & (1&4576) & (2&1022) &  \multicolumn{4}{c}{} \\ 
    $RV_{\text{wh}} + GPT$ & 0&0141 & 0&1075$^{***}$ & 0&8319$^{***}$ & $-$0&0444 & 0&0127 & 0&0056$^{***}$ & 0&1878 & 6&1433$^{***}$ & 5&2278 & 47.2573 & $-$11065.55 & 22147.88 & 22208.18 \\ 
    & (0&0263) & (0&0159) & (0&0245) & (0&0280) & (1&2618) & (0&0008) & (0&2725) & (1&3973) & (18&9358) &  \multicolumn{4}{c}{} \\ 
    $RV_{\text{wh}} + GPA$ & 0&0131 & 0&1074$^{***}$ & 0&8324$^{***}$ & $-$0&0456 & 1&0167$^{***}$ & 0&0054$^{***}$ & $-$0&0277 & 6&3779$^{***}$ & 1&0007 & 47.1019 & $-$11066.47 & 22149.73 & 22210.03 \\ 
    & (0&0263) & (0&0161) & (0&0253) & (0&0286) & (0&2904) & (0&0008) & (0&0599) & (1&4341) & (1&2927) &  \multicolumn{4}{c}{} \\
    $RV_{\text{wh}} + GPR_{\text{GLB}}$ & 0&0132 & 0&0847$^{***}$ & 0&8649$^{***}$ & $-$0&0280 & 1&0522$^{***}$ & 0&0055$^{***}$ & 0&1435 & 5&6481$^{***}$ & 3&4158 & 43.5451 & $-$11068.26 & 22153.31 & 22213.61 \\
	& (0&0261) & (0&0130) & (0&0221) & (0&0236) & (0&2667) & (0&0009) & (0&1831) & (1&4557) & (5&9728) &  \multicolumn{4}{c}{} \\
    $RV_{\text{wh}} + GPR_{\text{NA}}$ & 0&0137 & 0&1078$^{***}$ & 0&8300$^{***}$ & $-$0&0449 & 0&8901$^{***}$ & 0&0057$^{***}$ & 0&1964 & 6&2551$^{***}$ & 2&1640 & 47.7363 & $-$11065.58 & 22147.96 & 22208.26 \\
    & (0&0263) & (0&0159) & (0&0247) & (0&0284) & (0&0974) & (0&0008) & (0&1967) & (1&4768) & (1&4227) &  \multicolumn{4}{c}{} \\
    $RV_{\text{wh}} + GPR_{\text{SA}}$ & 0&0129 & 0&1085$^{***}$ & 0&8279$^{***}$ & $-$0&0468 & 1&5713$^{***}$ & 0&0055$^{***}$ & 0&1986$^{*}$ & 6&2834$^{***}$ & 4&5952 & 49.9053 & $-$11063.89 & 22144.57 & 22204.87 \\
    & (0&0263) & (0&0158) & (0&0243) & (0&0285) & (0&4233) & (0&0008) & (0&1183) & (1&4905) & (5&0388) &  \multicolumn{4}{c}{} \\
    $RV_{\text{wh}} + GPR_{\text{ENE}}$ & 0&0137 & 0&1078$^{***}$ & 0&8318$^{***}$ & $-$0&0451 & 1&0243$^{***}$ & 0&0056$^{***}$ & 0&0916 & 6&2931$^{***}$ & 1&5725 & 47.2709 & $-$11066.03 & 22148.85 & 22209.15 \\
    & (0&0263) & (0&0161) & (0&0249) & (0&0284) & (0&2449) & (0&0008) & (0&1398) & (1&4586) & (1&4348) &  \multicolumn{4}{c}{} \\
    $RV_{\text{wh}} + GPR_{\text{ESW}}$ & 0&0129 & 0&1070$^{***}$ & 0&8334$^{***}$ & $-$0&0462$^{*}$ & 1&0138$^{***}$ & 0&0057$^{***}$ & 0&0957 & 6&0190$^{***}$ & 7&5095 & 48.2683 & $-$11066.00 & 22148.80 & 22209.10 \\
    & (0&0262) & (0&0160) & (0&0250) & (0&0279) & (0&1895) & (0&0008) & (0&0969) & (1&6880) & (25&4385) &  \multicolumn{4}{c}{} \\
    $RV_{\text{wh}} + GPR_{\text{MEA}}$ & 0&0096 & 0&1016$^{***}$ & 0&8288$^{***}$ & $-$0&0425 & $-$0&3321 & 0&0039$^{***}$ & $-$0&8306$^{**}$ & 7&3917$^{***}$ & 1&0002 & 55.6751 & $-$11057.76 & 22132.31 & 22192.61 \\
    & (0&0262) & (0&0158) & (0&0279) & (0&0293) & (0&4847) & (0&0010) & (0&3303) & (2&1964) & (0&7362) &  \multicolumn{4}{c}{} \\
    $RV_{\text{wh}} + GPR_{\text{AO}}$ & 0&0122 & 0&1074$^{***}$ & 0&8320$^{***}$ & $-$0&0427 & 1&2131$^{***}$ & 0&0056$^{***}$ & 0&1794 & 6&1384$^{***}$ & 15&2607$^{*}$ & 47.4700 & $-$11064.73 & 22146.26 & 22206.56 \\
    & (0&0262) & (0&0158) & (0&0241) & (0&0280) & (0&2680) & (0&0008) & (0&1283) & (1&4482) & (8&9640) &  \multicolumn{4}{c}{} \\
    $RV_{\text{wh}} + GPR_{\text{wh,prod}}$ & 0&0138 & 0&1053$^{***}$ & 0&8360$^{***}$ & $-$0&0429 & 0&9318$^{***}$ & 0&0055$^{***}$ & 0&0854 & 6&1500$^{***}$ & 17&8511 & 46.7193 & $-$11066.02 & 22148.83 & 22209.13 \\
    & (0&0263) & (0&0157) & (0&0238) & (0&0275) & (0&1277) & (0&0008) & (0&1350) & (1&3977) & (19&3100) &  \multicolumn{4}{c}{} \\
    $RV_{\text{wh}} + GPR_{\text{wh,imp}}$ & 0&0141 & 0&1072$^{***}$ & 0&8303$^{***}$ & $-$0&0432 & 1&1470$^{**}$ & 0&0057$^{***}$ & 0&1572 & 5&9877$^{***}$ & 7&4120 & 48.2592 & $-$11065.45 & 22147.70 & 22208.00 \\
    & (0&0262) & (0&0158) & (0&0258) & (0&0284) & (0&4720) & (0&0008) & (0&2587) & (1&4169) & (26&4809) &  \multicolumn{4}{c}{} \\
    $RV_{\text{wh}} + GPR_{\text{wh,exp}}$ & 0&0135 & 0&1077$^{***}$ & 0&8310$^{***}$ & $-$0&0446 & 0&9605$^{***}$ & 0&0056$^{***}$ & 0&1601 & 6&2653$^{***}$ & 1&8587 & 47.4522 & $-$11065.82 & 22148.43 & 22208.73 \\
    & (0&0263) & (0&0160) & (0&0248) & (0&0283) & (0&1416) & (0&0008) & (0&1936) & (1&4800) & (1&3608) &  \multicolumn{4}{c}{} 
    \vspace{2mm}\\
    
    \multicolumn{23}{l}{\textit{Panel B: Two-factor models with rolling window}} \\
    $RV_{\text{wh}}^{(\text{rw})} + GPR^{(\text{rw})}$ & 0&0151 & 0&1062$^{***}$ & 0&8359$^{***}$ & $-$0&0410 & $-$0&1209 & 0&0052$^{***}$ & 0&2220 & 5&8228$^{***}$ & 3&5777 & 37.2662 & $-$10999.91 & 22016.56 & 22076.86 \\
    & (0&0265) & (0&0173) & (0&0296) & (0&0305) & (1&1107) & (0&0009) & (0&2416) & (1&5392) & (10&1322) &  \multicolumn{4}{c}{} \\
    $RV_{\text{wh}}^{(\text{rw})} + GPT^{(\text{rw})}$ & 0&0147 & 0&1032$^{***}$ & 0&8410$^{***}$ & $-$0&0380 & $-$0&0887 & 0&0051$^{***}$ & 0&2158 & 5&6128$^{***}$ & 7&6532 & 36.2373 & $-$10999.32 & 22015.36 & 22075.67 \\ 
    & (0&0264) & (0&0168) & (0&0280) & (0&0292) & (0&9286) & (0&0008) & (0&2034) & (1&4436) & (24&5326) &  \multicolumn{4}{c}{} \\ 
    $RV_{\text{wh}}^{(\text{rw})} + GPA^{(\text{rw})}$ & 0&0150 & 0&1058$^{***}$ & 0&8363$^{***}$ & $-$0&0414 & 0&6998 & 0&0051$^{***}$ & 0&0459 & 6&0472$^{***}$ & 2&6971 & 37.0087 & $-$11000.30 & 22017.32 & 22077.62 \\ 
    & (0&0265) & (0&0174) & (0&0298) & (0&0305) & (0&6329) & (0&0008) & (0&1380) & (1&5527) & (2&4314) &  \multicolumn{4}{c}{} \\
    $RV_{\text{wh}}^{(\text{rw})} + GPR_{\text{GLB}}^{(\text{rw})}$ & 0&0155 & 0&0909$^{***}$ & 0&8621$^{***}$ & $-$0&0303 & 1&0911$^{***}$ & 0&0050$^{***}$ & 0&1336 & 5&2786$^{***}$ & 4&6548 & 33.0773 & $-$11000.80 & 22018.32 & 22078.62 \\
    & (0&0263) & (0&0162) & (0&0296) & (0&0272) & (0&3044) & (0&0011) & (0&2116) & (1&8618) & (34&7388) &  \multicolumn{4}{c}{} \\
    $RV_{\text{wh}}^{(\text{rw})} + GPR_{\text{NA}}^{(\text{rw})}$ & 0&0152 & 0&1055$^{***}$ & 0&8363$^{***}$ & $-$0&0401 & 0&9211$^{***}$ & 0&0052$^{***}$ & 0&1964 & 5&8345$^{***}$ & 3&0138 & 36.8616 & $-$10999.56 & 22015.85 & 22076.15 \\
    & (0&0265) & (0&0171) & (0&0291) & (0&0302) & (0&1168) & (0&0008) & (0&2013) & (1&5527) & (3&8439) &  \multicolumn{4}{c}{} \\
    $RV_{\text{wh}}^{(\text{rw})} + GPR_{\text{SA}}^{(\text{rw})}$ & 0&0141 & 0&1052$^{***}$ & 0&8353$^{***}$ & $-$0&0419 & 1&6898$^{***}$ & 0&0051$^{***}$ & 0&2240$^{*}$ & 5&8001$^{***}$ & 4&5311 & 39.3651 & $-$10997.73 & 22012.18 & 22072.48 \\
    & (0&0265) & (0&0168) & (0&0288) & (0&0306) & (0&4490) & (0&0008) & (0&1228) & (1&5554) & (4&7580) &  \multicolumn{4}{c}{} \\
    $RV_{\text{wh}}^{(\text{rw})} + GPR_{\text{ENE}}^{(\text{rw})}$ & 0&0156 & 0&1056$^{***}$ & 0&8377$^{***}$ & $-$0&0408 & 1&0363$^{***}$ & 0&0051$^{***}$ & 0&0801 & 5&9163$^{***}$ & 2&2753 & 36.5286 & $-$10999.98 & 22016.68 & 22076.98 \\
    & (0&0264) & (0&0175) & (0&0299) & (0&0304) & (0&2458) & (0&0008) & (0&1326) & (1&5706) & (3&4032) &  \multicolumn{4}{c}{} \\
    $RV_{\text{wh}}^{(\text{rw})} + GPR_{\text{ESW}}^{(\text{rw})}$ & 0&0156 & 0&1052$^{***}$ & 0&8391$^{***}$ & $-$0&0412 & 1&1825$^{***}$ & 0&0053$^{***}$ & 0&1821 & 5&3604$^{***}$ & 23&6820$^{**}$ & 37.4498 & $-$10998.74 & 22014.21 & 22074.51 \\
    & (0&0265) & (0&0171) & (0&0283) & (0&0299) & (0&2392) & (0&0009) & (0&1342) & (1&3561) & (10&0689) &  \multicolumn{4}{c}{} \\
    $RV_{\text{wh}}^{(\text{rw})} + GPR_{\text{MEA}}^{(\text{rw})}$ & 0&0114 & 0&1024$^{***}$ & 0&8321$^{***}$ & $-$0&0408 & $-$0&5909 & 0&0032$^{***}$ & $-$1&0278$^{***}$ & 7&2459$^{***}$ & 1&0000$^{***}$ & 47.9703 & $-$10989.83 & 21996.39 & 22056.69 \\
    & (0&0265) & (0&0183) & (0&0338) & (0&0327) & (0&4813) & (0&0010) & (0&3356) & (2&2379) & (0&3773) &  \multicolumn{4}{c}{} \\
    $RV_{\text{wh}}^{(\text{rw})} + GPR_{\text{AO}}^{(\text{rw})}$ & 0&0142 & 0&1046$^{***}$ & 0&8395$^{***}$ & $-$0&0383 & 1&2306$^{***}$ & 0&0051$^{***}$ & 0&1724 & 5&7038$^{***}$ & 9&6566 & 36.4487 & $-$10998.99 & 22014.70 & 22075.00 \\
    & (0&0263) & (0&0172) & (0&0286) & (0&0297) & (0&2961) & (0&0008) & (0&1367) & (1&5291) & (13&6955) &  \multicolumn{4}{c}{} \\
    $RV_{\text{wh}}^{(\text{rw})} + GPR_{\text{wh,prod}}^{(\text{rw})}$ & 0&0162 & 0&1059$^{***}$ & 0&8373$^{***}$ & $-$0&0409 & 0&9714$^{***}$ & 0&0052$^{***}$ & 0&1144 & 5&8952$^{***}$ & 3&1446 & 36.7835 & $-$10999.93 & 22016.59 & 22076.89 \\
    & (0&0264) & (0&0174) & (0&0295) & (0&0303) & (0&1577) & (0&0009) & (0&1831) & (1&5603) & (11&4845) &  \multicolumn{4}{c}{} \\
    $RV_{\text{wh}}^{(\text{rw})} + GPR_{\text{wh,imp}}^{(\text{rw})}$ & 0&0149 & 0&1057$^{***}$ & 0&8370$^{***}$ & $-$0&0408 & 1&1630$^{**}$ & 0&0052$^{***}$ & 0&1487 & 5&8822$^{***}$ & 5&8290 & 37.1702 & $-$10999.57 & 22015.86 & 22076.16 \\
    & (0&0263) & (0&0173) & (0&0294) & (0&0302) & (0&5651) & (0&0008) & (0&3048) & (1&6390) & (27&4088) &  \multicolumn{4}{c}{} \\
    $RV_{\text{wh}}^{(\text{rw})} + GPR_{\text{wh,exp}}^{(\text{rw})}$ & 0&0152 & 0&1057$^{***}$ & 0&8370$^{***}$ & $-$0&0406 & 0&9823$^{***}$ & 0&0052$^{***}$ & 0&1453 & 5&8480$^{***}$ & 2&4950 & 36.7165 & $-$10999.80 & 22016.32 & 22076.62 \\
    & (0&0264) & (0&0174) & (0&0296) & (0&0303) & (0&1559) & (0&0008) & (0&1896) & (1&5648) & (2&9820) &  \multicolumn{4}{c}{} \\
  \bottomrule
    \end{tabular}
    }%
  \begin{flushleft}
    \footnotesize
    \justifying Note: This table reports the results of different two-factor GJR-GARCH-MIDAS models for the wheat market, where Panels A and B denote the model estimation with fixed time span and rolling window. Standard errors of the parameter estimates are given in parentheses.
    \end{flushleft} 
  \label{Tab:Two_Factor_Estimation_Wheat_Futures}%
\end{table}%

Combined with Table~\ref{Tab:Two_Factor_Estimation_Maize_Futures}, we note that the impact of realized volatility on the long-term component in the maize market is smaller relative to the wheat market. Only some of the two-factor models exhibit statistically significant estimates of $\theta_{RV}$, such as those that include both $RV_{\text{mz}}^{(\text{rw})}$ and $GPR^{(\text{rw})}$, $GPT^{(\text{rw})}$, or $GPA^{(\text{rw})}$. Meanwhile, compared to the estimated results of single-factor models for maize in Table~\ref{Tab:Single_Factor_Estimation_Maize_Futures}, the statistical significance of geopolitical risk factors in some two-factor models changes after removing the influence of realized volatility. For instance, the estimates of $\theta_{MV}$ for $GPT^{(\text{rw})}$, $GPR_{\text{NA}}^{(\text{rw})}$, $GPR_{\text{MEA}}^{(\text{rw})}$, $GPR_{\text{AO}}^{(\text{rw})}$, and $GPR_{\text{mz,imp}}^{(\text{rw})}$ shift from being significantly negative to being insignificant. Moreover, $GPR_{\text{SA}}^{(\text{rw})}$ experiences a change in the estimate of $\theta_{MV}$ from insignificant to significantly positive, similar to the results observed in the wheat market. Notably, there exist several geopolitical risk factors whose sign and significance remain consistent between single- and two-factor models. 

At the global level, the RMT-based $GPR_{\text{GLB}}^{(\text{rw})}$ always displays significantly negative coefficients, indicating that the composite geopolitical risk of 44 major economies has a significantly and robustly negative impact on long-term maize market volatility. At the regional level, the estimates of $\theta_{MV}$ for $GPR_{\text{ENE}}^{(\text{rw})}$ and $GPR_{\text{ESW}}^{(\text{rw})}$ remain significantly negative in both single- and two-factor models, suggesting the remarkably negative influence of geopolitical risk faced by Europe. For one thing, economies like Ukraine, France, and Russia in Europe are important maize producers globally, with Ukraine, France, Hungary, Russia, and Poland ranking among the world's top ten maize exporters. For another, economies such as Spain, Italy, and the Netherlands in Europe are major maize importers, whose import volumes are always at the forefront. Hence, the geopolitical risk of Europe prominently affects the supply-demand dynamics as well as the long-term volatility in the international maize market. In addition, the estimates associated with $GPR_{\text{mz,prod}}^{(\text{rw})}$ and $GPR_{\text{mz,exp}}^{(\text{rw})}$ reveal the significantly negative influences of composite geopolitical risks from combinations of important economies with global influence in maize production and export. In terms of goodness-of-fit and explanatory power, the two-factor models incorporating both $RV_{\text{mz}}^{(\text{rw})}$ and $GPR_{\text{ESW}}^{(\text{rw})}$ or $GPR_{\text{ENE}}^{(\text{rw})}$ perform well, which implies that the region-based division is more suitable for studying maize market volatility compared to other categorizations of geopolitical risks.

\begin{table}[!ht]
  \centering
  \setlength{\abovecaptionskip}{0pt}
  \setlength{\belowcaptionskip}{10pt}
  \caption{Estimation of two-factor models with fixed time span and rolling window for maize}
  \setlength\tabcolsep{3pt}   \resizebox{\textwidth}{!}{ 
    \begin{tabular}{l r@{.}l r@{.}l r@{.}l r@{.}l r@{.}l r@{.}l r@{.}l r@{.}l r@{.}l rccc}
    \toprule
         & \multicolumn{2}{c}{$\mu$} & \multicolumn{2}{c}{$\alpha$} & \multicolumn{2}{c}{$\beta$} & \multicolumn{2}{c}{$\gamma$} & \multicolumn{2}{c}{$m$} & \multicolumn{2}{c}{$\theta_{RV}$} & \multicolumn{2}{c}{$\theta_{MV}$} & \multicolumn{2}{c}{$\omega_{RV}$} & \multicolumn{2}{c}{$\omega_{MV}$} & VR (\%) & LLF & AIC & BIC \\
    \midrule
    \multicolumn{23}{l}{\textit{Panel A: Two-factor models with fixed time span}} \\
    $RV_{\text{mz}} + GPR$ & 0&0241 & 0&1781$^{***}$ & 0&7790$^{***}$ & $-$0&0412 & 0&6133 & 0&0066$^{**}$ & 0&0703 & 1&2671 & 1&0000 & 13.6273 & $-$10358.17 & 20733.13 & 20793.43 \\
    & (0&0222) & (0&0589) & (0&0552) & (0&0468) & (0&4399) & (0&0027) & (0&0759) & (0&9380) & (0&9502) &  \multicolumn{4}{c}{} \\ 
    $RV_{\text{mz}} + GPT$ & 0&0221 & 0&1707$^{***}$ & 0&7893$^{***}$ & $-$0&0394 & 0&6183$^{*}$ & 0&0065$^{**}$ & 0&0695 & 1&2968 & 1&0002$^{**}$ & 13.0275 & $-$10359.05 & 20734.89 & 20795.19 \\ 
    & (0&0222) & (0&0573) & (0&0533) & (0&0451) & (0&3751) & (0&0028) & (0&0607) & (0&9746) & (0&4796) &  \multicolumn{4}{c}{} \\ 
    $RV_{\text{mz}} + GPA$ & 0&0245 & 0&1748$^{***}$ & 0&7822$^{***}$ & $-$0&0407 & $-$0&4297 & 0&0072$^{**}$ & 0&2886 & 1&2830$^{*}$ & 1&0000 & 16.2235 & $-$10355.06 & 20726.91 & 20787.21 \\ 
    & (0&0224) & (0&0581) & (0&0543) & (0&0458) & (2&3647) & (0&0029) & (0&4921) & (0&7559) & (2&0368) &  \multicolumn{4}{c}{} \\
    $RV_{\text{mz}} + GPR_{\text{GLB}}$ & 0&0219 & 0&1738$^{***}$ & 0&7673$^{***}$ & $-$0&0334 & 0&0754 & 0&0031 & $-$0&8029$^{**}$ & 1&2656 & 6&7950 & 28.7533 & $-$10345.82 & 20708.43 & 20768.73 \\
    & (0&0221) & (0&0596) & (0&0614) & (0&0471) & (0&4569) & (0&0026) & (0&3480) & (1&5884) & (4&5772) &  \multicolumn{4}{c}{} \\
    $RV_{\text{mz}} + GPR_{\text{NA}}$ & 0&0227 & 0&1764$^{***}$ & 0&7775$^{***}$ & $-$0&0396 & 1&0092$^{***}$ & 0&0047 & $-$0&4309 & 1&2982 & 12&2788 & 16.8994 & $-$10355.71 & 20728.21 & 20788.51 \\
    & (0&0225) & (0&0602) & (0&0593) & (0&0473) & (0&3040) & (0&0034) & (0&3500) & (1&3443) & (66&3903) &  \multicolumn{4}{c}{} \\
    $RV_{\text{mz}} + GPR_{\text{SA}}$ & 0&0204 & 0&1636$^{***}$ & 0&7756$^{***}$ & $-$0&0364 & 3&5292$^{*}$ & 0&0117$^{***}$ & 0&8775 & 1&0000 & 1&0000 & 28.6674 & $-$10339.94 & 20696.67 & 20756.97 \\
    & (0&0220) & (0&0576) & (0&0552) & (0&0452) & (2&1454) & (0&0033) & (0&6753) & (1&0892) & (1&9942) &  \multicolumn{4}{c}{} \\
    $RV_{\text{mz}} + GPR_{\text{ENE}}$ & 0&0220 & 0&1777$^{***}$ & 0&7589$^{***}$ & $-$0&0320 & $-$0&3400 & 0&0013 & $-$0&9968 & 1&0004 & 1&2964 & 36.2171 & $-$10339.17 & 20695.13 & 20755.43 \\
    & (0&0221) & (0&0581) & (0&0577) & (0&0476) & (0&7848) & (0&0048) & (0&6335) & (2&8344) & (1&4121) &  \multicolumn{4}{c}{} \\
    $RV_{\text{mz}} + GPR_{\text{ESW}}$ & 0&0217 & 0&1610$^{**}$ & 0&7836$^{***}$ & $-$0&0219 & $-$1&0054 & $-$0&0024 & $-$1&5644$^{***}$ & 5&4498$^{**}$ & 2&0393$^{*}$ & 42.1397 & $-$10333.39 & 20683.56 & 20743.86 \\
    & (0&0219) & (0&0633) & (0&0698) & (0&0460) & (0&6709) & (0&0021) & (0&4445) & (2&7257) & (1&0468) &  \multicolumn{4}{c}{} \\
    $RV_{\text{mz}} + GPR_{\text{MEA}}$ & 0&0215 & 0&1622$^{***}$ & 0&7859$^{***}$ & $-$0&0284 & $-$0&4495 & 0&0062$^{***}$ & $-$0&8302$^{**}$ & 1&0000 & 6&2059 & 26.5465 & $-$10344.81 & 20706.42 & 20766.72 \\
    & (0&0221) & (0&0567) & (0&0559) & (0&0450) & (0&6443) & (0&0021) & (0&3766) & (1&3720) & (6&8917) &  \multicolumn{4}{c}{} \\
    $RV_{\text{mz}} + GPR_{\text{AO}}$ & 0&0232 & 0&1773$^{***}$ & 0&7759$^{***}$ & $-$0&0384 & 0&5926 & 0&0045 & $-$0&2593 & 1&6423$^{**}$ & 6&1228$^{**}$ & 15.0596 & $-$10357.49 & 20731.77 & 20792.07 \\
    & (0&0222) & (0&0583) & (0&0553) & (0&0470) & (0&3957) & (0&0037) & (0&2630) & (0&7508) & (2&4265) &  \multicolumn{4}{c}{} \\
    $RV_{\text{mz}} + GPR_{\text{mz,prod}}$ & 0&0227 & 0&1798$^{***}$ & 0&7673$^{***}$ & $-$0&0385 & 0&6534$^{**}$ & 0&0042 & $-$0&6168 & 1&3278 & 10&2604 & 21.6555 & $-$10351.80 & 20720.38 & 20780.68 \\
    & (0&0223) & (0&0604) & (0&0602) & (0&0481) & (0&3234) & (0&0029) & (0&4308) & (1&1334) & (28&3151) &  \multicolumn{4}{c}{} \\
    $RV_{\text{mz}} + GPR_{\text{mz,imp}}$ & 0&0229 & 0&1752$^{***}$ & 0&7777$^{***}$ & $-$0&0390 & 0&4719 & 0&0037 & $-$0&3729 & 1&6681$^{**}$ & 6&0413$^{**}$ & 16.1461 & $-$10356.28 & 20729.35 & 20789.65 \\
    & (0&0222) & (0&0588) & (0&0563) & (0&0471) & (0&3969) & (0&0037) & (0&2780) & (0&7924) & (2&8635) &  \multicolumn{4}{c}{} \\
    $RV_{\text{mz}} + GPR_{\text{mz,exp}}$ & 0&0228 & 0&1792$^{***}$ & 0&7585$^{***}$ & $-$0&0356 & 0&5933$^{**}$ & 0&0036 & $-$0&7429$^{**}$ & 1&3017 & 6&2666 & 31.1457 & $-$10343.74 & 20704.28 & 20764.58 \\
    & (0&0222) & (0&0603) & (0&0616) & (0&0483) & (0&2727) & (0&0027) & (0&3235) & (0&9805) & (5&4159) &  \multicolumn{4}{c}{} 
    \vspace{2mm}\\
    
    \multicolumn{23}{l}{\textit{Panel B: Two-factor models with rolling window}} \\
    $RV_{\text{mz}}^{(\text{rw})} + GPR^{(\text{rw})}$ & 0&0263 & 0&1875$^{***}$ & 0&7638$^{***}$ & $-$0&0445 & $-$0&5431 & 0&0070$^{***}$ & 0&3099 & 1&1461 & 1&0002 & 11.7133 & $-$10305.75 & 20628.23 & 20688.53 \\
    & (0&0226) & (0&0619) & (0&0603) & (0&0492) & (2&0843) & (0&0025) & (0&4359) & (0&8242) & (1&4124) &  \multicolumn{4}{c}{} \\ 
    $RV_{\text{mz}}^{(\text{rw})} + GPT^{(\text{rw})}$ & 0&0223 & 0&1595$^{***}$ & 0&7969$^{***}$ & $-$0&0332 & 0&8296$^{***}$ & 0&0061$^{**}$ & 0&0201 & 1&0629 & 1&0005$^{**}$ & 11.0850 & $-$10307.72 & 20632.17 & 20692.47 \\ 
    & (0&0223) & (0&0548) & (0&0553) & (0&0423) & (0&2628) & (0&0026) & (0&0199) & (0&7785) & (0&4322) &  \multicolumn{4}{c}{} \\ 
    $RV_{\text{mz}}^{(\text{rw})} + GPA^{(\text{rw})}$ & 0&0238 & 0&1328$^{***}$ & 0&8260$^{***}$ & $-$0&0213 & $-$0&4883 & 0&0065$^{**}$ & 0&2958 & 1&0474$^{*}$ & 1&0049 & 12.7412 & $-$10307.74 & 20632.20 & 20692.50 \\ 
    & (0&0223) & (0&0480) & (0&0501) & (0&0363) & (3&4118) & (0&0033) & (0&7063) & (0&6206) & (4&1323) &  \multicolumn{4}{c}{} \\
    $RV_{\text{mz}}^{(\text{rw})} + GPR_{\text{GLB}}^{(\text{rw})}$ & 0&0231 & 0&1776$^{***}$ & 0&7629$^{***}$ & $-$0&0356 & 0&1081 & 0&0026 & $-$0&8046$^{**}$ & 1&1990 & 6&3231$^{*}$ & 22.9024 & $-$10294.84 & 20606.41 & 20666.71 \\
    & (0&0224) & (0&0619) & (0&0652) & (0&0484) & (0&4648) & (0&0025) & (0&3588) & (1&4063) & (3&6650) &  \multicolumn{4}{c}{}  \\
    $RV_{\text{mz}}^{(\text{rw})} + GPR_{\text{NA}}^{(\text{rw})}$ & 0&0232 & 0&1781$^{***}$ & 0&7722$^{***}$ & $-$0&0389 & 1&0092$^{***}$ & 0&0045$^{*}$ & $-$0&3730 & 1&1859 & 10&3151 & 12.8739 & $-$10305.03 & 20626.79 & 20687.09 \\
    & (0&0225) & (0&0609) & (0&0626) & (0&0474) & (0&2629) & (0&0026) & (0&3660) & (1&2023) & (7&0161) &  \multicolumn{4}{c}{} \\
    $RV_{\text{mz}}^{(\text{rw})} + GPR_{\text{SA}}^{(\text{rw})}$ & 0&0223 & 0&1799$^{***}$ & 0&7546$^{***}$ & $-$0&0427 & 3&5305$^{**}$ & 0&0107$^{***}$ & 0&8624$^{*}$ & 1&0011 & 1&0000 & 22.7328 & $-$10285.22 & 20587.17 & 20647.47 \\
    & (0&0223) & (0&0608) & (0&0601) & (0&0495) & (1&5883) & (0&0029) & (0&4880) & (0&6526) & (1&4877) &  \multicolumn{4}{c}{} \\
    $RV_{\text{mz}}^{(\text{rw})} + GPR_{\text{ENE}}^{(\text{rw})}$ & 0&0221 & 0&1692$^{**}$ & 0&7753$^{***}$ & $-$0&0252 & $-$0&7888 & $-$0&0019 & $-$1&4376$^{***}$ & 7&2793 & 1&0435$^{*}$ & 37.4587 & $-$10282.96 & 20582.64 & 20642.94 \\
    & (0&0224) & (0&0753) & (0&0972) & (0&0509) & (0&6440) & (0&0032) & (0&5052) & (4&8113) & (0&5693) &  \multicolumn{4}{c}{} \\
    $RV_{\text{mz}}^{(\text{rw})} + GPR_{\text{ESW}}^{(\text{rw})}$ & 0&0198 & 0&1603$^{**}$ & 0&7852$^{***}$ & $-$0&0227 & $-$1&2690$^{*}$ & $-$0&0033 & $-$1&7672$^{***}$ & 5&3698 & 1&7289 & 39.5597 & $-$10279.07 & 20574.86 & 20635.16 \\
    & (0&0220) & (0&0739) & (0&0935) & (0&0494) & (0&7216) & (0&0026) & (0&4985) & (7&3258) & (1&1223) &  \multicolumn{4}{c}{} \\
    $RV_{\text{mz}}^{(\text{rw})} + GPR_{\text{MEA}}^{(\text{rw})}$ & 0&0233 & 0&1749$^{***}$ & 0&7697$^{***}$ & $-$0&0386 & $-$0&1395 & 0&0056 & $-$0&6551 & 1&0000 & 6&1594 & 16.9829 & $-$10294.67 & 20606.07 & 20666.37 \\
    & (0&0244) & (0&0618) & (0&0652) & (0&0517) & (2&3469) & (0&0057) & (1&1886) & (5&8060) & (7&7238) &  \multicolumn{4}{c}{} \\
    $RV_{\text{mz}}^{(\text{rw})} + GPR_{\text{AO}}^{(\text{rw})}$ & 0&0231 & 0&1743$^{***}$ & 0&7759$^{***}$ & $-$0&0365 & 0&6154 & 0&0041 & $-$0&2502 & 1&5503 & 7&1923$^{**}$ & 11.8845 & $-$10305.67 & 20628.07 & 20688.37 \\
    & (0&0224) & (0&0592) & (0&0598) & (0&0466) & (0&3999) & (0&0038) & (0&2800) & (1&0521) & (3&5577) &  \multicolumn{4}{c}{} \\
    $RV_{\text{mz}}^{(\text{rw})} + GPR_{\text{mz,prod}}^{(\text{rw})}$ & 0&0234 & 0&1825$^{***}$ & 0&7635$^{***}$ & $-$0&0400 & 0&6805$^{**}$ & 0&0038 & $-$0&6029$^{*}$ & 1&2447 & 9&4260 & 16.9090 & $-$10301.00 & 20618.72 & 20679.02 \\
    & (0&0225) & (0&0625) & (0&0641) & (0&0489) & (0&3096) & (0&0025) & (0&3424) & (1&1572) & (7&1659) &  \multicolumn{4}{c}{} \\
    $RV_{\text{mz}}^{(\text{rw})} + GPR_{\text{mz,imp}}^{(\text{rw})}$ & 0&0231 & 0&1744$^{***}$ & 0&7743$^{***}$ & $-$0&0356 & 0&4526 & 0&0030 & $-$0&4050 & 1&7140$^{*}$ & 6&0452$^{**}$ & 13.1195 & $-$10304.30 & 20625.33 & 20685.63 \\
    & (0&0224) & (0&0603) & (0&0616) & (0&0472) & (0&3982) & (0&0037) & (0&2921) & (1&0286) & (2&6382) &  \multicolumn{4}{c}{} \\
    $RV_{\text{mz}}^{(\text{rw})} + GPR_{\text{mz,exp}}^{(\text{rw})}$ & 0&0225 & 0&1722$^{***}$ & 0&7670$^{***}$ & $-$0&0330 & 0&6056$^{**}$ & 0&0030 & $-$0&7776$^{**}$ & 1&2070 & 5&4487 & 25.7445 & $-$10292.36 & 20601.46 & 20661.76 \\
    & (0&0223) & (0&0603) & (0&0634) & (0&0472) & (0&2786) & (0&0029) & (0&3780) & (0&9576) & (5&4133) &  \multicolumn{4}{c}{} \\
  \bottomrule
    \end{tabular}
    }%
  \begin{flushleft}
    \footnotesize
    \justifying Note: This table presents the estimated results of different two-factor GJR-GARCH-MIDAS models for the maize market, where Panels A and B refer to the model estimation with fixed time span and rolling window, respectively.
    \end{flushleft} 
  \label{Tab:Two_Factor_Estimation_Maize_Futures}%
\end{table}%

According to Table~\ref{Tab:Two_Factor_Estimation_Soybean_Futures}, similar to the maize market, the long-term volatility of the soybean market is relatively less affected by realized volatility, with the coefficients of $\theta_{RV}$ only showing significance at the 10\% level in the two-factor models associated with three global geopolitical risk indices. As for geopolitical risk factors, the estimates of $\theta_{MV}$ for the composite geopolitical risk indices based on 44 major economies, economies in different regions, and economies playing important roles in soybean production, import, and export all become insignificant after excluding the influence of realized volatility. However, the estimate of $\theta_{MV}$ for $GPR_{\text{NA}}^{(\text{rw})}$ remains significantly negative, indicating a substantially and robustly negative impact of geopolitical risk in North America on long-run soybean market volatility. The United States and Canada in North America are the main soybean producers and exporters, collectively representing 35.71\% of global production and 37.42\% of total exports in 2021, and Mexico in North America ranks as the third largest soybean importer in the world. As a result, the relationship between soybean supply and demand, along with fluctuations in the soybean market, is sensitive to the geopolitical risk faced by North America. Corresponding to the statistical significance, the GJR-GARCH-MIDAS model containing both $RV_{\text{sb}}^{(\text{rw})}$ and $GPR_{\text{NA}}^{(\text{rw})}$ exhibits the best fitting effect and strongest explanatory power among the various models, accounting for about 82.19\% of the overall volatility in the international soybean market. This finding is also consistent with the estimated results of single-factor models for soybean, where the model with $GPR_{\text{NA}}^{(\text{rw})}$ performs the best.

\begin{table}[!ht]
  \centering
  \setlength{\abovecaptionskip}{0pt}
  \setlength{\belowcaptionskip}{10pt}
  \caption{Estimation of two-factor models with fixed time span and rolling window for soybean}
  \setlength\tabcolsep{3pt}   \resizebox{\textwidth}{!}{ 
    \begin{tabular}{l r@{.}l r@{.}l r@{.}l r@{.}l r@{.}l r@{.}l r@{.}l r@{.}l r@{.}l rccc}
    \toprule
         & \multicolumn{2}{c}{$\mu$} & \multicolumn{2}{c}{$\alpha$} & \multicolumn{2}{c}{$\beta$} & \multicolumn{2}{c}{$\gamma$} & \multicolumn{2}{c}{$m$} & \multicolumn{2}{c}{$\theta_{RV}$} & \multicolumn{2}{c}{$\theta_{MV}$} & \multicolumn{2}{c}{$\omega_{RV}$} & \multicolumn{2}{c}{$\omega_{MV}$} & VR (\%) & LLF & AIC & BIC \\
    \midrule
    \multicolumn{23}{l}{\textit{Panel A: Two-factor models with fixed time span}} \\
    $RV_{\text{sb}} + GPR$ & 0&0315$^{*}$ & 0&0783$^{***}$ & 0&9179$^{***}$ & $-$0&0130 & 0&8385$^{***}$ & 0&0008 & 0&0462 & 3&3760 & 1&0201 & 0.1634 & $-$9405.53 & 18827.84 & 18888.14 \\
    & (0&0188) & (0&0108) & (0&0089) & (0&0164) & (0&2849) & (0&0007) & (0&0390) & (3&2574) & (0&9749) &  \multicolumn{4}{c}{} \\ 
    $RV_{\text{sb}} + GPT$ & 0&0303 & 0&0769$^{***}$ & 0&9218$^{***}$ & $-$0&0130 & 5&0849 & $-$0&0031 & $-$0&8158 & 7&1605 & 14&4626 & 12.3386 & $-$9401.17 & 18819.13 & 18879.43 \\ 
    & (0&0199) & (0&0107) & (0&0179) & (0&0279) & (28&9718) & (0&0071) & (6&2390) & (4&6019) & (256&0742) &  \multicolumn{4}{c}{} \\ 
    $RV_{\text{sb}} + GPA$ & 0&0314 & 0&0790$^{***}$ & 0&9189$^{***}$ & $-$0&0146 & 0&7083 & $-$0&0015 & 0&1054 & 5&0601 & 1&0072 & 0.5660 & $-$9405.29 & 18827.37 & 18887.67 \\ 
    & (0&0194) & (0&0118) & (0&0120) & (0&0167) & (1&0568) & (0&0073) & (0&2078) & (13&9240) & (2&0249) &  \multicolumn{4}{c}{} \\
    $RV_{\text{sb}} + GPR_{\text{GLB}}$ & 0&0297 & 0&0770$^{***}$ & 0&9213$^{***}$ & $-$0&0119 & $-$0&6563 & $-$0&0045 & $-$1&5372 & 6&9847 & 2&5614$^{**}$ & 40.1902 & $-$9399.03 & 18814.85 & 18875.15 \\
    & (0&0186) & (0&0109) & (0&0085) & (0&0158) & (1&2050) & (0&0082) & (1&1486) & (4&9840) & (1&1318) &  \multicolumn{4}{c}{} \\
    $RV_{\text{sb}} + GPR_{\text{NA}}$ & 0&0300 & 0&0750$^{***}$ & 0&9263$^{***}$ & $-$0&0104 & 1&5381$^{***}$ & $-$0&0120 & $-$3&3926$^{**}$ & 4&6078$^{***}$ & 2&3428$^{**}$ & 78.8148 & $-$9395.19 & 18807.18 & 18867.48 \\
    & (0&0185) & (0&0094) & (0&0068) & (0&0146) & (0&4303) & (0&0080) & (1&7063) & (1&5643) & (0&9187) &  \multicolumn{4}{c}{} \\
    $RV_{\text{sb}} + GPR_{\text{SA}}$ & 0&0309$^{*}$ & 0&0740$^{***}$ & 0&9233$^{***}$ & $-$0&0097 & $-$1&6041 & $-$0&0057 & $-$0&8669 & 5&8787$^{**}$ & 4&7099 & 31.0265 & $-$9397.49 & 18811.77 & 18872.07 \\
    & (0&0185) & (0&0103) & (0&0084) & (0&0151) & (1&6205) & (0&0074) & (0&5443) & (2&8513) & (4&4636) &  \multicolumn{4}{c}{} \\
    $RV_{\text{sb}} + GPR_{\text{ENE}}$ & 0&0304 & 0&0778$^{***}$ & 0&9184$^{***}$ & $-$0&0129 & $-$0&3075 & $-$0&0021 & $-$0&9257$^{*}$ & 7&0077 & 1&0006 & 18.6447 & $-$9399.56 & 18815.91 & 18876.21 \\
    & (0&0186) & (0&0109) & (0&0085) & (0&0163) & (0&8613) & (0&0047) & (0&4976) & (4&3083) & (1&5181) &  \multicolumn{4}{c}{} \\
    $RV_{\text{sb}} + GPR_{\text{ESW}}$ & 0&0306$^{*}$ & 0&0784$^{***}$ & 0&9186$^{***}$ & $-$0&0131 & $-$0&3498 & $-$0&0031 & $-$0&9848$^{*}$ & 6&6908$^{*}$ & 1&9463$^{**}$ & 18.4190 & $-$9400.80 & 18818.39 & 18878.69 \\
    & (0&0185) & (0&0110) & (0&0086) & (0&0160) & (0&8417) & (0&0051) & (0&5654) & (3&7300) & (0&9800) &  \multicolumn{4}{c}{} \\
    $RV_{\text{sb}} + GPR_{\text{MEA}}$ & 0&0303 & 0&0766$^{***}$ & 0&9220$^{***}$ & $-$0&0126 & $-$0&2226 & $-$0&0034 & $-$0&9467 & 6&9369$^{**}$ & 3&2127$^{***}$ & 13.2858 & $-$9402.44 & 18821.67 & 18881.97 \\
    & (0&0187) & (0&0110) & (0&0087) & (0&0159) & (1&0178) & (0&0052) & (0&6963) & (2&7785) & (1&0256) &  \multicolumn{4}{c}{} \\
    $RV_{\text{sb}} + GPR_{\text{AO}}$ & 0&0308$^{*}$ & 0&0754$^{***}$ & 0&9221$^{***}$ & $-$0&0113 & $-$0&1233 & $-$0&0051 & $-$0&7985 & 6&0542$^{*}$ & 5&9584 & 20.3736 & $-$9400.75 & 18818.29 & 18878.59 \\
    & (0&0187) & (0&0105) & (0&0079) & (0&0156) & (0&8687) & (0&0070) & (0&5407) & (3&2805) & (9&7292) &  \multicolumn{4}{c}{} \\
    $RV_{\text{sb}} + GPR_{\text{sb,prod}}$ & 0&0296 & 0&0760$^{***}$ & 0&9222$^{***}$ & $-$0&0111 & 0&2356 & $-$0&0056 & $-$1&6896 & 6&0984$^{**}$ & 2&1502$^{**}$ & 49.7028 & $-$9399.34 & 18815.47 & 18875.77 \\
    & (0&0186) & (0&0104) & (0&0083) & (0&0154) & (0&6112) & (0&0081) & (1&2160) & (3&0655) & (0&9948) &  \multicolumn{4}{c}{} \\
    $RV_{\text{sb}} + GPR_{\text{sb,imp}}$ & 0&0301 & 0&0755$^{***}$ & 0&9207$^{***}$ & $-$0&0105 & $-$0&4373 & $-$0&0038 & $-$1&0477 & 6&9867 & 3&4209$^{**}$ & 22.0647 & $-$9400.73 & 18818.24 & 18878.54 \\
    & (0&0186) & (0&0106) & (0&0082) & (0&0158) & (1&0449) & (0&0072) & (0&7652) & (4&9295) & (1&5930) &  \multicolumn{4}{c}{} \\
    $RV_{\text{sb}} + GPR_{\text{sb,exp}}$ & 0&0295 & 0&0763$^{***}$ & 0&9213$^{***}$ & $-$0&0114 & 0&2164 & $-$0&0048 & $-$1&5227 & 6&1256$^{**}$ & 1&9577$^{**}$ & 40.2934 & $-$9399.63 & 18816.06 & 18876.36 \\
    & (0&0185) & (0&0105) & (0&0083) & (0&0155) & (0&5638) & (0&0072) & (0&9842) & (3&0280) & (0&9683) &  \multicolumn{4}{c}{} 
    \vspace{2mm}\\
    
    \multicolumn{23}{l}{\textit{Panel B: Two-factor models with rolling window}} \\
    $RV_{\text{sb}}^{(\text{rw})} + GPR^{(\text{rw})}$ & 0&0305 & 0&0780$^{***}$ & 0&9178$^{***}$ & $-$0&0132 & 0&2373 & 0&0057$^{*}$ & 0&1136 & 1&0000 & 1&0003 & 3.6213 & $-$9349.47 & 18715.67 & 18775.97 \\
    & (0&0190) & (0&0109) & (0&0090) & (0&0165) & (0&6102) & (0&0033) & (0&1318) & (0&8050) & (1&1138) &  \multicolumn{4}{c}{} \\
    $RV_{\text{sb}}^{(\text{rw})} + GPT^{(\text{rw})}$ & 0&0296 & 0&0770$^{***}$ & 0&9183$^{***}$ & $-$0&0118 & 0&8941$^{***}$ & 0&0054$^{*}$ & $-$0&0252 & 1&0013 & 1&7911 & 3.2348 & $-$9349.25 & 18715.22 & 18775.52 \\ 
     & (0&0190) & (0&0106) & (0&0089) & (0&0164) & (0&3091) & (0&0032) & (0&0324) & (0&8112) & (1&7547) &  \multicolumn{4}{c}{} \\
    $RV_{\text{sb}}^{(\text{rw})} + GPA^{(\text{rw})}$ & 0&0307 & 0&0780$^{***}$ & 0&9189$^{***}$ & $-$0&0139 & 1&8597 & 0&0069$^{*}$ & $-$0&2532 & 1&0000 & 4&6316$^{**}$ & 4.8827 & $-$9348.52 & 18713.77 & 18774.07 \\ 
    & (0&0190) & (0&0108) & (0&0091) & (0&0164) & (2&1428) & (0&0035) & (0&4662) & (0&7617) & (1&8116) &  \multicolumn{4}{c}{} \\
    $RV_{\text{sb}}^{(\text{rw})} + GPR_{\text{GLB}}^{(\text{rw})}$ & 0&0282 & 0&0728$^{***}$ & 0&9253$^{***}$ & $-$0&0082 & $-$1&5058 & $-$0&0099 & $-$2&3774 & 5&0398$^{**}$ & 2&6122$^{***}$ & 71.6675 & $-$9342.50 & 18701.73 & 18762.03 \\
    & (0&0186) & (0&0099) & (0&0074) & (0&0149) & (1&5390) & (0&0116) & (1&5145) & (2&1383) & (0&9932) &  \multicolumn{4}{c}{} \\
    $RV_{\text{sb}}^{(\text{rw})} + GPR_{\text{NA}}^{(\text{rw})}$ & 0&0315$^{*}$ & 0&0750$^{***}$ & 0&9261$^{***}$ & $-$0&0094 & 1&5590$^{***}$ & $-$0&0137 & $-$3&9429$^{**}$ & 4&9538$^{***}$ & 2&5165$^{***}$ & 82.1928 & $-$9336.73 & 18690.19 & 18750.49 \\
    & (0&0186) & (0&0094) & (0&0065) & (0&0145) & (0&4572) & (0&0083) & (1&5850) & (1&3652) & (0&9090) &  \multicolumn{4}{c}{} \\
    $RV_{\text{sb}}^{(\text{rw})} + GPR_{\text{SA}}^{(\text{rw})}$ & 0&0297 & 0&0719$^{***}$ & 0&9247$^{***}$ & $-$0&0080 & $-$1&5730 & $-$0&0058 & $-$0&8573 & 6&5935$^{**}$ & 5&4813 & 29.9535 & $-$9341.76 & 18700.25 & 18760.55 \\
    & (0&0186) & (0&0106) & (0&0089) & (0&0151) & (1&7929) & (0&0099) & (0&6172) & (3&1965) & (5&7948) &  \multicolumn{4}{c}{} \\
    $RV_{\text{sb}}^{(\text{rw})} + GPR_{\text{ENE}}^{(\text{rw})}$ & 0&0281 & 0&0749$^{***}$ & 0&9187$^{***}$ & $-$0&0101 & $-$0&3732 & $-$0&0013 & $-$0&9035 & 7&2019$^{*}$ & 1&0002 & 15.3487 & $-$9344.90 & 18706.53 & 18766.83 \\
    & (0&0185) & (0&0109) & (0&0092) & (0&0159) & (0&8886) & (0&0058) & (0&5562) & (4&2281) & (1&3517) &  \multicolumn{4}{c}{} \\
    $RV_{\text{sb}}^{(\text{rw})} + GPR_{\text{ESW}}^{(\text{rw})}$ & 0&0278 & 0&0746$^{***}$ & 0&9206$^{***}$ & $-$0&0102 & $-$0&7599 & $-$0&0033 & $-$1&2142 & 5&5914 & 2&2207$^{*}$ & 25.5578 & $-$9345.18 & 18707.10 & 18767.40 \\
    & (0&0185) & (0&0109) & (0&0095) & (0&0158) & (0&9238) & (0&0085) & (0&7460) & (4&6501) & (1&1682) &  \multicolumn{4}{c}{} \\
    $RV_{\text{sb}}^{(\text{rw})} + GPR_{\text{MEA}}^{(\text{rw})}$ & 0&0283 & 0&0750$^{***}$ & 0&9227$^{***}$ & $-$0&0113 & $-$0&3439 & $-$0&0031 & $-$0&9953 & 7&1368$^{***}$ & 3&3234$^{***}$ & 13.0145 & $-$9347.53 & 18711.79 & 18772.09 \\
    & (0&0187) & (0&0116) & (0&0099) & (0&0159) & (1&1668) & (0&0074) & (0&8462) & (2&4470) & (0&9381) &  \multicolumn{4}{c}{} \\
    $RV_{\text{sb}}^{(\text{rw})} + GPR_{\text{AO}}^{(\text{rw})}$ & 0&0291 & 0&0730$^{***}$ & 0&9241$^{***}$ & $-$0&0091 & $-$0&3306 & $-$0&0072 & $-$0&9703 & 5&5177$^{**}$ & 6&4662 & 30.4201 & $-$9345.03 & 18706.79 & 18767.09 \\
    & (0&0188) & (0&0108) & (0&0083) & (0&0156) & (1&0829) & (0&0107) & (0&7308) & (2&4609) & (12&0240) &  \multicolumn{4}{c}{} \\
    $RV_{\text{sb}}^{(\text{rw})} + GPR_{\text{sb,prod}}^{(\text{rw})}$ & 0&0283 & 0&0730$^{***}$ & 0&9250$^{***}$ & $-$0&0088 & 0&0660 & $-$0&0081 & $-$2&1077 & 5&5411$^{***}$ & 2&3662$^{**}$ & 66.4342 & $-$9343.42 & 18703.56 & 18763.86 \\
    & (0&0186) & (0&0101) & (0&0081) & (0&0150) & (0&6600) & (0&0117) & (1&5247) & (2&0681) & (1&0300) &  \multicolumn{4}{c}{} \\
    $RV_{\text{sb}}^{(\text{rw})} + GPR_{\text{sb,imp}}^{(\text{rw})}$ & 0&0288 & 0&0734$^{***}$ & 0&9234$^{***}$ & $-$0&0092 & $-$0&9690 & $-$0&0073 & $-$1&4869 & 5&2515$^{*}$ & 3&7744$^{**}$ & 43.5217 & $-$9344.45 & 18705.63 & 18765.94 \\
    & (0&0187) & (0&0107) & (0&0086) & (0&0157) & (1&5421) & (0&0132) & (1&2752) & (2&8521) & (1&8716) &  \multicolumn{4}{c}{} \\
    $RV_{\text{sb}}^{(\text{rw})} + GPR_{\text{sb,exp}}^{(\text{rw})}$ & 0&0280 & 0&0734$^{***}$ & 0&9238$^{***}$ & $-$0&0089 & 0&1188 & $-$0&0067 & $-$1&7842 & 5&8282$^{**}$ & 2&2299$^{**}$ & 55.7476 & $-$9344.03 & 18704.79 & 18765.09 \\
    & (0&0186) & (0&0105) & (0&0087) & (0&0153) & (0&6052) & (0&0113) & (1&3180) & (2&3236) & (1&0331) &  \multicolumn{4}{c}{} \\
  \bottomrule
    \end{tabular}
    }%
  \begin{flushleft}
    \footnotesize
    \justifying Note: This table provides the parameter estimates and goodness-of-fit statistics of different two-factor GJR-GARCH-MIDAS models for the soybean market, where Panels A and B denote the model estimation with fixed time span and rolling window, respectively.
    \end{flushleft} 
  \label{Tab:Two_Factor_Estimation_Soybean_Futures}%
\end{table}%

As can be seen from Table~\ref{Tab:Two_Factor_Estimation_Rice_Futures}, the estimated results of two-factor models for the rice market are mixed. Similar to the coefficients of realized volatility being insignificantly negative in single-factor models, the estimates of $\theta_{RV}$ remain negative in all of the two-factor models, but this negative impact is shown to be statistically significant in the model that includes both $RV_{\text{rc}}^{(\text{rw})}$ and $GPR_{\text{MEA}}^{(\text{rw})}$. Accordingly, realized volatility negatively influences the long-term volatility of the international rice market, meaning that the higher the current realized volatility, the lower the long-term rice market volatility. This finding is quite different from the results in the wheat, maize, and soybean markets, which can be attributed to the peculiarity of the rice market. Due to factors such as climate and dietary habits, the majority of the main producers and consumers of rice are concentrated in the Asian region. According to USDA statistics, rice production and consumption in Asia account for 89.46\% and 84.25\% of global production and consumption in 2021, respectively. In other words, the self-sufficiency rate of rice in Asian economies is generally high, leading to a relatively simple consumption structure of rice with stable production and consumption. Moreover, rice is less internationalized and financialized than the other three staple grains, primarily reflected in the low trading volume of rice futures and weak market liquidity. 

Further analysis of geopolitical risk factors reveals that in these two-factor models, the estimates of $\theta_{MV}$ associated with global geopolitical risk indices are no longer significant, and only the coefficients of $GPR_{\text{NA}}^{(\text{rw})}$ and $GPR_{\text{MEA}}^{(\text{rw})}$ exhibit statistical significance. However, geopolitical risks in the two regions have opposite effects on the long-term volatility of the international rice market, in which the geopolitical risk in North America exacerbates long-run rice market volatility while the geopolitical risk in the Middle East and Africa tends to reduce long-term volatility. The United States in North America is one of the world's major rice exporters, whereas the Middle East and Africa heavily rely on rice imports, making up 47.38\% of global imports in 2021. This result is similar to the different impacts of geopolitical risks in South America as well as the Middle East and Africa on wheat market fluctuations. Additionally, in line with the significance of parameters, the two-factor model with both $RV_{\text{rc}}^{(\text{rw})}$ and $GPR_{\text{MEA}}^{(\text{rw})}$ provides the best goodness-of-fit as well as the strongest explanatory power, followed by the model with both $RV_{\text{rc}}^{(\text{rw})}$ and $GPR_{\text{NA}}^{(\text{rw})}$.

\begin{table}[!ht]
  \centering
  \setlength{\abovecaptionskip}{0pt}
  \setlength{\belowcaptionskip}{10pt}
  \caption{Estimation of two-factor models with fixed time span and rolling window for rice}
  \setlength\tabcolsep{3pt}   \resizebox{\textwidth}{!}{ 
    \begin{tabular}{l r@{.}l r@{.}l r@{.}l r@{.}l r@{.}l r@{.}l r@{.}l r@{.}l r@{.}l rccc}
    \toprule
         & \multicolumn{2}{c}{$\mu$} & \multicolumn{2}{c}{$\alpha$} & \multicolumn{2}{c}{$\beta$} & \multicolumn{2}{c}{$\gamma$} & \multicolumn{2}{c}{$m$} & \multicolumn{2}{c}{$\theta_{RV}$} & \multicolumn{2}{c}{$\theta_{MV}$} & \multicolumn{2}{c}{$\omega_{RV}$} & \multicolumn{2}{c}{$\omega_{MV}$} & VR (\%) & LLF & AIC & BIC \\
    \midrule
    \multicolumn{23}{l}{\textit{Panel A: Two-factor models with fixed time span}} \\
    $RV_{\text{rc}} + GPR$ & 0&0506$^{***}$ & 0&0700$^{***}$ & 0&9122$^{***}$ & 0&0060 & $-$5&2930 & $-$0&0025 & 1&4418 & 7&6254$^{**}$ & 1&2757$^{**}$ & 22.4529 & $-$9660.38 & 19337.55 & 19397.85 \\
    & (0&0195) & (0&0129) & (0&0129) & (0&0202) & (5&3029) & (0&0026) & (1&1547) & (3&1888) & (0&5188) &  \multicolumn{4}{c}{} \\ 
    $RV_{\text{rc}} + GPT$ & 0&0519$^{***}$ & 0&0687$^{***}$ & 0&9179$^{***}$ & 0&0055 & $-$0&8887 & $-$0&0035 & 0&5173 & 6&2027$^{**}$ & 1&0933$^{**}$ & 12.7252 & $-$9667.83 & 19352.44 & 19412.74 \\ 
    & (0&0195) & (0&0123) & (0&0127) & (0&0194) & (4&2704) & (0&0034) & (0&9380) & (2&8367) & (0&4466) &  \multicolumn{4}{c}{} \\ 
    $RV_{\text{rc}} + GPA$ & 0&0516$^{***}$ & 0&0694$^{***}$ & 0&9108$^{***}$ & 0&0068 & $-$3&0386 & $-$0&0021 & 0&9476$^{*}$ & 8&2742$^{**}$ & 1&5838 & 31.2164 & $-$9662.83 & 19342.44 & 19402.74 \\ 
    & (0&0197) & (0&0128) & (0&0141) & (0&0204) & (2&2979) & (0&0026) & (0&5171) & (3&4165) & (1&3891) &  \multicolumn{4}{c}{} \\
    $RV_{\text{rc}} + GPR_{\text{GLB}}$ & 0&0522$^{***}$ & 0&0689$^{***}$ & 0&9185$^{***}$ & 0&0044 & 2&8771 & $-$0&0037 & 1&0326 & 5&7666$^{**}$ & 1&0000 & 20.8734 & $-$9666.07 & 19348.93 & 19409.23 \\
    & (0&0195) & (0&0124) & (0&0134) & (0&0189) & (2&0631) & (0&0037) & (1&2871) & (2&4506) & (0&8998) &  \multicolumn{4}{c}{} \\
    $RV_{\text{rc}} + GPR_{\text{NA}}$ & 0&0494$^{**}$ & 0&0697$^{***}$ & 0&9135$^{***}$ & 0&0062 & 1&5917$^{***}$ & $-$0&0032 & 1&7345$^{*}$ & 6&5957$^{**}$ & 1&2307$^{***}$ & 37.5774 & $-$9659.93 & 19336.64 & 19396.94 \\
    & (0&0195) & (0&0132) & (0&0132) & (0&0197) & (0&4557) & (0&0030) & (0&9615) & (2&7696) & (0&3878) &  \multicolumn{4}{c}{} \\
    $RV_{\text{rc}} + GPR_{\text{SA}}$ & 0&0491$^{**}$ & 0&0667$^{***}$ & 0&9175$^{***}$ & 0&0080 & 4&1198 & $-$0&0046 & 0&7431 & 4&5122 & 1&0000$^{*}$ & 13.6107 & $-$9665.91 & 19348.60 & 19408.90 \\
    & (0&0195) & (0&0131) & (0&0129) & (0&0198) & (2&6230) & (0&0037) & (0&6978) & (3&1184) & (0&5119) &  \multicolumn{4}{c}{} \\
    $RV_{\text{rc}} + GPR_{\text{ENE}}$ & 0&0533$^{***}$ & 0&0682$^{***}$ & 0&9196$^{***}$ & 0&0052 & 2&1225$^{**}$ & $-$0&0040 & 0&3584 & 5&6938$^{**}$ & 1&0000 & 18.2009 & $-$9668.87 & 19354.53 & 19414.83 \\
    & (0&0198) & (0&0125) & (0&0123) & (0&0195) & (1&0438) & (0&0039) & (0&5739) & (2&5600) & (2&3151) &  \multicolumn{4}{c}{} \\
    $RV_{\text{rc}} + GPR_{\text{ESW}}$ & 0&0524$^{***}$ & 0&0689$^{***}$ & 0&9199$^{***}$ & 0&0045 & 2&9257$^{*}$ & $-$0&0041 & 0&8266 & 5&6491$^{**}$ & 1&0554 & 24.9687 & $-$9667.22 & 19351.24 & 19411.54 \\
    & (0&0197) & (0&0124) & (0&0140) & (0&0187) & (1&7505) & (0&0044) & (0&8336) & (2&6986) & (0&7234) &  \multicolumn{4}{c}{} \\
    $RV_{\text{rc}} + GPR_{\text{MEA}}$ & 0&0530$^{***}$ & 0&0609$^{***}$ & 0&9305$^{***}$ & 0&0108 & $-$0&7891 & $-$0&0164 & $-$2&1833$^{*}$ & 2&0446$^{**}$ & 4&3901$^{**}$ & 65.3965 & $-$9661.66 & 19340.11 & 19400.41 \\
    & (0&0200) & (0&0113) & (0&0125) & (0&0182) & (1&0723) & (0&0144) & (1&3106) & (0&9854) & (2&1656) &  \multicolumn{4}{c}{} \\
    $RV_{\text{rc}} + GPR_{\text{AO}}$ & 0&0527$^{***}$ & 0&0683$^{***}$ & 0&9189$^{***}$ & 0&0060 & 2&0706 & $-$0&0040 & 0&2812 & 5&6136 & 1&0003 & 14.4384 & $-$9669.11 & 19355.02 & 19415.32 \\
    & (0&0196) & (0&0137) & (0&0178) & (0&0193) & (2&5112) & (0&0058) & (1&0063) & (4&0661) & (6&4248) &  \multicolumn{4}{c}{} \\
    $RV_{\text{rc}} + GPR_{\text{rc,prod}}$ & 0&0504$^{***}$ & 0&0694$^{***}$ & 0&9168$^{***}$ & 0&0055 & 2&6623$^{**}$ & $-$0&0036 & 1&4677 & 6&0516$^{**}$ & 1&0544$^{**}$ & 26.8682 & $-$9663.56 & 19343.92 & 19404.22 \\
    & (0&0195) & (0&0130) & (0&0132) & (0&0193) & (1&1139) & (0&0033) & (1&0452) & (2&6754) & (0&4526) &  \multicolumn{4}{c}{} \\
    $RV_{\text{rc}} + GPR_{\text{rc,imp}}$ & 0&0530$^{***}$ & 0&0684$^{***}$ & 0&9202$^{***}$ & 0&0042 & 2&8944 & $-$0&0040 & 0&7452 & 5&7170$^{**}$ & 1&2143 & 19.5436 & $-$9667.95 & 19352.70 & 19413.00 \\
    & (0&0197) & (0&0122) & (0&0135) & (0&0189) & (2&1941) & (0&0041) & (1&0502) & (2&6562) & (0&7737) &  \multicolumn{4}{c}{} \\
    $RV_{\text{rc}} + GPR_{\text{rc,exp}}$ & 0&0516$^{***}$ & 0&0690$^{***}$ & 0&9180$^{***}$ & 0&0052 & 2&3982$^{*}$ & $-$0&0037 & 1&3586 & 5&9454$^{**}$ & 1&0000 & 25.4850 & $-$9664.91 & 19346.62 & 19406.92 \\
    & (0&0195) & (0&0126) & (0&0136) & (0&0190) & (1&2864) & (0&0037) & (1&3537) & (2&6562) & (0&7036) &  \multicolumn{4}{c}{} 
    \vspace{2mm}\\
    
    \multicolumn{23}{l}{\textit{Panel B: Two-factor models with rolling window}} \\
    $RV_{\text{rc}}^{(\text{rw})} + GPR^{(\text{rw})}$ & 0&0501$^{**}$ & 0&0684$^{***}$ & 0&9163$^{***}$ & 0&0066 & $-$3&3795 & $-$0&0030 & 1&0472 & 8&2439$^{*}$ & 1&2716$^{*}$ & 18.1697 & $-$9588.29 & 19193.30 & 19253.60 \\
    & (0&0197) & (0&0131) & (0&0161) & (0&0199) & (8&8711) & (0&0036) & (1&8727) & (4&2940) & (0&6616) &  \multicolumn{4}{c}{} \\
    $RV_{\text{rc}}^{(\text{rw})} + GPT^{(\text{rw})}$ & 0&0512$^{***}$ & 0&0660$^{***}$ & 0&9229$^{***}$ & 0&0062 & $-$0&5259 & $-$0&0049 & 0&4787 & 6&0801$^{*}$ & 1&0001 & 22.6859 & $-$9592.44 & 19201.60 & 19261.91 \\ 
     & (0&0196) & (0&0129) & (0&0155) & (0&0189) & (7&3827) & (0&0057) & (1&6783) & (3&4382) & (2&8086) &  \multicolumn{4}{c}{} \\
    $RV_{\text{rc}}^{(\text{rw})} + GPA^{(\text{rw})}$ & 0&0498$^{**}$ & 0&0681$^{***}$ & 0&9146$^{***}$ & 0&0071 & $-$2&2942 & $-$0&0026 & 0&8043 & 9&0207$^{**}$ & 1&7194 & 26.6632 & $-$9588.88 & 19194.48 & 19254.78 \\ 
    & (0&0197) & (0&0130) & (0&0138) & (0&0202) & (2&7290) & (0&0030) & (0&5870) & (3&6892) & (1&1348) &  \multicolumn{4}{c}{} \\
    $RV_{\text{rc}}^{(\text{rw})} + GPR_{\text{GLB}}^{(\text{rw})}$ & 0&0502$^{**}$ & 0&0666$^{***}$ & 0&9217$^{***}$ & 0&0066 & 2&7871$^{*}$ & $-$0&0044 & 0&8644 & 6&6833$^{**}$ & 1&0000$^{*}$ & 24.2209 & $-$9591.61 & 19199.95 & 19260.25 \\
    & (0&0196) & (0&0128) & (0&0138) & (0&0188) & (1&5872) & (0&0043) & (0&9858) & (3&0979) & (0&5687) &  \multicolumn{4}{c}{} \\
    $RV_{\text{rc}}^{(\text{rw})} + GPR_{\text{NA}}^{(\text{rw})}$ & 0&0480$^{**}$ & 0&0671$^{***}$ & 0&9166$^{***}$ & 0&0085 & 1&6578$^{***}$ & $-$0&0035 & 1&6796$^{*}$ & 7&7362$^{**}$ & 1&2968$^{***}$ & 36.3308 & $-$9586.19 & 19189.11 & 19249.41 \\
    & (0&0195) & (0&0132) & (0&0131) & (0&0195) & (0&4585) & (0&0033) & (0&9887) & (3&4771) & (0&3635) &  \multicolumn{4}{c}{} \\
    $RV_{\text{rc}}^{(\text{rw})} + GPR_{\text{SA}}^{(\text{rw})}$ & 0&0466$^{**}$ & 0&0641$^{***}$ & 0&9207$^{***}$ & 0&0112 & 3&6313 & $-$0&0046 & 0&5811 & 6&1158$^{*}$ & 1&0788$^{*}$ & 16.3323 & $-$9591.07 & 19198.87 & 19259.17 \\
    & (0&0196) & (0&0130) & (0&0132) & (0&0197) & (2&3595) & (0&0038) & (0&6586) & (3&2710) & (0&6026) &  \multicolumn{4}{c}{} \\
    $RV_{\text{rc}}^{(\text{rw})} + GPR_{\text{ENE}}^{(\text{rw})}$ & 0&0519$^{***}$ & 0&0648$^{***}$ & 0&9247$^{***}$ & 0&0066 & 2&0649$^{**}$ & $-$0&0053 & 0&2112 & 5&8488$^{*}$ & 1&0075 & 28.5879 & $-$9593.58 & 19203.88 & 19264.18 \\
    & (0&0197) & (0&0129) & (0&0152) & (0&0187) & (1&0519) & (0&0061) & (0&4455) & (3&1449) & (0&6969) &  \multicolumn{4}{c}{} \\
    $RV_{\text{rc}}^{(\text{rw})} + GPR_{\text{ESW}}^{(\text{rw})}$ & 0&0510$^{***}$ & 0&0664$^{***}$ & 0&9228$^{***}$ & 0&0061 & 2&7022$^{*}$ & $-$0&0047 & 0&6293 & 6&3636$^{**}$ & 1&3517$^{**}$ & 28.1194 & $-$9592.38 & 19201.48 & 19261.78 \\
    & (0&0197) & (0&0129) & (0&0146) & (0&0187) & (1&4166) & (0&0050) & (0&6870) & (3&1487) & (0&6729) &  \multicolumn{4}{c}{} \\
    $RV_{\text{rc}}^{(\text{rw})} + GPR_{\text{MEA}}^{(\text{rw})}$ & 0&0487$^{**}$ & 0&0583$^{***}$ & 0&9324$^{***}$ & 0&0156 & $-$1&2434 & $-$0&0193$^{**}$ & $-$2&9670$^{***}$ & 2&3370$^{***}$ & 4&3333$^{***}$ & 76.8306 & $-$9582.38 & 19181.49 & 19241.79 \\
    & (0&0199) & (0&0108) & (0&0098) & (0&0181) & (1&2379) & (0&0098) & (1&1180) & (0&6370) & (1&3827) &  \multicolumn{4}{c}{} \\
    $RV_{\text{rc}}^{(\text{rw})} + GPR_{\text{AO}}^{(\text{rw})}$ & 0&0487$^{**}$ & 0&0658$^{***}$ & 0&9228$^{***}$ & 0&0072 & 2&1684$^{**}$ & $-$0&0048 & 0&2588 & 6&2629$^{*}$ & 1&0002 & 22.3138 & $-$9593.33 & 19203.39 & 19263.69 \\
    & (0&0197) & (0&0129) & (0&0139) & (0&0189) & (0&8544) & (0&0047) & (0&3693) & (3&2782) & (2&5749) &  \multicolumn{4}{c}{} \\
    $RV_{\text{rc}}^{(\text{rw})} + GPR_{\text{rc,prod}}^{(\text{rw})}$ & 0&0501$^{**}$ & 0&0688$^{***}$ & 0&9191$^{***}$ & 0&0053 & 2&7234$^{**}$ & $-$0&0040 & 1&4272 & 7&0497$^{**}$ & 1&0790$^{**}$ & 27.4505 & $-$9589.23 & 19195.19 & 19255.49 \\
    & (0&0195) & (0&0136) & (0&0139) & (0&0191) & (1&1097) & (0&0036) & (1&0665) & (3&1862) & (0&5467) &  \multicolumn{4}{c}{} \\
    $RV_{\text{rc}}^{(\text{rw})} + GPR_{\text{rc,imp}}^{(\text{rw})}$ & 0&0521$^{***}$ & 0&0684$^{***}$ & 0&9218$^{***}$ & 0&0035 & 2&9050 & $-$0&0045 & 0&6890 & 6&5698$^{**}$ & 1&2324$^{*}$ & 23.9401 & $-$9592.90 & 19202.52 & 19262.82 \\
    & (0&0197) & (0&0132) & (0&0144) & (0&0189) & (2&1851) & (0&0045) & (1&0670) & (3&0662) & (0&6711) &  \multicolumn{4}{c}{} \\
    $RV_{\text{rc}}^{(\text{rw})} + GPR_{\text{rc,exp}}^{(\text{rw})}$ & 0&0493$^{**}$ & 0&0674$^{***}$ & 0&9211$^{***}$ & 0&0050 & 2&3598$^{**}$ & $-$0&0043 & 1&1561 & 6&8134$^{**}$ & 1&0015$^{**}$ & 26.4931 & $-$9590.88 & 19198.49 & 19258.79 \\
    & (0&0195) & (0&0131) & (0&0140) & (0&0187) & (1&0218) & (0&0043) & (1&0640) & (3&3477) & (0&4323) &  \multicolumn{4}{c}{} \\
  \bottomrule
    \end{tabular}
    }%
  \begin{flushleft}
    \footnotesize
    \justifying Note: This table shows the estimated results of different two-factor GJR-GARCH-MIDAS models for the rice market, where Panels A and B refer to the model estimation based on fixed time span and rolling window, respectively.
    \end{flushleft} 
  \label{Tab:Two_Factor_Estimation_Rice_Futures}%
\end{table}%

Further comparing the estimated results of single-factor models in Tables~\ref{Tab:Single_Factor_Estimation_Wheat_Futures}--\ref{Tab:Single_Factor_Estimation_Rice_Futures} with the estimated results of two-factor models in Tables~\ref{Tab:Two_Factor_Estimation_Wheat_Futures}--\ref{Tab:Two_Factor_Estimation_Rice_Futures}, respectively, we find that there is no clear superiority or inferiority between the single- and two-factor models, but in most cases, the two-factor models tend to provide higher explanatory power.

\section{Conclusions}
\label{S1:Conclude}

The current international situation is highly unpredictable, with escalating geopolitical risks globally and locally that have a huge impact on financial markets. Focusing on the food futures market, this paper empirically examines the influence of geopolitical risks across different dimensions on the volatility of the international wheat, maize, soybean, and rice markets, respectively. We utilize the global geopolitical risk, geopolitical threats, and geopolitical acts indices constructed by \cite{Caldara-Iacoviello-2022-AmEconRev}, as well as the individual sub-indices of geopolitical risk for 44 economies, to conduct our research. In general, food production exhibits distinct regional characteristics, yet these indices only measure geopolitical risks at global and economy-specific levels. To encompass different dimensions of geopolitical risks in our analysis, we first employ the random matrix theory to build regional geopolitical risk indices for six major geographic regions, namely North America, South America, Europe (North and East), Europe (South and West), Middle East and Africa, as well as Asia and Oceania. In addition, considering that production, import, and export are key aspects of food trade, we identify the top ten producers, importers, and exporters of the four staple food crops and then construct their respective composite geopolitical risk indices to measure the geopolitical risks faced by combinations of economies with global influence in production, import, and export.

Given the mixed frequencies between monthly geopolitical risk indices and daily agricultural futures prices, we adopt the MIDAS method to make full use of the valuable information in the data of different frequencies. Furthermore, the GJR-GARCH model is introduced into the MIDAS framework to take into account the possible asymmetries in short-term market fluctuations. We construct single-factor GJR-GARCH-MIDAS models based on fixed time span and rolling window, respectively, to decompose the overall volatility of agricultural markets into short- and long-run components. The results indicate that the single-factor models with rolling window outperform those with fixed time span in terms of goodness of fit. Regarding short-term fluctuations, the returns of wheat, maize, soybean, and rice show obvious volatility clustering, with their short-run volatility demonstrating high persistence but no significant asymmetry. Concerning long-term volatility, the realized volatility of wheat, maize, and soybean markets significantly intensifies the long-run volatility of their respective markets, whereas this is not the case for the rice market. Moreover, the performance of these four major grain markets varies when the explanatory variables in the single-factor models are designated as different geopolitical risk factors.

Specifically, a single geopolitical risk factor has limited explanatory power for long-term wheat market volatility, but performs well in explaining the long-term components of the maize and soybean markets. From a global perspective, both the global geopolitical threats index and our constructed global geopolitical risk index exert significant negative influences on long-run maize and soybean market volatility. This conclusion seems counterintuitive. One possible reason is that the occurrence of adverse geopolitical events leads to the growth of global risk aversion, which in turn affects investment behavior and reduces long-run fluctuations in the international food market. Additionally, the geopolitical risk indices we use are quantified based on newspapers, so the amount of information, the level of noise, and the accuracy of political signals in news tend to influence investors' decisions in the agricultural futures market, thereby affecting market volatility. At the regional level, composite geopolitical risks in North America, Europe, the Middle East and Africa, as well as Asia and Oceania considerably and negatively impact long-term maize market volatility, while composite geopolitical risks in North America, South America, and Europe exhibit significantly negative effects on the long-run fluctuations of the soybean market. Furthermore, both the wheat and maize markets are susceptible to geopolitical risks faced by various combinations of economies with global influence in production, import, and export. The composite geopolitical risks of major wheat exporters and major maize producers better explain volatility in the international maize and soybean markets, respectively. For rice, increasing global geopolitical risks and acts significantly exacerbate the long-term rice market volatility, with the geopolitical risk in North America serving as an important macro-factor in explaining fluctuations in the international rice market.

We further develop two-factor GJR-GARCH-MIDAS models based on fixed time span and rolling window, respectively, to simultaneously assess the impacts of realized volatility and geopolitical risk on the long-term volatility of international grain markets. Similar to single-factor models, the two-factor models with rolling window perform better. There is no clear superiority between single- and two-factor models, but in most cases, two-factor models provide higher explanatory power for overall market volatility. For wheat, realized volatility always plays an important role in increasing long-term wheat market volatility. After excluding the effect of realized volatility, geopolitical risks in South America and the Middle East and Africa exert markedly positive and negative influences on the long-run component of the wheat market, respectively, suggesting that the former significantly intensifies the long-term fluctuations while the latter does the opposite. For maize and soybean, only some coefficients of realized volatility in the two-factor models show significantly positive effects on long-term market volatility. Moreover, the influences of the RMT-based global geopolitical risk index, the regional geopolitical risk in Europe, and the composite geopolitical risks faced by combinations of major maize producers and exporters remain significantly negative, consistent with the results of single-factor models. Similar to the wheat market, elevated geopolitical risk in South America contributes to increasing long-term maize market volatility. In addition, the geopolitical risk faced by North America consistently exhibits a significantly and robustly negative effect on the long-run fluctuations of the international soybean market, with the two-factor model incorporating both realized volatility and geopolitical risk in North America demonstrating the strongest explanatory power for the expected volatility. For rice, realized volatility negatively impacts the long-term market volatility, which may be attributed to the uniqueness of the rice market. Furthermore, geopolitical risks in North America and the Middle East and Africa have equally significant but directionally opposite influences on the rice market volatility, with the former aggravating while the latter reducing the long-term volatility.

The global geopolitical landscape is experiencing profound shifts. As food is a fundamental necessity for human survival, food security is crucial for global stability, human well-being, and sustainable development. Therefore, effectively safeguarding global and local food security in the complex global situation is of utmost significance, where reducing the drastic fluctuations in food prices and maintaining the stable operation of food markets are important links. Faced with severe food insecurity, the international community, individual economies, and stakeholders should promote efforts to address the problem of high food price volatility, reject food trade protectionism, advocate food trade liberalization, and advance the transformation of agri-food systems in order to restore and ensure the efficient and smooth flow of food as soon as possible after adverse geopolitical events. From a regional perspective, there are huge differences in the degree of food insecurity and the level of geopolitical risk faced by different regions around the world. Relevant international organizations, such as FAO and IFAD, are expected to pay great attention to regions seriously affected by geopolitical conflicts and prone to sudden and severe food insecurity, call for immediate internationally coordinated action, and promote the in-depth development of multilateral and bilateral cooperation between countries and regions to help fragile economies cope with the growing threats to food security. Furthermore, futures exchanges ought to strengthen market research, enhance market supervision, and jointly curb excessive speculation in agricultural futures markets, so as to better play the unique role of agriculture futures in price discovery and risk hedging, especially in times of geopolitical turmoil. Considering the close connection between geopolitical risk and food market fluctuations, investors should keep up with current events and understand food market volatility when utilizing agricultural futures for asset allocation or risk management.

The factors affecting fluctuations in food prices are multifaceted, with complex underlying mechanisms. Closely linked to the global context, our study places particular emphasis on the impact of geopolitical risks on the volatility within the international food market. In fact, trade policy uncertainty and sluggish global economic recovery, along with natural disasters and extreme weather events caused by climate change, also contribute to both global and local food insecurity, potentially exerting a strong influence on the expected volatility of the international agricultural market. Hence, other macro-factors can be introduced into the analytical framework in future research to empirically investigate the effects of diverse low-frequency macro-factors, which is conducive to clarifying the sources of food market volatility. In addition, this paper decomposes the overall fluctuations in the food market into short- and long-term components, and constructs single- and two-factor GJR-GARCH-MIDAS models to analyze market volatility. Nevertheless, there is always room for improvement in models, such as changing the assumption of normal distribution for the error term or simultaneously incorporating more factors into the model.

\section*{Acknowledgment}

This work was supported by the National Natural Science Foundation of China (Grant Numbers: 72201099, 72171083), the China Postdoctoral Science Foundation (Grant Number: 2023T160217), and the Fundamental Research Funds for the Central Universities.

\section*{Data availability}

The dataset in this article is available at the \href{https://www.wind.com.cn}{Wind database}, the webpage of \href{https://www.matteoiacoviello.com/gpr.htm}{Geopolitical Risk (GPR) Index}, and the official website of \href{https://apps.fas.usda.gov/psdonline/app/index.html\#/app/downloads}{the United States Department of Agriculture}.

%

\begin{thebibliography}{66}
\expandafter\ifx\csname natexlab\endcsname\relax\def\natexlab#1{#1}\fi
\providecommand{\url}[1]{\texttt{#1}}
\providecommand{\href}[2]{#2}
\providecommand{\path}[1]{#1}
\providecommand{\DOIprefix}{doi:}
\providecommand{\ArXivprefix}{arXiv:}
\providecommand{\URLprefix}{URL: }
\providecommand{\Pubmedprefix}{pmid:}
\providecommand{\doi}[1]{\href{http://dx.doi.org/#1}{\path{#1}}}
\providecommand{\Pubmed}[1]{\href{pmid:#1}{\path{#1}}}
\providecommand{\bibinfo}[2]{#2}
\ifx\xfnm\relax \def\xfnm[#1]{\unskip,\space#1}\fi
\bibitem[{Abid et~al.(2023)Abid, Dhaoui, Kaabia and
  Tarchella}]{Abid-Dhaoui-Kaabia-Tarchella-2023-ResourPolicy}
\bibinfo{author}{Abid, I.}, \bibinfo{author}{Dhaoui, A.},
  \bibinfo{author}{Kaabia, O.}, \bibinfo{author}{Tarchella, S.},
  \bibinfo{year}{2023}.
\newblock \bibinfo{title}{Geopolitical risk on energy, agriculture, livestock,
  precious and industrial metals: {N}ew from a {M}arkov model}.
\newblock \bibinfo{journal}{Resour. Policy} \bibinfo{volume}{85},
  \bibinfo{pages}{103925}.
\newblock \DOIprefix\doi{10.1016/j.resourpol.2023.103925}.
\bibitem[{Ahmed et~al.(2022)Ahmed, Hasan and
  Kamal}]{Ahmed-Hasan-Kamal-2022-EurFinancManag}
\bibinfo{author}{Ahmed, S.}, \bibinfo{author}{Hasan, M.M.},
  \bibinfo{author}{Kamal, M.R.}, \bibinfo{year}{2022}.
\newblock \bibinfo{title}{Russia-{U}kraine crisis: {T}he effects on the
  {E}uropean stock market}.
\newblock \bibinfo{journal}{Eur. Financ. Manag.}
  \DOIprefix\doi{10.1111/eufm.12386}.
\bibitem[{Asgharian et~al.(2013)Asgharian, Hou and
  Javed}]{Asgharian-Hou-Javed-2013-JForecast}
\bibinfo{author}{Asgharian, H.}, \bibinfo{author}{Hou, A.J.},
  \bibinfo{author}{Javed, F.}, \bibinfo{year}{2013}.
\newblock \bibinfo{title}{The importance of the macroeconomic variables in
  forecasting stock return variance: {A} {GARCH}-{MIDAS} approach}.
\newblock \bibinfo{journal}{J. Forecast.} \bibinfo{volume}{32},
  \bibinfo{pages}{600--612}.
\newblock \DOIprefix\doi{10.1002/for.2256}.
\bibitem[{Baumeister et~al.(2022)Baumeister, Korobilis and
  Lee}]{Baumeister-Korobilis-Lee-2022-RevEconStat}
\bibinfo{author}{Baumeister, C.}, \bibinfo{author}{Korobilis, D.},
  \bibinfo{author}{Lee, T.K.}, \bibinfo{year}{2022}.
\newblock \bibinfo{title}{Energy markets and global economic conditions}.
\newblock \bibinfo{journal}{Rev. Econ. Stat.} \bibinfo{volume}{104},
  \bibinfo{pages}{828--844}.
\newblock \DOIprefix\doi{10.1162/rest_a_00977}.
\bibitem[{Bialkowski et~al.(2022)Bialkowski, Dang and
  Wei}]{Bialkowski-Dang-Wei-2022-JFinancEcon}
\bibinfo{author}{Bialkowski, J.}, \bibinfo{author}{Dang, H.D.},
  \bibinfo{author}{Wei, X.}, \bibinfo{year}{2022}.
\newblock \bibinfo{title}{High policy uncertainty and low implied market
  volatility: an academic puzzle?}
\newblock \bibinfo{journal}{J. Financ. Econ.} \bibinfo{volume}{143},
  \bibinfo{pages}{1185--1208}.
\newblock \DOIprefix\doi{10.1016/j.jfineco.2021.05.011}.
\bibitem[{Bollerslev(1986)}]{Bollerslev-1986-JEconom}
\bibinfo{author}{Bollerslev, T.}, \bibinfo{year}{1986}.
\newblock \bibinfo{title}{Generalized autoregressive conditional
  heteroskedasticity}.
\newblock \bibinfo{journal}{J. Econom.} \bibinfo{volume}{31},
  \bibinfo{pages}{307--327}.
\newblock \DOIprefix\doi{10.1016/0304-4076(86)90063-1}.
\bibitem[{Caldara and Iacoviello(2022)}]{Caldara-Iacoviello-2022-AmEconRev}
\bibinfo{author}{Caldara, D.}, \bibinfo{author}{Iacoviello, M.},
  \bibinfo{year}{2022}.
\newblock \bibinfo{title}{{Measuring Geopolitical Risk}}.
\newblock \bibinfo{journal}{Am. Econ. Rev.} \bibinfo{volume}{112},
  \bibinfo{pages}{1194--1225}.
\newblock \DOIprefix\doi{10.1257/aer.20191823}.
\bibitem[{Conrad and Kleen(2020)}]{Conrad-Kleen-2020-JApplEconom}
\bibinfo{author}{Conrad, C.}, \bibinfo{author}{Kleen, O.},
  \bibinfo{year}{2020}.
\newblock \bibinfo{title}{Two are better than one: {V}olatility forecasting
  using multiplicative component {GARCH}-{MIDAS} models}.
\newblock \bibinfo{journal}{J. Appl. Econom.} \bibinfo{volume}{35},
  \bibinfo{pages}{19--45}.
\newblock \DOIprefix\doi{10.1002/jae.2742}.
\bibitem[{Conrad and Loch(2015)}]{Conrad-Loch-2015-JApplEconom}
\bibinfo{author}{Conrad, C.}, \bibinfo{author}{Loch, K.}, \bibinfo{year}{2015}.
\newblock \bibinfo{title}{Anticipating long-term stock market volatility}.
\newblock \bibinfo{journal}{J. Appl. Econom.} \bibinfo{volume}{30},
  \bibinfo{pages}{1090--1114}.
\newblock \DOIprefix\doi{10.1002/jae.2404}.
\bibitem[{Conrad et~al.(2014)Conrad, Loch and
  Rittler}]{Conrad-Loch-Rittler-2014-JEmpirFinanc}
\bibinfo{author}{Conrad, C.}, \bibinfo{author}{Loch, K.},
  \bibinfo{author}{Rittler, D.}, \bibinfo{year}{2014}.
\newblock \bibinfo{title}{On the macroeconomic determinants of long-term
  volatilities and correlations in {US} stock and crude oil markets}.
\newblock \bibinfo{journal}{J. Empir. Financ.} \bibinfo{volume}{29},
  \bibinfo{pages}{26--40}.
\newblock \DOIprefix\doi{10.1016/j.jempfin.2014.03.009}.
\bibitem[{Conrad and Schienle(2020)}]{Conrad-Schienle-2020-JBusEconStat}
\bibinfo{author}{Conrad, C.}, \bibinfo{author}{Schienle, M.},
  \bibinfo{year}{2020}.
\newblock \bibinfo{title}{Testing for an omitted multiplicative long-term
  component in {GARCH} models}.
\newblock \bibinfo{journal}{J. Bus. Econ. Stat.} \bibinfo{volume}{38},
  \bibinfo{pages}{229--242}.
\newblock \DOIprefix\doi{10.1080/07350015.2018.1482759}.
\bibitem[{Dai et~al.(2022a)Dai, Xiong, Zhang and
  Zhou}]{Dai-Xiong-Zhang-Zhou-2022-ResourPolicy}
\bibinfo{author}{Dai, P.F.}, \bibinfo{author}{Xiong, X.},
  \bibinfo{author}{Zhang, J.}, \bibinfo{author}{Zhou, W.X.},
  \bibinfo{year}{2022}a.
\newblock \bibinfo{title}{The role of global economic policy uncertainty in
  predicting crude oil futures volatility: evidence from a two-factor
  {GARCH-MIDAS} model}.
\newblock \bibinfo{journal}{Resour. Policy} \bibinfo{volume}{78},
  \bibinfo{pages}{102849}.
\newblock \DOIprefix\doi{10.1016/j.resourpol.2022.102849}.
\bibitem[{Dai et~al.(2021)Dai, Xiong and
  Zhou}]{Dai-Xiong-Zhou-2021-FinancResLett}
\bibinfo{author}{Dai, P.F.}, \bibinfo{author}{Xiong, X.},
  \bibinfo{author}{Zhou, W.X.}, \bibinfo{year}{2021}.
\newblock \bibinfo{title}{A global economic policy uncertainty index from
  principal component analysis}.
\newblock \bibinfo{journal}{Financ. Res. Lett.} \bibinfo{volume}{40},
  \bibinfo{pages}{101686}.
\newblock \DOIprefix\doi{10.1016/j.frl.2020.101686}.
\bibitem[{Dai et~al.(2016)Dai, Xie, Jiang, Jiang and
  Zhou}]{Dai-Xie-Jiang-Jiang-Zhou-2016-EmpirEcon}
\bibinfo{author}{Dai, Y.H.}, \bibinfo{author}{Xie, W.J.},
  \bibinfo{author}{Jiang, Z.Q.}, \bibinfo{author}{Jiang, G.J.},
  \bibinfo{author}{Zhou, W.X.}, \bibinfo{year}{2016}.
\newblock \bibinfo{title}{Correlation structure and principal components in the
  global crude oil market}.
\newblock \bibinfo{journal}{Empir. Econ.} \bibinfo{volume}{51},
  \bibinfo{pages}{1501--1519}.
\newblock \DOIprefix\doi{10.1007/s00181-015-1057-1}.
\bibitem[{Dai et~al.(2023)Dai, Dai and
  Zhou}]{Dai-Dai-Zhou-2023-JIntFinancMarkInstMoney}
\bibinfo{author}{Dai, Y.S.}, \bibinfo{author}{Dai, P.F.},
  \bibinfo{author}{Zhou, W.X.}, \bibinfo{year}{2023}.
\newblock \bibinfo{title}{Tail dependence structure and extreme risk spillover
  effects between the international agricultural futures and spot markets}.
\newblock \bibinfo{journal}{J. Int. Financ. Mark. Inst. Money}
  \bibinfo{volume}{88}, \bibinfo{pages}{101820}.
\newblock \DOIprefix\doi{10.1016/j.intfin.2023.101820}.
\bibitem[{Dai et~al.(2022b)Dai, Huynh, Zheng and
  Zhou}]{Dai-Huynh-Zheng-Zhou-2022-ResIntBusFinanc}
\bibinfo{author}{Dai, Y.S.}, \bibinfo{author}{Huynh, N.Q.A.},
  \bibinfo{author}{Zheng, Q.H.}, \bibinfo{author}{Zhou, W.X.},
  \bibinfo{year}{2022}b.
\newblock \bibinfo{title}{Correlation structure analysis of the global
  agricultural futures market}.
\newblock \bibinfo{journal}{Res. Int. Bus. Financ.} \bibinfo{volume}{61},
  \bibinfo{pages}{101677}.
\newblock \DOIprefix\doi{10.1016/j.ribaf.2022.101677}.
\bibitem[{Engle(1982)}]{Engle-1982-Econometrica}
\bibinfo{author}{Engle, R.F.}, \bibinfo{year}{1982}.
\newblock \bibinfo{title}{Autoregressive conditional heteroscedasticity with
  estimates of the variance of {U}nited {K}ingdom inflation}.
\newblock \bibinfo{journal}{Econometrica} \bibinfo{volume}{50},
  \bibinfo{pages}{987--1007}.
\bibitem[{Engle et~al.(2013)Engle, Ghysels and
  Sohn}]{Engle-Ghysels-Sohn-2013-RevEconStat}
\bibinfo{author}{Engle, R.F.}, \bibinfo{author}{Ghysels, E.},
  \bibinfo{author}{Sohn, B.}, \bibinfo{year}{2013}.
\newblock \bibinfo{title}{Stock market volatility and macroeconomic
  fundamentals}.
\newblock \bibinfo{journal}{Rev. Econ. Stat.} \bibinfo{volume}{95},
  \bibinfo{pages}{776--797}.
\newblock \DOIprefix\doi{10.1162/REST_a_00300}.
\bibitem[{Engle et~al.(1987)Engle, Lilien and
  Robins}]{Engle-Lilien-Robins-1987-Econometrica}
\bibinfo{author}{Engle, R.F.}, \bibinfo{author}{Lilien, D.M.},
  \bibinfo{author}{Robins, R.P.}, \bibinfo{year}{1987}.
\newblock \bibinfo{title}{Estimating time varying risk premia in the term
  structure: {T}he {A}rch-{M} model}.
\newblock \bibinfo{journal}{Econometrica} \bibinfo{volume}{55},
  \bibinfo{pages}{391--407}.
\bibitem[{Engle and Rangel(2008)}]{Engle-Rangel-2008-RevFinancStud}
\bibinfo{author}{Engle, R.F.}, \bibinfo{author}{Rangel, J.G.},
  \bibinfo{year}{2008}.
\newblock \bibinfo{title}{The spline-{GARCH} model for low-frequency volatility
  and its global macroeconomic causes}.
\newblock \bibinfo{journal}{Rev. Financ. Stud.} \bibinfo{volume}{21},
  \bibinfo{pages}{1187--1222}.
\newblock \DOIprefix\doi{10.1093/rfs/hhn004}.
\bibitem[{Fang et~al.(2018)Fang, Chen, Yu and
  Qian}]{Fang-Chen-Yu-Qian-2018-JFuturesMark}
\bibinfo{author}{Fang, L.}, \bibinfo{author}{Chen, B.}, \bibinfo{author}{Yu,
  H.}, \bibinfo{author}{Qian, Y.}, \bibinfo{year}{2018}.
\newblock \bibinfo{title}{The importance of global economic policy uncertainty
  in predicting gold futures market volatility: {A GARCH-MIDAS} approach}.
\newblock \bibinfo{journal}{J. Futures Mark.} \bibinfo{volume}{38},
  \bibinfo{pages}{413--422}.
\newblock \DOIprefix\doi{10.1002/fut.21897}.
\bibitem[{Fang et~al.(2020)Fang, Lee and Su}]{Fang-Lee-Su-2020-JEmpirFinanc}
\bibinfo{author}{Fang, T.}, \bibinfo{author}{Lee, T.H.}, \bibinfo{author}{Su,
  Z.}, \bibinfo{year}{2020}.
\newblock \bibinfo{title}{Predicting the long-term stock market volatility: {A
  GARCH-MIDAS} model with variable selection}.
\newblock \bibinfo{journal}{J. Empir. Financ.} \bibinfo{volume}{58},
  \bibinfo{pages}{36--49}.
\newblock \DOIprefix\doi{10.1016/j.jempfin.2020.05.007}.
\bibitem[{Flannery and
  Protopapadakis(2002)}]{Flannery-Protopapadakis-2002-RevFinancStud}
\bibinfo{author}{Flannery, M.}, \bibinfo{author}{Protopapadakis, A.},
  \bibinfo{year}{2002}.
\newblock \bibinfo{title}{Macroeconomic factors do influence aggregate stock
  returns}.
\newblock \bibinfo{journal}{Rev. Financ. Stud.} \bibinfo{volume}{15},
  \bibinfo{pages}{751--782}.
\newblock \DOIprefix\doi{10.1093/rfs/15.3.751}.
\bibitem[{Francq et~al.(2023)Francq, Kandji and
  Zakoian}]{Francq-Kandji-Zakoian-2023-EconometTheory}
\bibinfo{author}{Francq, C.}, \bibinfo{author}{Kandji, B.M.},
  \bibinfo{author}{Zakoian, J.M.}, \bibinfo{year}{2023}.
\newblock \bibinfo{title}{Inference on {GARCH-MIDAS} models without any
  small-order moment}.
\newblock \bibinfo{journal}{Economet. Theory} , \bibinfo{pages}{PII
  S0266466623000142}\DOIprefix\doi{10.1017/S0266466623000142}.
\bibitem[{Fu(2009)}]{Fu-2009-JFinancEcon}
\bibinfo{author}{Fu, F.}, \bibinfo{year}{2009}.
\newblock \bibinfo{title}{Idiosyncratic risk and the cross-section of expected
  stock returns}.
\newblock \bibinfo{journal}{J. Financ. Econ.} \bibinfo{volume}{91},
  \bibinfo{pages}{24--37}.
\newblock \DOIprefix\doi{10.1016/j.jfineco.2008.02.003}.
\bibitem[{Ghysels et~al.(2005)Ghysels, Santa-Clara and
  Valkanov}]{Ghysels-SantaClara-Valkanov-2005-JFinancEcon}
\bibinfo{author}{Ghysels, E.}, \bibinfo{author}{Santa-Clara, P.},
  \bibinfo{author}{Valkanov, R.}, \bibinfo{year}{2005}.
\newblock \bibinfo{title}{There is a risk-return trade-off after all}.
\newblock \bibinfo{journal}{J. Financ. Econ.} \bibinfo{volume}{76},
  \bibinfo{pages}{509--548}.
\newblock \DOIprefix\doi{10.1016/j.jfineco.2004.03.008}.
\bibitem[{Ghysels et~al.(2006)Ghysels, Santa-Clara and
  Valkanov}]{Ghysels-SantaClara-Valkanov-2006-JEconom}
\bibinfo{author}{Ghysels, E.}, \bibinfo{author}{Santa-Clara, P.},
  \bibinfo{author}{Valkanov, R.}, \bibinfo{year}{2006}.
\newblock \bibinfo{title}{Predicting volatility: {G}etting the most out of
  return data sampled at different frequencies}.
\newblock \bibinfo{journal}{J. Econom.} \bibinfo{volume}{131},
  \bibinfo{pages}{59--95}.
\newblock \DOIprefix\doi{10.1016/j.jeconom.2005.01.004}.
\bibitem[{Ghysels et~al.(2007)Ghysels, Sinko and
  Valkanov}]{Ghysels-Sinko-Valkanov-2007-EconomRev}
\bibinfo{author}{Ghysels, E.}, \bibinfo{author}{Sinko, A.},
  \bibinfo{author}{Valkanov, R.}, \bibinfo{year}{2007}.
\newblock \bibinfo{title}{{MIDAS} regressions: {F}urther results and new
  directions}.
\newblock \bibinfo{journal}{Econom. Rev.} \bibinfo{volume}{26},
  \bibinfo{pages}{53--90}.
\newblock \DOIprefix\doi{10.1080/07474930600972467}.
\bibitem[{Girardi and Erguen(2013)}]{Girardi-Ergun-2013-JBankFinanc}
\bibinfo{author}{Girardi, G.}, \bibinfo{author}{Erguen, A.T.},
  \bibinfo{year}{2013}.
\newblock \bibinfo{title}{Systemic risk measurement: {M}ultivariate {GARCH}
  estimation of {C}o{V}a{R}}.
\newblock \bibinfo{journal}{J. Bank Financ.} \bibinfo{volume}{37},
  \bibinfo{pages}{3169--3180}.
\newblock \DOIprefix\doi{10.1016/j.jbankfin.2013.02.027}.
\bibitem[{Glosten et~al.(1993)Glosten, Jagannathan and
  Runkle}]{Glosten-Jagannathan-Runkle-1993-JFinanc}
\bibinfo{author}{Glosten, L.R.}, \bibinfo{author}{Jagannathan, R.},
  \bibinfo{author}{Runkle, D.E.}, \bibinfo{year}{1993}.
\newblock \bibinfo{title}{On the relation between the expected value and the
  volatility of the nominal excess return on stocks}.
\newblock \bibinfo{journal}{J. Financ.} \bibinfo{volume}{48},
  \bibinfo{pages}{1779--1801}.
\newblock \DOIprefix\doi{10.1111/j.1540-6261.1993.tb05128.x}.
\bibitem[{Gong et~al.(2022)Gong, Wang and
  Shao}]{Gong-Wang-Shao-2022-IntJFinancEcon}
\bibinfo{author}{Gong, X.}, \bibinfo{author}{Wang, M.}, \bibinfo{author}{Shao,
  L.}, \bibinfo{year}{2022}.
\newblock \bibinfo{title}{The impact of macro economy on the oil price
  volatility from the perspective of mixing frequency}.
\newblock \bibinfo{journal}{Int. J. Financ. Econ.} \bibinfo{volume}{27},
  \bibinfo{pages}{4487--4514}.
\newblock \DOIprefix\doi{10.1002/ijfe.2383}.
\bibitem[{Gong and Xu(2022)}]{Gong-Xu-2022-EnergyEcon}
\bibinfo{author}{Gong, X.}, \bibinfo{author}{Xu, J.}, \bibinfo{year}{2022}.
\newblock \bibinfo{title}{Geopolitical risk and dynamic connectedness between
  commodity markets}.
\newblock \bibinfo{journal}{Energy Econ.} \bibinfo{volume}{110},
  \bibinfo{pages}{106028}.
\newblock \DOIprefix\doi{10.1016/j.eneco.2022.106028}.
\bibitem[{Li et~al.(2023a)Li, Zhang and Li}]{Li-Zhang-Li-2023-IntRevFinancAnal}
\bibinfo{author}{Li, D.}, \bibinfo{author}{Zhang, L.}, \bibinfo{author}{Li,
  L.}, \bibinfo{year}{2023}a.
\newblock \bibinfo{title}{Forecasting stock volatility with economic policy
  uncertainty: {A} smooth transition {GARCH-MIDAS} model}.
\newblock \bibinfo{journal}{Int. Rev. Financ. Anal.} \bibinfo{volume}{88},
  \bibinfo{pages}{102708}.
\newblock \DOIprefix\doi{10.1016/j.irfa.2023.102708}.
\bibitem[{Li et~al.(2023b)Li, Bouri, Gupta and
  Fang}]{Li-Bouri-Gupta-Fang-2023-JCleanProd}
\bibinfo{author}{Li, H.}, \bibinfo{author}{Bouri, E.}, \bibinfo{author}{Gupta,
  R.}, \bibinfo{author}{Fang, L.}, \bibinfo{year}{2023}b.
\newblock \bibinfo{title}{Return volatility, correlation, and hedging of green
  and brown stocks: {I}s there a role for climate risk factors?}
\newblock \bibinfo{journal}{J. Clean Prod.} \bibinfo{volume}{414},
  \bibinfo{pages}{137594}.
\newblock \DOIprefix\doi{10.1016/j.jclepro.2023.137594}.
\bibitem[{Li et~al.(2022)Li, Wei, Chen, Ma, Liang and
  Chen}]{Li-Wei-Chen-Ma-Liang-Chen-2022-IntJFinancEcon}
\bibinfo{author}{Li, X.}, \bibinfo{author}{Wei, Y.}, \bibinfo{author}{Chen,
  X.}, \bibinfo{author}{Ma, F.}, \bibinfo{author}{Liang, C.},
  \bibinfo{author}{Chen, W.}, \bibinfo{year}{2022}.
\newblock \bibinfo{title}{Which uncertainty is powerful to forecast crude oil
  market volatility? {N}ew evidence}.
\newblock \bibinfo{journal}{Int. J. Financ. Econ.} \bibinfo{volume}{27},
  \bibinfo{pages}{4279--4297}.
\newblock \DOIprefix\doi{10.1002/ijfe.2371}.
\bibitem[{Li et~al.(2023c)Li, Ye, Bhuiyan and
  Huang}]{Li-Ye-Bhuiyan-Huang-2023-JForecast}
\bibinfo{author}{Li, X.}, \bibinfo{author}{Ye, C.}, \bibinfo{author}{Bhuiyan,
  M.A.}, \bibinfo{author}{Huang, S.}, \bibinfo{year}{2023}c.
\newblock \bibinfo{title}{Volatility forecasting with an extended {GARCH-MIDAS}
  approach}.
\newblock \bibinfo{journal}{J. Forecast.} ,
  \bibinfo{pages}{MISSING}\DOIprefix\doi{10.1002/for.3023}.
\bibitem[{Liang et~al.(2021)Liang, Ma, Wang and
  Zeng}]{Liang-Ma-Wang-Zeng-2021-JForecast}
\bibinfo{author}{Liang, C.}, \bibinfo{author}{Ma, F.}, \bibinfo{author}{Wang,
  L.}, \bibinfo{author}{Zeng, Q.}, \bibinfo{year}{2021}.
\newblock \bibinfo{title}{The information content of uncertainty indices for
  natural gas futures volatility forecasting}.
\newblock \bibinfo{journal}{J. Forecast.} \bibinfo{volume}{40},
  \bibinfo{pages}{1310--1324}.
\newblock \DOIprefix\doi{10.1002/for.2769}.
\bibitem[{Liang et~al.(2022)Liang, Zhang, Li and
  Ma}]{Liang-Zhang-Li-Ma-2022-IntJFinancEcon}
\bibinfo{author}{Liang, C.}, \bibinfo{author}{Zhang, Y.}, \bibinfo{author}{Li,
  X.}, \bibinfo{author}{Ma, F.}, \bibinfo{year}{2022}.
\newblock \bibinfo{title}{Which predictor is more predictive for {B}itcoin
  volatility? {A}nd why?}
\newblock \bibinfo{journal}{Int. J. Financ. Econ.} \bibinfo{volume}{27},
  \bibinfo{pages}{1947--1961}.
\newblock \DOIprefix\doi{10.1002/ijfe.2252}.
\bibitem[{Liu et~al.(2019)Liu, Ma, Tang and
  Zhang}]{Liu-Ma-Tang-Zhang-2019-EnergyEcon}
\bibinfo{author}{Liu, J.}, \bibinfo{author}{Ma, F.}, \bibinfo{author}{Tang,
  Y.}, \bibinfo{author}{Zhang, Y.}, \bibinfo{year}{2019}.
\newblock \bibinfo{title}{Geopolitical risk and oil volatility: {A} new
  insight}.
\newblock \bibinfo{journal}{Energy Econ.} \bibinfo{volume}{84},
  \bibinfo{pages}{104548}.
\newblock \DOIprefix\doi{10.1016/j.eneco.2019.104548}.
\bibitem[{Liu et~al.(2021)Liu, Han and Xu}]{Liu-Han-Xu-2021-IntRevFinancAnal}
\bibinfo{author}{Liu, Y.}, \bibinfo{author}{Han, L.}, \bibinfo{author}{Xu, Y.},
  \bibinfo{year}{2021}.
\newblock \bibinfo{title}{The impact of geopolitical uncertainty on energy
  volatility}.
\newblock \bibinfo{journal}{Int. Rev. Financ. Anal.} \bibinfo{volume}{75},
  \bibinfo{pages}{101743}.
\newblock \DOIprefix\doi{10.1016/j.irfa.2021.101743}.
\bibitem[{Lorente et~al.(2023)Lorente, Mohammed, Cifuentes-Faura and
  Shahzad}]{Lorente-Mohammed-CifuentesFaura-Shahzad-2023-RenewEnergy}
\bibinfo{author}{Lorente, D.B.}, \bibinfo{author}{Mohammed, K.S.},
  \bibinfo{author}{Cifuentes-Faura, J.}, \bibinfo{author}{Shahzad, U.},
  \bibinfo{year}{2023}.
\newblock \bibinfo{title}{Dynamic connectedness among climate change index,
  green financial assets and renewable energy markets: {N}ovel evidence from
  sustainable development perspective}.
\newblock \bibinfo{journal}{Renew. Energy} \bibinfo{volume}{204},
  \bibinfo{pages}{94--105}.
\newblock \DOIprefix\doi{10.1016/j.renene.2022.12.085}.
\bibitem[{Mandelbrot(1963)}]{Mandelbrot-1963-JBus}
\bibinfo{author}{Mandelbrot, B.B.}, \bibinfo{year}{1963}.
\newblock \bibinfo{title}{The variation of certain speculative prices}.
\newblock \bibinfo{journal}{J. Bus.} \bibinfo{volume}{36},
  \bibinfo{pages}{394--419}.
\newblock \DOIprefix\doi{10.1086/294632}.
\bibitem[{Mandelbrot(1967)}]{Mandelbrot-1967-JBus}
\bibinfo{author}{Mandelbrot, B.B.}, \bibinfo{year}{1967}.
\newblock \bibinfo{title}{The variation of some other speculative prices}.
\newblock \bibinfo{journal}{J. Bus.} \bibinfo{volume}{40},
  \bibinfo{pages}{393--413}.
\newblock \DOIprefix\doi{10.1086/295006}.
\bibitem[{Mo et~al.(2018)Mo, Gupta, Li and
  Singh}]{Mo-Gupta-Li-Singh-2018-EconModel}
\bibinfo{author}{Mo, D.}, \bibinfo{author}{Gupta, R.}, \bibinfo{author}{Li,
  B.}, \bibinfo{author}{Singh, T.}, \bibinfo{year}{2018}.
\newblock \bibinfo{title}{The macroeconomic determinants of commodity futures
  volatility: {E}vidence from {C}hinese and {I}ndian markets}.
\newblock \bibinfo{journal}{Econ. Model.} \bibinfo{volume}{70},
  \bibinfo{pages}{543--560}.
\newblock \DOIprefix\doi{10.1016/j.econmod.2017.08.032}.
\bibitem[{Nelson(1991)}]{Nelson-1991-Econometrica}
\bibinfo{author}{Nelson, D.B.}, \bibinfo{year}{1991}.
\newblock \bibinfo{title}{Conditional heteroskedasticity in asset returns: {A}
  new approach}.
\newblock \bibinfo{journal}{Econometrica} \bibinfo{volume}{59},
  \bibinfo{pages}{347--370}.
\bibitem[{Pan et~al.(2020)Pan, Bu, Liu and
  Wang}]{Pan-Bu-Liu-Wang-2020-QuantFinanc}
\bibinfo{author}{Pan, Z.}, \bibinfo{author}{Bu, R.}, \bibinfo{author}{Liu, L.},
  \bibinfo{author}{Wang, Y.}, \bibinfo{year}{2020}.
\newblock \bibinfo{title}{Macroeconomic fundamentals, jump dynamics and
  expected volatility}.
\newblock \bibinfo{journal}{Quant. Financ.} \bibinfo{volume}{20},
  \bibinfo{pages}{1345--1371}.
\newblock \DOIprefix\doi{10.1080/14697688.2020.1736317}.
\bibitem[{Pan et~al.(2017)Pan, Wang, Wu and
  Yin}]{Pan-Wang-Wu-Yin-2017-JEmpirFinanc}
\bibinfo{author}{Pan, Z.}, \bibinfo{author}{Wang, Y.}, \bibinfo{author}{Wu,
  C.}, \bibinfo{author}{Yin, L.}, \bibinfo{year}{2017}.
\newblock \bibinfo{title}{Oil price volatility and macroeconomic fundamentals:
  {A} regime switching {GARCH-MIDAS} model}.
\newblock \bibinfo{journal}{J. Empir. Financ.} \bibinfo{volume}{43},
  \bibinfo{pages}{130--142}.
\newblock \DOIprefix\doi{10.1016/j.jempfin.2017.06.005}.
\bibitem[{Plerou et~al.(2002)Plerou, Gopikrishnan, Rosenow, Amaral, Guhr and
  Stanley}]{Plerou-Gopikrishnan-Rosenow-Amaral-Guhr-Stanley-2002-PhysRevE}
\bibinfo{author}{Plerou, V.}, \bibinfo{author}{Gopikrishnan, P.},
  \bibinfo{author}{Rosenow, B.}, \bibinfo{author}{Amaral, L.},
  \bibinfo{author}{Guhr, T.}, \bibinfo{author}{Stanley, H.},
  \bibinfo{year}{2002}.
\newblock \bibinfo{title}{Random matrix approach to cross correlations in
  financial data}.
\newblock \bibinfo{journal}{Phys. Rev. E} \bibinfo{volume}{65},
  \bibinfo{pages}{066126}.
\newblock \DOIprefix\doi{10.1103/PhysRevE.65.066126}.
\bibitem[{Poertner et~al.(2022)Poertner, Lambrecht, Springmann, Bodirsky,
  Gaupp, Freund, Lotze-Campen and
  Gabrysch}]{Poertner-Lambrecht-Springmann-Bodirsky-Gaupp-Freund-LotzeCampen-Gabrysch-2022-OneEarth}
\bibinfo{author}{Poertner, L.M.}, \bibinfo{author}{Lambrecht, N.},
  \bibinfo{author}{Springmann, M.}, \bibinfo{author}{Bodirsky, B.L.},
  \bibinfo{author}{Gaupp, F.}, \bibinfo{author}{Freund, F.},
  \bibinfo{author}{Lotze-Campen, H.}, \bibinfo{author}{Gabrysch, S.},
  \bibinfo{year}{2022}.
\newblock \bibinfo{title}{We need a food system transformation-{I}n the face of
  the {R}ussia-{U}kraine war, now more than ever}.
\newblock \bibinfo{journal}{One Earth} \bibinfo{volume}{5},
  \bibinfo{pages}{470--472}.
\newblock \DOIprefix\doi{10.1016/j.oneear.2022.04.004}.
\bibitem[{Raza et~al.(2023)Raza, Masood, Benkraiem and
  Urom}]{Raza-Masood-Benkraiem-Urom-2023-EnergyEcon}
\bibinfo{author}{Raza, S.A.}, \bibinfo{author}{Masood, A.},
  \bibinfo{author}{Benkraiem, R.}, \bibinfo{author}{Urom, C.},
  \bibinfo{year}{2023}.
\newblock \bibinfo{title}{Forecasting the volatility of precious metals prices
  with global economic policy uncertainty in pre and during the {COVID}-19
  period: {N}ovel evidence from the {GARCH-MIDAS} approach}.
\newblock \bibinfo{journal}{Energy Econ.} \bibinfo{volume}{120},
  \bibinfo{pages}{106591}.
\newblock \DOIprefix\doi{10.1016/j.eneco.2023.106591}.
\bibitem[{Salisu et~al.(2023)Salisu, Demirer and
  Gupta}]{Salisu-Demirer-Gupta-2023-JForecast}
\bibinfo{author}{Salisu, A.A.}, \bibinfo{author}{Demirer, R.},
  \bibinfo{author}{Gupta, R.}, \bibinfo{year}{2023}.
\newblock \bibinfo{title}{Policy uncertainty and stock market volatility
  revisited: {T}he predictive role of signal quality}.
\newblock \bibinfo{journal}{J. Forecast.} \bibinfo{volume}{42},
  \bibinfo{pages}{2307--2321}.
\newblock \DOIprefix\doi{10.1002/for.3016}.
\bibitem[{Segnon et~al.(2024)Segnon, Gupta and
  Wilfling}]{Segnon-Gupta-Wilfling-2024-IntJForecast}
\bibinfo{author}{Segnon, M.}, \bibinfo{author}{Gupta, R.},
  \bibinfo{author}{Wilfling, B.}, \bibinfo{year}{2024}.
\newblock \bibinfo{title}{Forecasting stock market volatility with
  regime-switching {GARCH-MIDAS}: {T}he role of geopolitical risks}.
\newblock \bibinfo{journal}{Int. J. Forecast.} \bibinfo{volume}{40},
  \bibinfo{pages}{29--43}.
\newblock \DOIprefix\doi{10.1016/j.ijforecast.2022.11.007}.
\bibitem[{Su et~al.(2017)Su, Fang and Yin}]{Su-Fang-Yin-2017-EconLett}
\bibinfo{author}{Su, Z.}, \bibinfo{author}{Fang, T.}, \bibinfo{author}{Yin,
  L.}, \bibinfo{year}{2017}.
\newblock \bibinfo{title}{The role of news-based implied volatility among {US}
  financial markets}.
\newblock \bibinfo{journal}{Econ. Lett.} \bibinfo{volume}{157},
  \bibinfo{pages}{24--27}.
\newblock \DOIprefix\doi{10.1016/j.econlet.2017.05.028}.
\bibitem[{Tse and Tsui(2002)}]{Tse-Tsui-2002-JBusEconStat}
\bibinfo{author}{Tse, Y.}, \bibinfo{author}{Tsui, A.}, \bibinfo{year}{2002}.
\newblock \bibinfo{title}{A multivariate generalized autoregressive conditional
  heteroscedasticity model with time-varying correlations}.
\newblock \bibinfo{journal}{J. Bus. Econ. Stat.} \bibinfo{volume}{20},
  \bibinfo{pages}{351--362}.
\newblock \DOIprefix\doi{10.1198/073500102288618496}.
\bibitem[{Walther et~al.(2019)Walther, Klein and
  Bouri}]{Walther-Klein-Bouri-2019-JIntFinancMarkInstMoney}
\bibinfo{author}{Walther, T.}, \bibinfo{author}{Klein, T.},
  \bibinfo{author}{Bouri, E.}, \bibinfo{year}{2019}.
\newblock \bibinfo{title}{Exogenous drivers of {B}itcoin and {C}ryptocurrency
  volatility - {A} mixed data sampling approach to forecasting}.
\newblock \bibinfo{journal}{J. Int. Financ. Mark. Inst. Money}
  \bibinfo{volume}{63}, \bibinfo{pages}{101133}.
\newblock \DOIprefix\doi{10.1016/j.intfin.2019.101133}.
\bibitem[{Wang et~al.(2020)Wang, Ma, Liu and
  Yang}]{Wang-Ma-Liu-Yang-2020-IntJForecast}
\bibinfo{author}{Wang, L.}, \bibinfo{author}{Ma, F.}, \bibinfo{author}{Liu,
  J.}, \bibinfo{author}{Yang, L.}, \bibinfo{year}{2020}.
\newblock \bibinfo{title}{Forecasting stock price volatility: {N}ew evidence
  from the {GARCH}-{MIDAS} model}.
\newblock \bibinfo{journal}{Int. J. Forecast.} \bibinfo{volume}{36},
  \bibinfo{pages}{684--694}.
\newblock \DOIprefix\doi{10.1016/j.ijforecast.2019.08.005}.
\bibitem[{Wang et~al.(2023)Wang, Wu, Ma and
  Xu}]{Wang-Wu-Ma-Xu-2023-IntRevFinancAnal}
\bibinfo{author}{Wang, L.}, \bibinfo{author}{Wu, R.}, \bibinfo{author}{Ma, W.},
  \bibinfo{author}{Xu, W.}, \bibinfo{year}{2023}.
\newblock \bibinfo{title}{Examining the volatility of soybean market in the
  {MIDAS} framework: {T}he importance of bagging-based weather information}.
\newblock \bibinfo{journal}{Int. Rev. Financ. Anal.} \bibinfo{volume}{89},
  \bibinfo{pages}{102720}.
\newblock \DOIprefix\doi{10.1016/j.irfa.2023.102720}.
\bibitem[{Wei et~al.(2017)Wei, Liu, Lai and
  Hu}]{Wei-Liu-Lai-Hu-2017-EnergyEcon}
\bibinfo{author}{Wei, Y.}, \bibinfo{author}{Liu, J.}, \bibinfo{author}{Lai,
  X.}, \bibinfo{author}{Hu, Y.}, \bibinfo{year}{2017}.
\newblock \bibinfo{title}{Which determinant is the most informative in
  forecasting crude oil market volatility: {F}undamental, speculation, or
  uncertainty?}
\newblock \bibinfo{journal}{Energy Econ.} \bibinfo{volume}{68},
  \bibinfo{pages}{141--150}.
\newblock \DOIprefix\doi{10.1016/j.eneco.2017.09.016}.
\bibitem[{Wu and Liu(2023)}]{Wu-Liu-2023-JEnvironManage}
\bibinfo{author}{Wu, R.}, \bibinfo{author}{Liu, B.Y.}, \bibinfo{year}{2023}.
\newblock \bibinfo{title}{Do climate policy uncertainty and investor sentiment
  drive the dynamic spillovers among green finance markets?}
\newblock \bibinfo{journal}{J. Environ. Manage.} \bibinfo{volume}{347},
  \bibinfo{pages}{119008}.
\newblock \DOIprefix\doi{10.1016/j.jenvman.2023.119008}.
\bibitem[{Xu et~al.(2019)Xu, Bo, Jiang and
  Liu}]{Xu-Bo-Jiang-Liu-2019-Knowledge-BasedSyst}
\bibinfo{author}{Xu, Q.}, \bibinfo{author}{Bo, Z.}, \bibinfo{author}{Jiang,
  C.}, \bibinfo{author}{Liu, Y.}, \bibinfo{year}{2019}.
\newblock \bibinfo{title}{Does {G}oogle search index really help predicting
  stock market volatility? {E}vidence from a modified mixed data sampling model
  on volatility}.
\newblock \bibinfo{journal}{Knowledge-Based Syst.} \bibinfo{volume}{166},
  \bibinfo{pages}{170--185}.
\newblock \DOIprefix\doi{10.1016/j.knosys.2018.12.025}.
\bibitem[{You and Liu(2020)}]{You-Liu-2020-JBankFinanc}
\bibinfo{author}{You, Y.}, \bibinfo{author}{Liu, X.}, \bibinfo{year}{2020}.
\newblock \bibinfo{title}{Forecasting short-run exchange rate volatility with
  monetary fundamentals: {A GARCH-MIDAS} approach}.
\newblock \bibinfo{journal}{J. Bank Financ.} \bibinfo{volume}{116},
  \bibinfo{pages}{105849}.
\newblock \DOIprefix\doi{10.1016/j.jbankfin.2020.105849}.
\bibitem[{Zakoian(1994)}]{Zakoian-1994-JEconDynControl}
\bibinfo{author}{Zakoian, J.M.}, \bibinfo{year}{1994}.
\newblock \bibinfo{title}{Threshold heteroscedastic models}.
\newblock \bibinfo{journal}{J. Econ. Dyn. Control} \bibinfo{volume}{18},
  \bibinfo{pages}{931--955}.
\newblock \DOIprefix\doi{10.1016/0165-1889(94)90039-6}.
\bibitem[{Zhang et~al.(2023)Zhang, Hong and Ding}]{Zhang-Hong-Ding-2023-Energy}
\bibinfo{author}{Zhang, H.}, \bibinfo{author}{Hong, H.}, \bibinfo{author}{Ding,
  S.}, \bibinfo{year}{2023}.
\newblock \bibinfo{title}{The role of climate policy uncertainty on the
  long-term correlation between crude oil and clean energy}.
\newblock \bibinfo{journal}{Energy} \bibinfo{volume}{284},
  \bibinfo{pages}{128529}.
\newblock \DOIprefix\doi{10.1016/j.energy.2023.128529}.
\bibitem[{Zhang et~al.(2024)Zhang, Xiang, Zou and
  Guo}]{Zhang-Xiang-Zou-Guo-2024-IntRevFinancAnal}
\bibinfo{author}{Zhang, J.}, \bibinfo{author}{Xiang, Y.}, \bibinfo{author}{Zou,
  Y.}, \bibinfo{author}{Guo, S.}, \bibinfo{year}{2024}.
\newblock \bibinfo{title}{Volatility forecasting of {C}hinese energy market:
  {W}hich uncertainty have better performance?}
\newblock \bibinfo{journal}{Int. Rev. Financ. Anal.} \bibinfo{volume}{91},
  \bibinfo{pages}{102952}.
\newblock \DOIprefix\doi{10.1016/j.irfa.2023.102952}.
\bibitem[{Zhou et~al.(2024)Zhou, Dai, Duong and
  Dai}]{Zhou-Dai-Duong-Dai-2024-JEconBehavOrgan}
\bibinfo{author}{Zhou, W.X.}, \bibinfo{author}{Dai, Y.S.},
  \bibinfo{author}{Duong, K.T.}, \bibinfo{author}{Dai, P.F.},
  \bibinfo{year}{2024}.
\newblock \bibinfo{title}{The impact of the {R}ussia-{U}kraine conflict on the
  extreme risk spillovers between agricultural futures and spots}.
\newblock \bibinfo{journal}{J. Econ. Behav. Organ.} \bibinfo{volume}{217},
  \bibinfo{pages}{91--111}.
\newblock \DOIprefix\doi{10.1016/j.jebo.2023.11.004}.
\bibitem[{Zhou et~al.(2023)Zhou, Lu, Xu, Yan, Khu, Yang and
  Zhao}]{Zhou-Lu-Xu-Yan-Khu-Yang-Zhao-2023-ResourConservRecycl}
\bibinfo{author}{Zhou, X.Y.}, \bibinfo{author}{Lu, G.}, \bibinfo{author}{Xu,
  Z.}, \bibinfo{author}{Yan, X.}, \bibinfo{author}{Khu, S.T.},
  \bibinfo{author}{Yang, J.}, \bibinfo{author}{Zhao, J.}, \bibinfo{year}{2023}.
\newblock \bibinfo{title}{Influence of {R}ussia-{U}kraine war on the global
  energy and food security}.
\newblock \bibinfo{journal}{Resour. Conserv. Recycl.} \bibinfo{volume}{188},
  \bibinfo{pages}{106657}.
\newblock \DOIprefix\doi{10.1016/j.resconrec.2022.106657}.

\end{thebibliography}

\end{document}